\documentclass[a4paper,11pt,oneside,titlepage]{book}
\usepackage[T1]{fontenc}		
\usepackage[utf8]{inputenc}		
\usepackage[english]{babel}		

 \usepackage[a4paper]{geometry}
\usepackage{geometry}
\if@twoside\fi

\usepackage{setspace}			
\usepackage{fancyhdr}			
\usepackage{afterpage} 			
\usepackage{hyperref}			
\usepackage{url}
\usepackage{color}			
\usepackage{xcolor}			
\usepackage{enumerate}			
\usepackage{enumitem}			
\usepackage{graphicx}			
\usepackage{epstopdf}			
\usepackage{pstool}			
\usepackage{psfrag}
\usepackage{subfig}		
\usepackage{array}			
\usepackage{tabularx}			
\usepackage[bottom]{footmisc}		
\usepackage{booktabs}			
\usepackage{longtable}			
\usepackage{caption}			
\usepackage{float}			
\usepackage{rotating}			
\usepackage{rotfloat}			
\usepackage{amsmath} 			
\usepackage{amssymb}			
\usepackage{cancel}                     
\usepackage{mleftright}
\usepackage{listings}                   
\definecolor{lstgreen}{rgb}{0,0.6,0}
\definecolor{lstgray}{rgb}{0.5,0.5,0.5}
\definecolor{lstmauve}{rgb}{0.58,0,0.82}
\usepackage[swapnames,norules,nouppercase]{frontespizio}
\usepackage[intoc]{nomencl}
\usepackage[acronym,toc,nomain,nopostdot,nonumberlist]{glossaries}
\usepackage[square, numbers, comma, sort&compress]{natbib}
\usepackage{verbatim}                   
\usepackage{wrapfig}                 
\usepackage{blindtext}
\fancypagestyle{plain}{
                        \fancyhf{}                              
                        \fancyfoot[R]{\thepage}                 
                        \renewcommand{\headrulewidth}{0pt}      
                        \renewcommand{\footrulewidth}{0.4pt}    
                        }                                       
\hypersetup{colorlinks=true, linkcolor=black, citecolor=black, filecolor=black, urlcolor=black}
\definecolor{red-univaq}{RGB}{130,36,51} 
\definecolor{univaq}{RGB}{242,195,80} 
\makenomenclature 
\makeglossaries 
\newenvironment{dedication}
{
  \phantom{.}
  \vspace{13cm}
  \begin{quote} \begin{flushright}}
{\end{flushright} \end{quote}}
\usepackage{calligra}

\usepackage{multirow}
\usepackage{xcolor}
\usepackage{graphicx}
\usepackage{adjustbox}
\usepackage{tabularx}
\usepackage{lscape}
\usepackage{todonotes}
\usepackage{lipsum}
\usepackage{csquotes}
\usepackage{rotating}
\usepackage[paper=portrait,pagesize]{typearea}

\usepackage[T1]{fontenc}
\usepackage{helvet}

\begin{document}

\frontmatter 	   
\pagestyle{empty}  


\begin{center}
  \begin{figure*}[!h]
	\centering
	\makebox[\textwidth]
	{
	    \includegraphics[width=0.85\paperwidth]
	    {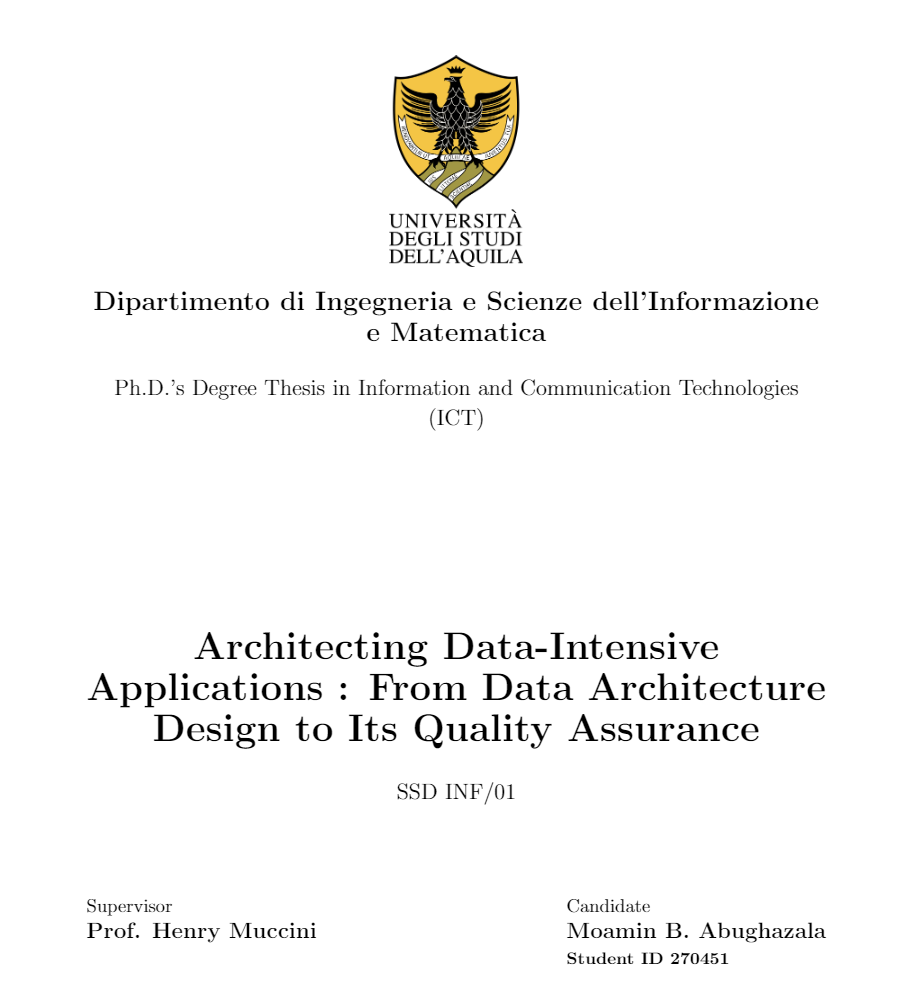}
	}
	\label{fig:dv_mm_s}
    \end{figure*}
\end{center} 
\clearpage

\begin{dedication}

{\fontfamily{calligra}\selectfont
{\Huge

The Sun Rises, Giving Light.
Data Rises, Giving Wisdom
}}

{\fontfamily{calligra}\selectfont
{\Large Feras M. Awaysheh}
}

{\fontfamily{calligra}\selectfont
{\Huge
By Having A Well Designed Data Architecture
}
}

{\fontfamily{calligra}\selectfont
{\Large Moamin Abughazala}
}

\end{dedication}

\newpage

\begin{center}
  \textbf{Declaration}  
\end{center}

\begin{table}[ht!]
    \centering
    \begin{tabular}{cccccccccc}
    &  &  &&  &  & &  &   & \\
    &  &  & &  &  &  &  & & \\

    \end{tabular}
    
    \label{tab:my_label}
\end{table}

I hereby declare that the thesis entitled "Architecting Data-Intensive Applications: From Data Architecture Design to Its Quality Assurance," which I am submitting as part of my doctoral program at the University of L'Aquila, is a bona fide representation of my work. This research was undertaken under the supervision of Professor Henry Muccini, in conjunction with my colleagues and co-authors. In the present thesis, the personal pronoun "I" is used instead of "We." I affirm that I have neither submitted nor will I submit this work, either in part or in its entirety, to obtain any other degree or diploma from this or any other academic institution.

\vfill

\textbf{Place: L'Aquila}  
\hfill       
\textbf{Moamin B. Abughazla}

Date: 04/01/2024
    
\newpage
\begin{center}
    \textbf{DEDICATION}
\end{center}

\begin{table}[ht!]
    \centering
    \begin{tabular}{cccccccccc}
    &  &  &&  &  & &  &   & \\
    &  &  & &  &  &  &  & & \\

    \end{tabular}
    
    \label{tab:my_label}
\end{table}

{\fontfamily{calligra}\selectfont
{\huge
To my dearest \textit{parents and beloved family members}, who have been the driving force behind my success and continue to inspire me with their unwavering love and support.

To my \textit{loving and supportive wife}, who has been my rock and my partner in every step of my journey. Her constant encouragement and belief in my dreams have been the fuel that keeps me going and striving toward excellence.

To my \textit{handsome and wonderful sons}, I wish nothing but the brightest and most fulfilling paths for your lives. 

}}

\newpage

\afterpage{\null\thispagestyle{empty}\clearpage} 
\section*{Abstract}
\textbf{Context -} 
The exponential growth of data is becoming a significant concern. Managing this data has become incredibly challenging, especially when dealing with various sources in different formats and speeds. Moreover, Ensuring data quality has become increasingly crucial for effective decision-making and operational processes. 
Data Architecture is crucial in describing, collecting, storing, processing, and analyzing data to meet business needs. Providing an abstract view of data-intensive applications is essential to ensure that the data is transformed into valuable information. We must take these challenges seriously to ensure we can effectively manage and use the data to our advantage.

\begin{flushleft}
\textbf{Objective - } To establish an architecture framework that enables a comprehensive description of the data architecture and effectively streamlines data quality monitoring.
\end{flushleft}

\begin{flushleft}
\textbf{Method -} 
The architecture framework utilizes Model Driven Engineering (MDE) techniques. Its backing of data-intensive architecture descriptions empowers with an automated generation for data quality checks.
\end{flushleft}
\begin{flushleft}
\textbf{Result -}  
The Framework offers a comprehensive solution for data-intensive applications to model their architecture efficiently and monitor the quality of their data. It automates the entire process and ensures precision and consistency in data. With DAT, architects and analysts gain access to a powerful tool that simplifies their workflow and empowers them to make informed decisions based on reliable data insights.
\end{flushleft}
\begin{flushleft}
\textbf{Conclusion -} 
We have evaluated the DAT on more than five cases within various
industry domains, demonstrating its exceptional adaptability and effectiveness.
\end{flushleft}
\vfill
\textbf{Keywords}: Data Architecture, Data Quality, Data-Intensive

\clearpage
\pagestyle{empty}



    

\section*{Publications}
 This thesis is firmly grounded in the publications described below:

\begin{enumerate}
    \item \textbf{Abughazala, M}, Muccini, H, Sharaf, M. "Architecting Data-intensive applications: from architectural design to data quality" \textit{Journal of Systems and Software} JSS (2023) (\textbf{Submitted)}
    
    \item Sharaf, M., \textbf{Abughazala, M}, Muccini, Mai Abusair,  "Architecting IoT applications: from architectural design to
realization" \textit{Journal of Systems and Software} JSS (2023) (\textbf{Submitted)}
    
    \item \textbf{Abughazala, Moamin}, Henry Muccini, Khitam, Qadri " \textit{PyDaQu: Python Data Quality Code Generation Based on Data Architecture }" Proceedings of the 26th International Conference on Model Driven Engineering Languages and Systems (\textbf{MODELS 2023}), \textit{Tool \& Demo}.
   
    \item \textbf{Abughazala, Moamin}, Henry Muccini, and M. Sharaf. “\textit{Architecture Description Framework For Data-Intensive Applications}” International Conference on Intelligent Data Science Technologies and Applications, (\textbf{IDSTA 2023}).
  
    \item \textbf{Abughazala, Moamin}, and Henry Muccini. "\textit{Modeling data analytics architecture for iot applications using dat.}" 2023 IEEE 20th International Conference on Software Architecture Companion (\textbf{ICSA-C}). IEEE, 2023.
 
    \item \textbf{Abughazala, Moamin}, Henry Muccini, and M. Sharaf. "\textit{Dat: Data architecture modeling tool for data-driven applications.}" European Conference on Software Architecture. (\textbf{ECSA 2022}), \textit{Tool \& Demo}.

    \item \textbf{Abughazala, Moamin}, and Henry Muccini. "\textit{Modeling Data Analytics Architecture for Smart Cities Data-Driven Applications using DAT.}" \textbf{I-Cities2023}.
 
    \item \textbf{Abughazala, Moamin}., Moghaddam, M. T., Muccini, H., \& Vaidhyanathan, K. (2021, August).\textit{ Human behavior-oriented architectural design. In European Conference on Software Architecture} (\textbf{ECSA 2021}) (pp. 134-143).

    \item \textbf{Abughazala, Moamin}, and Henry Muccini. "\textit{Data Architecture for Digital Object Space Management Service (DOSM) using DAT}." \textbf{I-Cities (2022)}.
    
\end{enumerate}
\clearpage
\pagestyle{empty}

\begin{singlespace}
 \tableofcontents 	
 \addcontentsline{toc}{chapter}{\listfigurename}
 \listoffigures
 \addcontentsline{toc}{chapter}{\listtablename}
 \listoftables

 \printnomenclature
 \printglossaries
\end{singlespace}

\mainmatter	  
\clearpage
\pagestyle{fancy}

\renewcommand{\chaptermark}[1]{\markright{\chaptername\ \thechapter.\ #1}{}}
\renewcommand{\sectionmark}[1]{\markright{\thesection.\ #1}}
\lhead{} 
\chead{}                   
\rhead{\slshape \rightmark} 
\lfoot{}
\cfoot{} 
\rfoot{\thepage}          
\renewcommand{\headrulewidth}{0.4pt} 
\renewcommand{\footrulewidth}{0.4pt} 

\chapter{Introduction}
\thispagestyle{plain}

\newacronym{dq}{DQ}{Data Quality}
\newacronym{dda}{DDA}{Data Driven Applications}
\newacronym{da}{DA}{Data Architecture}
\newacronym{mde}{MDE}{Model Driven Engineering}

Data's transformative impact will certainly extend to all areas of our lives, including our work and thoughts \cite{john2014big}. The quantity of data is rapidly increasing due to advancements in mobile and sensing devices, social media, and web technologies. For example, Twitter handles over 70 million tweets daily, generating over eight terabytes of data daily \cite{twiter}. Google generates around 2.5 million Terabytes per day. IDC expects data to reach 175 zettabytes by 2025 \cite{idc}. Data products are products designed to facilitate achieving a goal through the use of data. Google Analytics is an example of a data product whose primary purpose is to provide users with a quantitative understanding of online behavior. Instagram is not considered a data product, but its features, such as tagging and searching, can be classified as data products. These products include raw data, derived data, algorithms, decision support, and automated decision-making \cite{Towardsdatascience}. Data can be either raw or derived. Raw data is collected directly from the source and stored for later use, while derived data is processed from raw data. Algorithms are used to analyze data and provide insights. For instance, Google Image enables users to input an image and get similar images as output. The product matches images and provides similar ones. Decision support systems like Google Analytics help with decision-making, but users interpret results. Automated decision-making systems present the final output to the user. One instance of an automated decision-making system is the movie suggestions provided by Netflix \cite{Towardsdatascience}. 

Companies are utilizing the data they have gathered throughout the years for various purposes. Data enhances quality assurance and diagnostics, leading to a notable reduction in troubleshooting efforts. Additionally, some companies use data to improve features, functionality, and performance optimization.

\section{Data }
Data is a collection of information expressed in various forms, such as facts, measurements, or observations. It finds application in diverse fields, from business intelligence to analytics, artificial intelligence, and machine learning. The role of data in driving innovation and progress cannot be overstated. It is the foundation upon which IoT, personalization, healthcare, finance, transportation, social media, climate analysis, education, smart cities, scientific research, security, surveillance, and environmental monitoring are built. As technology advances, the importance and impact of data will only continue to grow, shaping our lives and transforming how we interact with the world.

\section{Data Architecture}

In the realm of information technology and data management, \emph{\gls{da}}  is a crucial concept. It involves {\em arranging, blueprinting, and managing an organization's data resources, encompassing how data is collected, stored, managed, and utilized} \cite{zheng2010data} \cite{INMON20191} \cite{10.5555/3165209}. A well-established \emph{\gls{da}} is essential for effectively managing and capitalizing on data to support an organization's strategic objectives, decision-making procedures, and overall operations \cite{sinan2022data}.

When designing a data architecture, it's important to consider factors such as the data's amount, variety, speed, and accuracy \cite{ji2020quality}
. Additionally, it's necessary to consider the technologies, tools, and platforms that will best fit the organization's requirements and align with its business goals. \emph{\gls{da}} is essential for managing internal data and integrating and exchanging data with external partners, vendors, and customers \cite{sinan2022data}.

Effective \emph{\gls{da}} is essential for advanced data analytics, business intelligence, and artificial intelligence applications \cite{zschornig2020iot}. It allows organizations to extract valuable insights from their data, make data-intensive decisions, and stay ahead of the competition in today's data-intensive world \cite{Rawat_2021} \cite{balusamy2021big}. \emph{\gls{da}} is essential in today's fast-paced digital world. It helps organizations utilize the power of data, fosters innovation, and guides them toward success \cite{almeida2013main}. It's an integral part of any comprehensive data management strategy and is constantly refined to adapt to changing business needs and emerging technologies  \cite{gray1996evolution}.

\section{Need For Data Architecture}
The significance of \emph{\gls{da}} lies in its ability to serve as a strategic framework that helps organizations \textit{manage their data} efficiently \cite{TalendDA} \cite{ibmDA}.

Effective \emph{\gls{da}} allows for \textit{integrating data} from various sources, systems, and applications, resulting in a cohesive data view. Such integration is essential for making informed decisions that rely on a comprehensive understanding of the available data \cite{ibmDA}.

A well-designed \emph{\gls{da}} is important because it establishes \textit{data governance} principles, \textit{data standards, and data validation rules}. This helps ensure the data is consistent, accurate, and high-quality, ultimately leading to better decision-making and more reliable insights \cite{INMON201933} \cite{ibmDA}. As the amount of data increases, proper \emph{\gls{da}} becomes essential to ensure that data storage, processing, and retrieval systems can handle the growing demands. By optimizing the flow of data, it leads to improved \textit{system performance} \cite{acceldataDA}.




A well-designed data architecture can \textit{reduce redundancies}, inefficiencies, and data silos, leading to significant cost savings in managing and maintaining data infrastructure \cite{ibmDA} \cite{acceldataDA}. It allows for smooth \textit{data sharing} and collaboration between various organizational departments and stakeholders. This promotes a culture of collaboration and knowledge exchange driven by data \cite{ibmDA} \cite{acceldataDA} \cite{techtargetDA}.
It gives the power to utilize \textit{data-intensive insights} to drive innovation and stay ahead of the competition in their respective markets.

\section{Data Architecture for Data-intensive }
\label{sec:Dat_state_Of_the_Art}
Many small and medium-sized companies are interested in implementing data-intensive applications, but they face challenges in the development phase that prevent them from reaching their full potential. The development of real-time, data-intensive applications is not a straightforward task, and differs from the traditional software engineering process \cite{maxim2021data}. There is an increasing demand to understand the development, deployment, and maintenance of data-intensive applications in real-world commercial settings. Big data analytics platforms, Real-time data processing systems, data warehousing solutions, content delivery networks (CDNs), and AI applications \cite{bosch2021engineering} are stated as data-intensive and need to handle data in each phase of their lifecycle \cite{lwakatare2019taxonomy} \cite{maxim2021data} \cite{kleppmann2019designing}.

The DICE project, as presented in the works of Guerriero et al. and Artac et al. \cite{guerriero2016towards} \cite{artac2018infrastructure} \cite{perez2019uml}, provides a domain-specific model (DSM) that offers big data design. This design includes data, computation, technology frameworks, and deployment concepts to design and deploy data-intensive applications. DICE proposes a model-driven engineering approach that automatically transforms application models into IaC (infrastructure as Code). Furthermore, DICE includes quality of service requirements associated with elements within the application, which are similar to QS (Quality of Service).

Accordant \cite{castellanos2021accordant} presented a solution for designing, deploying, and monitoring performance metrics in Big Data Analytics (BDA) applications. The proposed solution includes a domain-specific model (DSM) and DevOps practices. It defines a design process and a framework to specify architectural inputs, software components, and deployment strategies through high-level abstractions. This enables Quality of Service (QS) monitoring.

M. Gribaudo \cite{gribaudo2018performance} proposed a modeling framework using graph-based language to evaluate the performance of lambda architecture applications. Users can define stream, batch, storage, and computation nodes with performance indices.

Bosh .J and Raj R. \cite{raj2020modelling} developed a comprehensive \textit{conceptual model for a data pipeline}. The model consists of two fundamental components: nodes and connectors. Nodes are the pipeline's primary building blocks, representing an abstract data unit that can be processed or manipulated. On the other hand, connectors are the channels through which data is transmitted and carried between the nodes. They define how data flows through the pipeline, ensuring seamless data processing and efficient execution of tasks. The model provides a clear and concise approach to designing and implementing data pipelines, enabling businesses to optimize their data processing capabilities and enhance their overall performance.

Erraissi A. and Belangour A. \cite{erraissi2018data} proposed a \textit{meta-model} that caters to specific layers of the data life cycle. This model encompasses layers such as \textit{the data source layer, ingestion layer, and visualization layer}.

Nesi P. \cite{nesi2018auditing} proposed a \textit{solution consisting of tools} designed to collect and save data from IoT and Smart City devices in real-time. This data will be used to monitor the performance of individual devices and services, detect any issues, and keep track of data traffic.

Wei D. \cite{wei2021dataflow} reviewed IoT dataflow management, analyzing its challenges and techniques such as data sensing, mining, control, security, and privacy. He also compared different tools and platforms for IoT dataflow management. Additionally, They explained potential application scenarios for IoT dataflow management, such as smart cities, transportation, and manufacturing. 

Kalipe G. \cite{kalipe2019big} examined the major existing \textit{architectures for Big Data} (Lambda, Kappa, Zeta, ...). And explored the benefits and drawbacks of each architecture, as well as their hardware requirements, open source and proprietary software needs, and examples of real-world applications in various industries. Specifically, he identified common problems within each domain that can be addressed by leveraging the respective architecture. Lastly, he concluded by presenting a comparison of the different architectures, highlighting their trade-offs. Gillet A. \cite{gillet2021lambda+} has developed the \textit{Lambda+ Architecture}, an enhanced version of the Lambda Architecture, that enables both real-time and exploratory data analysis. She applied category theory to the investigation to ensure the preservation of component properties when combined.

\section{Limitations}
Many studies have discovered that practitioners face difficulties in managing data, as evidenced by research such as \cite{polyzotis2018data}, \cite{polyzotis2017data}, and \cite{amershi2019software}. However, these studies have primarily focused on identifying challenges rather than providing solutions. 
Several studies have not fully addressed the need for a \textit{holistic framework}; they cover specific stages of the data life cycle, such as ingestion \cite{erraissi2018data}. Architectures based on instruments and technologies are proposed to monitor the data flow for smart cities using IoT technology \cite{nesi2018auditing} \cite{borelli2020architectural}.
On the other hand, significant studies have made valuable contributions to data management. These studies have provided conceptual models for data pipelines \cite{raj2020modelling} and insights into the areas of DataOps \cite{munappy2020ad}.
Our goal is to improve how data is managed by offering a comprehensive overview of abstract \emph{\gls{da}}. This methodology facilitates the creation and upkeep of high-quality data products designed for data-intensive applications.

Data-intensive applications are software systems designed to handle large volumes of data efficiently. These applications are focused on managing and manipulating data to ensure scalability, especially with constantly increasing data loads. They include big data analytics platforms, real-time data processing systems, and data warehousing solutions. Even though creating top-notch data-intensive systems is still crucial, companies specializing in embedded systems now incorporate AI components and data products alongside their electronics and software. 
Improving data management practices is a critical need for businesses across diverse industries. It is an urgent issue that demands prompt attention. 
The primary objective of this thesis is to enhance the quality of data-intensive applications utilizing \emph{\gls{mde}}. This will be accomplished by creating models for \emph{\gls{da}} and simplifying the process of producing quality checks. 
\section{Contributions}
This thesis presents a multitude of valuable contributions. 
First, It provides valuable insights into the \emph{\gls{da}} challenges that are prevalent across various industry domains with this comprehensive overview. Second, it delves into the ability of data-intensive systems to adapt and cope with the evolving significance of data, as well as the transformations in \emph{\gls{da}} practices that occur over time. Third, it develops a meta-model for \emph{\gls{da}} suitable for different real cases like data analytical architecture, operation data warehouse, and data pipeline.  Fourth, the graphical modeling of \emph{\gls{da}} is based on a meta-model. Fifth, it simplifies data monitoring by utilizing the model to generate \emph{\gls{dq}} check code.

\section{Structure of the chapters}
This thesis is organized as follows. 

Chapter 1 introduces the topic of data architecture for data-intensive systems and presents the goals of this Ph.D. research,
 the research questions, and the key contributions. 

Chapter 2 presents the background of the study. Chapter 3 discusses the research methodology, questions, and the motivation for using each study method. 

Chapters 4, 5, 6, 7 and 8 are based on the publications and constitute the key contributions of this thesis. Chapter 4 proposes the Architecture Description Framework For Data-Intensive Applications. Chapter 5 presents the DAT, a graphical modeling tool for architecting data-intensive applications. Chapter 6  presents DAT as a modeling tool for overcoming challenges in data analytics architecture. Chapter 7 discusses data quality throughout the data journey. Chapter 8 proposes PyDAQu as a framework for generating \emph{\gls{dq}} codes that make it easy to monitor \emph{\gls{dq}}. 

In summary, Chapter 8  presents an overview of the primary research findings and offers recommendations for future investigations.
\chapter{Background}
\thispagestyle{plain}

\newacronym{ad}{AD}{Architecture description}
\newacronym{bda}{BDA}{Big Data Architecture}

In this thesis, \emph{\gls{da}} for data-intensive applications is studied. To help readers better understand the rest of the thesis, this section offers background information and describes related work.

In Section \ref{sec:dataintensive}, various data applications are discussed, allowing the reader to gain insight into the impact of data on different industries. 
3Vs of big data are discussed in section \ref{3Vs}, which helps understand why studying data architecture challenges is essential. IEEE/ISO/IEE 42010 Architecture Description is discussed in section \ref{BK_42010},
We utilized it as a conceptual foundation for constructing our metamodel. Section \ref{BDArch_Lambda_Kappa} discusses various big data architectures. \emph{\gls{dq}} is discussed in section \ref{DQ}. Finally, section \ref{BK_summary} summarizes the chapter. 

\section{Data-intensive Applications}
\label{sec:dataintensive}

Data-intensive applications are advanced software systems that effectively handle large amounts of data. These applications are specifically designed to manage and manipulate data, primarily focusing on storing and processing significant volumes of data \cite{chen2014data}. They can easily adapt to accommodate ever-increasing data loads and often rely on distributed computing, parallel processing, and other techniques to efficiently handle high data volume. Notable examples of data-intensive applications include big data analytics platforms, real-time data processing systems, data warehousing solutions, and content delivery networks (CDNs), Additionally, Creating AI applications in real-time environments poses unique challenges and requires a distinct development approach compared to traditional software engineering \cite{bosch2021engineering}. 


In today's world, data has become ubiquitous, and its significance has grown exponentially. The increasing usage of data has profoundly impacted our society, affecting various aspects of our lives in numerous ways. From businesses to governments, healthcare to entertainment, data is utilized in countless domains to streamline processes, make informed decisions, and enhance overall performance. The vast potential of data has made it a valuable resource that is being leveraged extensively, and its importance is only set to increase in the future \cite{john2014big}.

\section{4Vs of Big Data}
\label{3Vs}

Big data is characterized by four Vs: Volume, Velocity, Variety, and Veracity. Volume relates to the quantity of data, variety pertains to the diversity of data types, and velocity refers to the speed at which data is processed \cite{tole2013big}.

\textbf{Volume}: Volume is a crucial aspect of the 4 Vs framework, which organizations use to determine the size of the big data they store and manage. It refers to vast amounts of data often expressed in unfamiliar numerical terms. Human interactions with machines, networks on social media, sensors, and mobile devices generate this data.

\textbf{Velocity}: Velocity is a crucial aspect of the 4 Vs framework, which determines the rate at which big data volume grows and how easily it can be accessed. Another term for velocity is "data in motion". For example, approximately 200 million emails are sent every minute, 300000 tweets are posted, and 100 hours of YouTube videos are uploaded. This rapid increase in velocity poses new obstacles when collecting big data consistently and comprehensively.

\textbf{Variety}: The 4 Vs framework uses variety to classify the data types and categories in a big data repository, along with their associated management. Data is collected from various sources, such as sensors, social media, and wireless networks, in various formats, including website logs, astronomical data, and medical records. Extracting insights from such diverse data requires powerful processing capabilities and accurate algorithms. As technology advances and users find new ways to collect and process data, the amount of data available continues to grow, presenting challenges for effective big data management.

\textbf{Veracity}: is a paramount concern concerning the data's precision, dependability, and credibility under analysis. It is imperative to have a high level of confidence in the accuracy and integrity of the data, given that it is often derived from diverse sources such as sensors, social media, weblogs, and other sources. Therefore, the significance of data integrity cannot be overstated in today's data-intensive applications.

\section{IEEE/ISO/IEE 42010:2022 Architecture Description } 
\label{BK_42010}
In this paper, we build upon the conceptual foundations of the ISO/IEC/
IEEE 42010:2022, Systems and software engineering — \emph{\gls{ad}} \cite{42010} standard, to investigate the essential elements of \emph{\gls{ad}} for Data-Intensive applications.
\begin{figure}[h!]
	\centering
	\includegraphics[width=\linewidth]{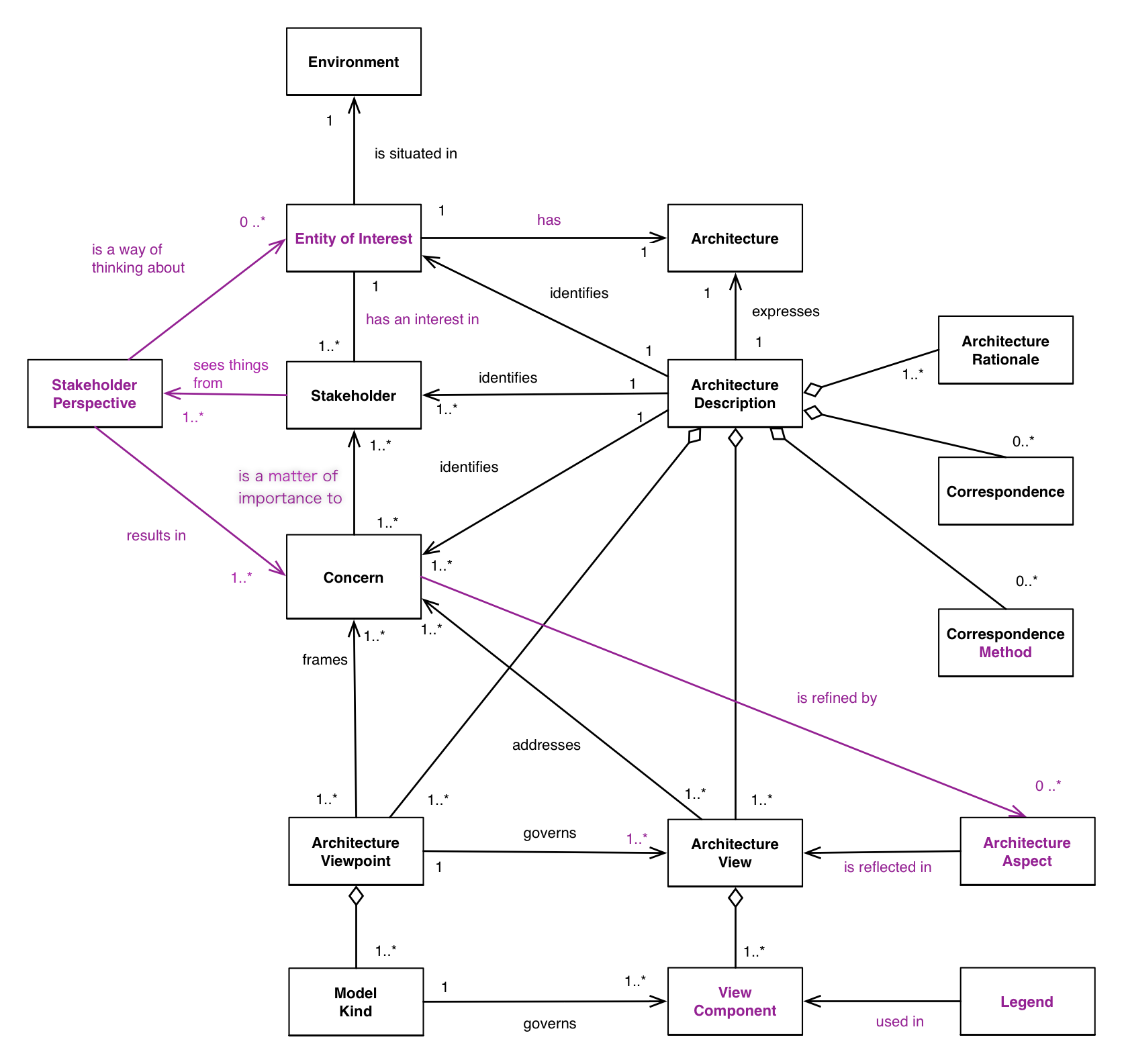}
	\caption{Content model of an \emph{\gls{ad}} (following ISO/IEC/IEEE 42010:2022)}
	\label{fig:ad}
\end{figure}
The standard addresses \emph{\gls{ad}}, the practices of
recording software, system, and enterprise architectures so that architectures
can be understood, documented, analyzed, and realized. Architecture descriptions take on many forms, from informal to rigorously specified models.
The content model for an \emph{\gls{ad}} is illustrated in Figure~\ref{fig:ad}. The architecture viewpoint is a fundamental building block representing
common ways of expressing recurring architectural concerns reusable across
projects and organizations. It encapsulates model kinds framing particular
concerns for a specific audience of system stakeholders. The concerns determine what the model kinds must be able to express: e.g., security, reliability,
cost, etc. A model determines the notations, conventions, methods, and
techniques. Viewpoints, defining the contents of each architecture view,
are built up from one or more model kinds and correspondence rules, linking
them together to maintain consistency.
\section{Big Data Architectures}
\label{BDArch_Lambda_Kappa}
Design patterns called \emph{\gls{bda}}s are utilized to handle, store, and manage massive amounts of data in data-intensive applications. Within this realm, two important architectures are Lambda Architecture and Kappa Architecture.

The \textbf{Lambda} Architecture has three layers: Batch, Speed, and Serving. Batch processes large-scale data offline for accurate analysis, while Speed handles real-time processing for up-to-date insights. Serving combines results from both layers for interactive queries, accommodating eventual consistency between batch and real-time views \cite{gillet2021lambda+} \cite{kiran2015lambda}.

The \textbf{Kappa} Architecture simplifies real-time data processing using a single pipeline for real-time and batch data. It leverages stream processing frameworks like Apache Kafka Streams or Apache Flink, treating all data as a continuous stream. This allows for low-latency processing and analysis but may require additional historical analysis and reprocessing mechanisms \cite{pathirage2020kappa}.

Both architectures have benefits and drawbacks, and the decision of which to choose depends on various factors, such as the type of data, the desired speed, and the tools and resources available. In addition, different architectures like the Zeta Architecture and the Mu Architecture have also been suggested to meet diverse data processing requirements in data-heavy applications.

\section{Data Quality}
\label{DQ}
\emph{\gls{dq}} is defined as fitness for use \cite{sidi2012data}. \emph{\gls{dq}} should be understood as the degree of data's correctness and usefulness \cite{liu2002evolutional}.
\emph{\gls{dq}} cannot be overstated when making informed decisions based on data. The accuracy and reliability of the insights and decisions derived from data are directly linked to the quality of the data itself. It is essential to ensure the data is accurate, up-to-date, and consistent to avoid any potential mistakes or errors in decision-making. By prioritizing data quality, organizations can ensure that they make the best use of their data and derive the most valuable insights possible \cite{karkouch2016data} \cite{ji2020quality}. It is imperative for organizations to maintain high-quality data, as poor data quality can have severe ramifications. Inaccurate conclusions, incorrect decision-making, and adverse outcomes are all potential consequences of poor data quality. Therefore, it is crucial to ensure that data is accurate, up-to-date, and reliable to avoid any negative impact on the organization \cite{cai2015challenges}.

\emph{\gls{dq}} refers to the reliability, accuracy, completeness, consistency, and overall usability of data for a given purpose. Several dimensions are used to assess data quality, and the six most commonly used data quality dimensions \cite{sidi2012data} \cite{jesiclevska2017data} \cite{BDQ}:
\begin{enumerate}

\item  \textbf{Accuracy} pertains to the correctness and precision of information. The correct information is devoid of any errors, discrepancies, or mistakes. 

\item  \textbf{Completeness}
The concept of completeness involves ensuring that all the necessary data elements are present. Incomplete data can lead to skewed analyses and a lack of insights. It is crucial to have data that encompasses all relevant aspects to prevent incomplete conclusions from being drawn.

\item \textbf{Consistency} Assessing consistency is key when dealing with data from various sources over a certain time frame. Inconsistent data can lead to misinterpretations and confusion, so it's imperative to guarantee data consistency to ensure precise analyses.

\item \textbf{Timeliness} refers to the freshness of data about its intended use. Using outdated data can result in uninformed decisions, but having timely data ensures that insights and actions are based on current and accurate information.

\item \textbf{Validity} refers to whether the information complies with established business rules, standards, and limitations. Information that does not meet the requirements can result in erroneous analyses. Validating data guarantees that it is consistent with the intended objective.

\item \textbf{Uniqueness} Ensuring the uniqueness of data records is crucial to avoid duplication and maintain analysis accuracy. Duplicates can cause misleading conclusions, leading to confusion and redundancy. Therefore, it's essential to maintain data uniqueness to prevent such errors.

\end{enumerate}

\emph{\gls{dq}} dimensions are essential for organizations that gather, store, and use data. These qualities are critical in assuring data precision, comprehensiveness, dependability, and consistency. Organizations should evaluate their data against external sources and use automated methods to detect inaccuracies to do this. Organizations may confidently manage their data's correctness, completeness, consistency, and dependability in this manner.

\section{Summary}
\label{BK_summary}
This chapter presents five main concepts discussed in this thesis, Data-intensive Applications, 4Vs of Big Data, IEEE 42010 Architecture Description \emph{\gls{ad}}, Big Data Architectures \emph{\gls{bda}},  and Data Quality \emph{\gls{dq}}.

\chapter{Research Methodology}
\thispagestyle{plain}

The research methodology is a systematic approach to resolving research problems. The focus of this study is to emphasize the significance of enhancing data architecture practices in data-intensive systems and its impact on the quality of data. Previous research has highlighted professionals' challenges in creating and maintaining data products. Therefore, the primary objective of this study is to establish a data architecture model that can efficiently manage data and produce high-quality data products. This chapter is divided into several crucial parts, which include research design selection, research questions, motivation, informant selection, data collection methods, data analysis, an overview of the research process, and potential threats to validity.

\section{Research Questions}
This research aims to empirically identify the data architecture challenges encountered during the development and maintenance of data-intensive systems, propose an improved data architecture approach, and empirically validate the proposed approach. We have adopted a qualitative research methodology to address the research goals. The research is focused on the five primary research questions.

\begin{itemize}
    \item \textbf{RQ1: What data architecture challenges are experienced in the industry while developing data-intensive systems?}
\item \textbf{RQ2: What essential components are needed for data-
intensive applications to model data-intensive systems?}
\item \textbf{RQ3: How can a comprehensive model, which includes
main concerns in various data-intensive related domains,
be designed?}

\item \textbf{RQ4: How can the framework be used in practice in software design?}

\item \textbf{RQ5: How can monitoring data quality be streamlined?}
\end{itemize}

The main aim of RQ1 is to identify the challenges in data architecture that professionals face due to current advancements. Data architecture is a topic under exploration, and a few papers discuss various data architecture practices. Therefore, conducting a thorough study showcasing the challenges experts face in light of technological progress is imperative. This study would effectively establish the urgency for an enhanced data architecture practice.  The second research question (RQ2) was set to analyze a list of essential data components practitioners could use to address data architecture challenges in data-intensive systems. The third question (RQ3) explicitly demonstrates that the main concepts and the list of analyzed challenges are firmly rooted in a holistic framework. The fourth question (RQ4) effectively demonstrates the framework's ability to deliver architectural solutions tailored to the demands of various industrial cases. The primary objective of the fifth research question (RQ5) is to establish the impact of improved data architecture practices on the quality of data products. It delves into how simplifying data quality can facilitate automation, ultimately leading to the speedy development and maintenance of top-notch data-intensive systems.

\section{Qualitative Research}
Qualitative research aims to gather, evaluate, and clarify non-numerical data like written words, recorded sounds, and filmed images to comprehend ideas, viewpoints, or personal encounters. It is crucial to understand that qualitative research centers on individuals constructing meaning through their interactions with the world around them \cite{merriam2019qualitative}. In qualitative research, results are presented using words instead of numbers. This type of research focuses on themes rather than numerical data \cite{lambert2012qualitative}.
We have strategically adopted this methodology as it not only allows us to effectively construct new and innovative ideas for improving or fine-tuning a product or practice but also empowers us to construct a comprehensive theoretical framework that emerges from the data gathered during the research and logically explains the results. 

Qualitative data sources encompass observation, participant observation (fieldwork), interviews, questionnaires, documents, texts, and the researcher's personal impressions and reactions. Qualitative research provides more comprehensive information, deeper insights into real-time practices, and a better understanding of underlying perceptions \cite{bogdan1997qualitative}. In this type of research, the primary techniques used are individual interviews, group interviews (also known as focus groups), observations, and document analysis \cite{seaman1999qualitative}. 

For our study, we opted for individual interviews, group interviews, and observations as they can provide comprehensive information about a limited number of cases. This approach enabled us to gain deeper insights into each case. Our objective was to examine how companies utilize certain practices, identify issues they are facing, and suggest better solutions. As such, using a qualitative methodology was appropriate for our study.

\section{Case Companies}
\label{section:RM_Cases}
In our research, we carefully chose companies with data-intensive systems demonstrating a high domain maturity level in adopting data architecture. For confidentiality reasons, the companies involved in the study have chosen to remain anonymous. The research highlights limitations, errors, and current development practices in technology. We have included a concise summary of the case companies examined in this study. The companies allowed collaboration through interviews, in-company workshops, action research, and weekly meetings.

Cases A, B, C, D, and E, actively participated in this research by allowing collaboration through interviews, workshops, interactive sessions, weekly meetings, and action research.

Case A and B, The company offers a next-generation frontline employee experience platform built for companies that have served at the heart of their business. The Company suite of talent attraction, workforce management, employee engagement, and compliance technologies enable organizations to intelligently attract, manage, engage, and retain the best talent to run and improve their business. It serves over 20,000 restaurant and hotel locations and four million hospitality employees globally, with emerging growth in retail and healthcare.

Case C, The company is a multinational information technology corporation with a notable presence in several technology industries. One of its main focuses is on Printers and Imaging Solutions, where it is recognized as a major player in the printing industry. It offers a range of inkjet, laser, and multifunction printers catering to home and office use. The context of this case involves managing error data that arises from these printers.

Case D, The company is An Internet Service Provider (ISP)  telecommunications company that enables individuals and businesses to connect to the Internet. ISPs offer various technologies like broadband, wireless, and satellite connections to facilitate data transmission between users and the global network of servers.

Case E, This research project examines the conversations happening in the health and food fields. Analyzing data from social networks aims to detect any emerging trends or issues as they happen. Although this project is challenging, it has the potential to provide valuable insights that could benefit these important aspects of human life.
Table \ref{tab:RM_Cases} shows more details about the cases and complexity (The ”complexity” column provides information on each case’s complexity level concerning
the primary data components. These components are typically measured regarding the
number of internal data elements that comprise a given case.).

\begin{table}[]
\caption{Cases Details}
\label{tab:RM_Cases}
 \makebox[\linewidth]{
\centering
\begin{tabular}{|c|c|c|c|l|}
\hline
\textbf{Company} & \textbf{Cases} & \textbf{Case Name}                                              & \textbf{Case Compixity} & \textbf{Experts Roles} \\ \hline
\multirow{6}{*}{\textbf{1}} &
  \multirow{3}{*}{A} &
  \multirow{3}{*}{\begin{tabular}[c]{@{}c@{}}Analytical \\  Data Architecture\end{tabular}} &
  \multirow{3}{*}{4 ( 48 )} &
  Big Data Team Lead \\ \cline{5-5} 
                 &                &                                                                 &                         & Data Architect         \\ \cline{5-5} 
                 &                &                                                                 &                         & Big Data Engineer      \\ \cline{2-5} 
 &
  \multirow{3}{*}{B} &
  \multirow{3}{*}{\begin{tabular}[c]{@{}c@{}}Operational \\  Data Warehouse\end{tabular}} &
  \multirow{3}{*}{9 ( 72 )} &
  Big Data Team Lead \\ \cline{5-5} 
                 &                &                                                                 &                         & Data Architect         \\ \cline{5-5} 
                 &                &                                                                 &                         & Big Data Engineer      \\ \hline
\multirow{5}{*}{\textbf{2}} &
  \multirow{5}{*}{C} &
  \multirow{5}{*}{Data Pipeline} &
  \multirow{5}{*}{5 ( 34 )} &
  Big Data Team Lead \\ \cline{5-5} 
                 &                &                                                                 &                         & Data Architect         \\ \cline{5-5} 
                 &                &                                                                 &                         & Big Data Engineer      \\ \cline{5-5} 
                 &                &                                                                 &                         & ETL Developer          \\ \cline{5-5} 
                 &                &                                                                 &                         & Data Quality Engineer  \\ \hline
\multirow{3}{*}{\textbf{3}} &
  \multirow{3}{*}{D} &
  \multirow{3}{*}{\begin{tabular}[c]{@{}c@{}}Operational \\  Data Architecture\end{tabular}} &
  \multirow{3}{*}{9 ( 66 )} &
  Big Data Team Lead \\ \cline{5-5} 
                 &                &                                                                 &                         & Data Architect         \\ \cline{5-5} 
                 &                &                                                                 &                         & Data Quality Team Lead \\ \hline
\textbf{4}       & E              & \begin{tabular}[c]{@{}c@{}}Lambda \\  Architecture\end{tabular} & 7 ( 67 )                & Researchers            \\ \hline
\end{tabular}%
}
\end{table}


\section{Research Methods}
Research methods play a crucial role in collecting and analyzing empirical data. They are an essential component of any research design. Qualitative research methods offer a distinct advantage as they compel the researcher to explore the intricacies of the problem rather than simplifying it \cite{seaman1999qualitative}. Empirical data is gathered through the meticulous observation and documentation of patterns and behavior, typically via experimentation. Such data is invaluable in answering research questions, as it provides a solid foundation of information based on sensory evidence \cite{seaman1999qualitative}. Qualitative and quantitative methods can be used to collect and analyze data \cite{sutton2015qualitative}. Collecting qualitative data involves producing a substantial volume of data \cite{singer2008software}. After transcription, the audio or video recording data collection method is employed for data analysis \cite{sutton2015qualitative}.
\subsection{Case Study}
A case study is a detailed examination of a specific occurrence or a few instances of a phenomenon \cite{hyde2000recognising}. The purpose of a case study is to explore contemporary phenomena in their real-life setting, particularly when it is unclear where the boundary between the phenomenon and the context lies \cite{perry2006case}. The advantages of using the case study method include the ability to collect and analyze data within the context of the phenomenon, combine qualitative and quantitative data during the analysis, and gain a deeper understanding of the complexities of real-life situations. However, the disadvantages of case studies include a lack of strict methodology, difficulties associated with analyzing data, and limited ability to generalize findings and conclusions \cite{perry2006case}.
Different research strategies are categorized based on purpose, including exploratory, descriptive, explanatory, and improvement. Case study is a research method initially used mainly for exploratory purposes, but it can also serve descriptive, improvement, and explanatory purposes \cite{runeson2009guidelines}. It is an ideal choice for exploratory research questions. However, for descriptive research questions, a case study may be feasible only if the sampling-based study's representativeness can be sacrificed for better realism in a case study. If representativeness is crucial, a survey is a better option.
When addressing explanatory research questions, case studies can be helpful. However, the evidence collected is not based on a statistically significant quantitative analysis of a representative sample. Instead, it provides a qualitative understanding of how phenomena operate in their specific context. If quantitative evidence is crucial, an experimental strategy is a better option. For improving research purposes, the action research strategy is a natural choice and a variant of case study research \cite{runeson2009guidelines}. In our case, we adopted an exploratory case study to identify the challenges associated with data architecture practices in a real-world company context. This study captured the complexities of data architecture in a data-intensive system company scenario.
\subsection{Action Research}
In software engineering, action research is a collaborative approach that involves working with concerned actors or problem owners within an organization. This method allows for the real-time proposal, implementation, and evaluation of solutions, with researchers engaging with the company over an extended period. Problem owners are crucial in action research, providing valuable insights and expertise based on their experiences and domain knowledge \cite{petersen2014action} \cite{mckay2001dual}. The primary goal of this approach is to address real-world issues while exploring the outcomes and learning experiences of problem-solving \cite{easterbrook2008selecting}. For our study, we chose to use the action research method as it allowed us to systematically identify and define issues related to data architecture practices and propose solutions in a participatory manner within the context of our investigation.
Additionally, we participated in the subsequent phases of implementing the solution, known as action \cite{petersen2014action} \cite{mckay2001dual}. The cycle of the action research process comprises five stages, namely (1) diagnosis, (2) action planning and design, (3) taking action, (4) evaluation, and (5) specifying learning \cite{petersen2014action} \cite{mckay2001dual}. Action research is beneficial as it can provide sturdy and applicable knowledge to various management and organization researchers \cite{coghlan2011action}.
\section{Research Techniques}
Researchers utilize research techniques to gather empirical data to analyze actions in real-world industrial settings. These techniques include semi-structured interviews and literature reviews, which are particularly useful for understanding behavior, the meanings and contexts of events, and the influence of values on decision-making in practical situations \cite{strauss1998basics}. 
\subsection{Interviews}
Interview-based data collection involves asking subjects questions about the case study's areas of interest. The researcher uses interview questions based on the research question to guide the dialogue with the subject(s). Interviews can be conducted with a group or individual practitioners and can be open-ended or closed. Interviews can be unstructured, semi-structured, or fully structured \cite{moran1994real}.

Interviews can be structured, semi-structured, or unstructured. Structured interviews have pre-planned questions in a fixed order, similar to a survey. Unstructured interviews are more conversational and based on the researcher's interests and subject. Semi-structured interviews have planned questions but can be asked in any order. We chose semi-structured interviews because they allow for more exploration and clarification of responses and create a rapport between respondents and researchers \cite{barriball1994collecting}.
\subsection{Observation}
Observation involves carefully observing and examining participants' behavior in a natural setting \cite{cowie2009observation}. This can be used to study how practitioners perform a specific task. There are various observation methods, such as monitoring a group of practitioners with a video recorder and analyzing the recording later using protocol analysis \cite{owen2006protocol}\cite{von1996identification}. 
An additional option is to utilize a "think aloud" protocol during research by frequently asking subjects questions like "What is your strategy?" and "What are you thinking?" to prompt them to vocalize their thoughts. Meetings can also generate information about the studied topic through attendees' interactions. Another approach involves using a sampling tool to gather participant feedback and data \cite{karahasanoviae2005collecting}.
Researchers at a research site must observe and take detailed notes on the people, concepts they discuss, and occurring interactions \cite{cowie2009observation}. These notes, called field notes, are an important part of the research process. During the action research, the researchers performed participant observation and took field notes. Observation can either be structured or unstructured. Structured observation involves collecting data using specific variables and a pre-defined schedule. Unstructured observation, on the other hand, is more open-ended and has no pre-determined variables or objectives \cite{runeson2009guidelines}. This research used unstructured observation mainly during weekly stand-up meetings, pair programming, and weekly result presentations.
\section{Data Analysis}
Qualitative research involves collecting unstructured text data from various sources, including interview transcripts, observation notes, diary entries, and medical records. In addition, multimedia materials such as pictures, audio, or video recordings may also be included in the data. Given the diverse nature of the data, analysis methods must be flexible, intuitive, and imaginative, involving inductive reasoning, critical thinking, and theorizing.
\subsection{Qualitative Data Analysis}
In qualitative research, data analysis involves systematically searching and arranging interview transcripts, observation notes, and other non-textual materials to understand a phenomenon better \cite{wong2008data}. The process primarily involves coding or categorizing the data. Coding involves breaking down a large amount of raw information and assigning it to categories. Two types of coding used in this thesis are thematic coding using the NVivo tool and open coding \cite{dey2003qualitative}. Thematic coding involves finding text themes by analyzing words' meanings and sentence structure. The NVivo software automates the data \cite{cruzes2011recommended}. Open coding is a qualitative data coding method that creates codes from scratch. The coder manually creates these codes to cover the entire transcript. The codes are then applied to the remaining transcripts, with necessary adjustments made to ensure that the codes apply to all transcripts in the study \cite{saldana2021coding}.

\section{Research Design}
The initial and most important research phase involved identifying an idea to investigate. The research idea originated from participating with one of the other team members to develop a data-intensive application.

Our primary intention was to empirically identify the data architecture challenges encountered during the development and maintenance of data-intensive systems, propose an improved data architecture approach, and empirically validate the proposed approach.
Our approach to formulating RQs involves problem identification, analyzing existing approaches, identifying challenges, proposing solutions, and validating them. We chose qualitative research to gain insight into practitioners' perspectives on the current data architecture approach and their willingness to transition to a new one. Additionally, it allows us to identify practitioners' needs, the difficulties they face with current approaches, and how these difficulties impact the final product.
Additionally, utilizing qualitative analysis can assist in generating ideas for enhancing the data architecture approach. While quantitative research methodology can measure behaviors and answer questions such as "how often" and "how many," our objective is to delve into data architecture challenges, current practices, and their evolution. Gathering data from informants is the most effective technique for comprehending embedded system industries' challenges. In quantitative methodology, free-text responses are not allowed, potentially missing vital contextual details. Thus, we opted for a qualitative research methodology for our study.


\begin{itemize}
\item \textbf{Case Study 1: Problem Identification}

A literature review was conducted to learn more about the topic under investigation, which enabled the researcher to identify similar studies conducted in the past. We identified papers on data architecture concepts and challenges through the literature review. 
However, all the previous studies were focused on data pipelines, big data architectures (Lambda, Kappa, and ...), real-time and batch processing, and architecture-based technologies. Therefore, the researcher decided to study data architecture challenges for data-intensive applications. To explore and identify the challenges encountered by practitioners during the development and maintenance of data products, a case study method was used as it allows for the examination of the phenomenon in depth using various kinds of evidence obtained from interviews with those involved, direct observation of events and analysis of documents and artifacts. Data collection primarily relied on interviews, as illustrated in Table \ref{tab:ex_case_study}.
 We utilized an exploratory case study method and conducted 6 in-depth interviews with practitioners for two different cases from Company 1, to collect comprehensive information. The selected informants also provided relevant documents to aid in understanding their use case descriptions and data formats, among others. The interviews were semi-structured, with open-ended questions, which provided a great opportunity to delve deeper into the issues. All interviews were recorded and transcribed with the practitioners' permission. We also made observations and analyzed the documents provided. The challenges were categorized, and the results were sent to the practitioners for review, with data analysis conducted using the NVivo tool for thematic coding. The codes were analyzed, and the results were formulated with confidence.

Participant reflections were meticulously recorded, and the researcher conducted follow-up interviews with three senior practitioners to clarify any uncertainties. The final results were then sent to two esteemed senior practitioners for review, and modifications were seamlessly made according to their invaluable feedback. The published paper is a testament to our research's thoroughness and professionalism.

\item \textbf{Case Study 2: Analysing the existing Practices}

We took the next step in our research by investigating the challenges of data architecture. To do this, we conducted a study that involved interviewing multiple teams from companies A, B, C, and D. We looked at 5 different use cases and conducted 6 in-depth semi-structured interviews. The researcher was granted permission to participate in the weekly stand-up calls and meetings. This allowed us to gain insight and contribute significantly to the study. We presented the results to the practitioners involved in the study, as well as the company's steering committee, for approval. We also incorporated the reflections and comments from the informants into the final results, which were published as a paper.

\item \textbf{Case Study 3: Propose a Solution}

We created a data architecture strategy for building high-quality systems that rely heavily on data. 
Our meta-model consisted of structural and behavioral components informed by previous research. 
We conducted extensive interviews with individuals and focus groups to understand better the current data architecture at one of our case companies. 

In addition, the researcher gathered information by observing various meetings inside and outside the company. They took notes during these meetings and also conducted interviews to gain insights. Using this information, we created a meta-model for data architecture. To ensure the model's accuracy, we presented it to teams involved in the study for feedback. They then presented it again to teams working on data products not involved in the study. The model was also presented at two other companies for external validation, and feedback was incorporated. The meta-model was eventually published as a paper and presented at several workshops.  Table \ref{tab:ex_case_study} provides a comprehensive overview of the empirical endeavors regarding exploratory case study research. Using our meta-model, we developed DAT, a graphical modeling tool specifically designed for data architecture. 

\item \textbf{Action Research: Implement and Validate the Soluation}

The study aimed to establish a robust data architecture using MDE. 
It was conducted across several companies, including B, C, and D, and the findings were thoroughly examined. Company B, specializing in technology solutions such as printers and imaging, served as a practical test-bed for 
framework implementation. 

A literature review was conducted to dive into the data-intensive systems. The review helped to identify diverse data architectures utilized across various industries. The study aimed to enhance the quality and structure of data via pragmatic research and analysis, which was carried out with utmost confidence in its potential to improve the field.

The researcher, a Team Lead, and a QA Engineer analyzed cases from companies B and C and developed a framework to enhance the quality of the existing data pipelines. During the development of the framework, it was realized that there was a need to improve the process for monitoring data quality. To address this issue, it was suggested to generate automatic data quality codes using the widely used Python library, "Great Expectation".
\end{itemize}


\newpage

\pagestyle{empty}

  \begin{landscape}
\begin{table}[ht!]
\centering
\caption{Overview of Exploratory Case Studies}
\label{tab:ex_case_study}
\resizebox{\paperwidth}{!}
{%
\begin{tabular}{|c|c|c|l|l|l|l|}
\hline
\textbf{Company} &
  \textbf{Cases} &
  \textbf{Case Name} &
  \textbf{Experts Roles} &
  \textbf{Research Objective} &
  \textbf{\begin{tabular}[c]{@{}l@{}}Research \\  Technique\end{tabular}} &
  \multicolumn{1}{c|}{\textbf{\begin{tabular}[c]{@{}c@{}}Data Collection \\  Technique\end{tabular}}} \\ \hline
\multirow{6}{*}{\textbf{1}} &
  \multirow{3}{*}{A} &
  \multirow{3}{*}{\begin{tabular}[c]{@{}c@{}}Analytical \\  Data Architecture\end{tabular}} &
  Big Data Team Lead &
  \multirow{6}{*}{\begin{tabular}[c]{@{}l@{}}1. To identify the DA Challenges\\  2. To develop a meta-model \\  3. To develop an Architecture tool\end{tabular}} &
  \multirow{6}{*}{\begin{tabular}[c]{@{}l@{}}1.Interview\\  2.Observation\end{tabular}} &
  \multirow{6}{*}{\begin{tabular}[c]{@{}l@{}}1. Semi-structured interviews\\ 2. Observation from \\ the weekly meeting\\ 3. Follow-up interviews\end{tabular}} \\ \cline{4-4}
 &                    &                                                                                         & Data Architect         &  &  &  \\ \cline{4-4}
 &                    &                                                                                         & Big Data Engineer      &  &  &  \\ \cline{2-4}
 & \multirow{3}{*}{B} & \multirow{3}{*}{\begin{tabular}[c]{@{}c@{}}Operational \\  Data Warehouse\end{tabular}} & Big Data Team Lead     &  &  &  \\ \cline{4-4}
 &                    &                                                                                         & Data Architect         &  &  &  \\ \cline{4-4}
 &                    &                                                                                         & Big Data Engineer      &  &  &  \\ \hline
\multirow{5}{*}{\textbf{2}} &
  \multirow{5}{*}{C} &
  \multirow{5}{*}{Data Pipeline} &
  Big Data Team Lead &
  \multirow{5}{*}{\begin{tabular}[c]{@{}l@{}}1. To identify the existing data pipeline\\  2. To identify data quality challenges\\  3. To Identify data quality approaches\end{tabular}} &
  \multirow{5}{*}{\begin{tabular}[c]{@{}l@{}}1.Interview\\  2.Observation\end{tabular}} &
  \multirow{5}{*}{\begin{tabular}[c]{@{}l@{}}1. Semi-structured interviews\\ 2. Observation from \\ the weekly meeting\\ 3. Focus group interview\\ 4. follow-up interview\end{tabular}} \\ \cline{4-4}
 &                    &                                                                                         & Data Architect         &  &  &  \\ \cline{4-4}
 &                    &                                                                                         & Big Data Engineer      &  &  &  \\ \cline{4-4}
 &                    &                                                                                         & ETL Developer          &  &  &  \\ \cline{4-4}
 &                    &                                                                                         & Data Quality Engineer  &  &  &  \\ \hline
\multirow{4}{*}{\textbf{3}} &
  \multirow{4}{*}{D} &
  \multirow{4}{*}{\begin{tabular}[c]{@{}c@{}}Operational \\  Data Architecture\end{tabular}} &
  Big Data Team Lead &
  \multirow{4}{*}{To Develop a Data Quality Tool} &
  \multirow{4}{*}{\begin{tabular}[c]{@{}l@{}}1.Interview\\  2.Observation\end{tabular}} &
  \multirow{4}{*}{\begin{tabular}[c]{@{}l@{}}1. Semi-structured interviews\\ 2. Observation from \\ the weekly meeting\\ 3. Follow-up interviews\end{tabular}} \\ \cline{4-4}
 &                    &                                                                                         & Data Architect         &  &  &  \\ \cline{4-4}
 &                    &                                                                                         & Data Quality Team Lead &  &  &  \\ \cline{4-4}
 &                    &                                                                                         & Analytics System Architect          &  &  &  \\ \hline
\textbf{4} &
  E &
  \begin{tabular}[c]{@{}c@{}}Lambda \\  Architecture\end{tabular} &
  Researchers &
  \begin{tabular}[c]{@{}l@{}}To identify \\  big data architectures\end{tabular} &
  Interview &
  Semi-structured interviews \\ \hline
\end{tabular}%
}
\end{table}
  \end{landscape}

\pagestyle{fancy}

\newpage

\section{Threats to Validity}
This section discusses threats to validity regarding how our research questions were answered.
\subsection{Construct Validity}
Construct Validity comprises two integral components: the measure must be all-encompassing and discriminatory in that it solely encompasses aspects of the intended theoretical construct. A few cases were excluded from the results due to interviewers not fully understanding discussed concepts to maintain construct validity. Our study has some limitations due to the screening process and a limited number of interviews. However, this limitation can be considered an opportunity for future inquiry. To minimize researcher bias, the researcher conducted the interviews. Before conducting the interviews, we sent the interviewees a semi-structured interview guide and a brief description of the topic to explore.

During the interview, we started by introducing the topic of the study. We rephrased questions if the interviewee's response seemed off-topic or unclear. Additionally, during our analysis of the interview transcripts, we contacted interviewees to clarify any confusion or lack of clarity.
\subsection{Internal Validity}
Internal validity refers to how accurately the observed outcome reflects the truth in the studied population without being affected by methodological errors \cite{petersen2014action}. 
It is important to note that a potential threat to the validity of this thesis could arise due to the fact that the strategies and results of RQ2 and RQ3 were developed within a specific company context.
The researcher only had limited access to the strategy descriptions, which makes it impossible to determine if other factors had a more significant impact on the final result than the proposed strategies. The co-author validated the findings with in-depth knowledge of company data to enhance internal validity. The respective company’s steering committee also reviewed the results.

\subsection{External Validity}
The presented work is derived from the cases studied with different teams in the domains of telecommunication, software development, and information technology. Some parts of the work can be seen differently in different parts of the company. All the terminologies used in the companies are normalized, and the implementation details are explained with the necessary level of abstraction \cite{maxwell2009sage}. We do not claim that the opportunities and challenges will be the same for industries from different disciplines.

\section{Summary}

This chapter delves into the research questions that were the focus of this thesis, providing a detailed examination of each. The research methods used to tackle these questions are discussed comprehensively, particularly their effectiveness. The research design section further elaborates on how these methods successfully addressed each research question.
\chapter{Architecture Description Framework For Data-Intensive  Applications}
\thispagestyle{plain}
\newacronym{daf}{DAF}{Architecture Description Framework for Data-intensive Applications}
\newacronym{daml}{DAML}{Data Modeling Language}
\newacronym{hla}{HLA}{high-level architecture} 
\newacronym{lla}{LLA}{low-level architecture}
\newacronym{adw}{ADW}{Analytics Data Warehouse }
\newacronym{odw}{ODW}{Operational Data Warehouse }
\newacronym{dat}{DAT}{Data Architecture Modeling Tool}

Managing data has become increasingly difficult due to its exponential growth and multiple sources. In the business world, \emph{\gls{da}} provides a systematic approach to describing, collecting, storing, processing, and analyzing data to meet business needs. It plays a crucial role in transforming data into valuable information by providing an abstract view of data-intensive applications. A framework for data-intensive applications is presented in this article, which utilizes \emph{\gls{mde}} to achieve this goal.

The framework's effectiveness was evaluated through five thorough case studies, with valuable feedback provided by seven practitioners and two prominent companies on its capabilities.

\section{Introduction}
According to the International Data Corporation (IDC) \cite{idc}, the amount of useful data is predicted to exceed 175 zettabytes by 2025, with a compounded annual growth rate of 61\%. About 90 zettabytes of this data will come from IoT devices, and roughly 30\% of the generated data will be consumed in real-time. Additionally, digital transformation programs are estimated to cost around \$7.4 trillion between 2020 and 2023.

Data is the backbone that plays a decisive role in various aspects such as decision-making, machine learning, generating insights, and creating reports \cite{bibri2019anatomy}. To become data-intensive, it is essential to follow a strategic process involving effectively utilizing data insights. Therefore, adopting the most efficient methods for collecting, storing, analyzing, and safeguarding valuable data is crucial.

A well-designed data architecture \emph{\gls{da}} is crucial for any organization to manage and utilize its data assets effectively. Providing a comprehensive set of specification artifacts enables businesses to align their data investments with their overall strategy and seamlessly integrate different data sources. With a series of master blueprints at various levels of abstraction \cite{10.5555/3165209}, \emph{\gls{da}} empowers organizations to take a top-level view of their data collection, storage, analysis, and presentation, unlocking endless possibilities for growth and success \cite{inmon2019data}.

We have discovered an urgent demand for a comprehensive blueprint to guide data-intensive projects through our discussions with researchers and practitioners across various industry domains. While recent research has offered meta-models and conceptual models for specific layers of these projects, such as data sources and ingestion, as well as technology-based architecture, these contributions are limited in scope. Our findings confirm the importance of a more holistic approach to data-intensive projects; this study addresses this need.

We developed an innovative architectural description to assist in creating efficient data-intensive applications. Built-in accordance with the IEEE/ISO/IEC 42010 standard \cite{42010}, this framework provides comprehensive support for every stage of the architecture description, reasoning, and design decision process. Our research is proud to offer a modeling and analysis framework that will help facilitate an architecture-driven development of data-intensive applications.

The framework is called \textit{DAF}\footnote{\emph{\gls{daf}}} \cite{abughazala2023architecture}; it is based on an architectural approach to describe a data flow for data-intensive Applications.
The \emph{\gls{daf}} framework is realized by exploiting the advanced Model-Driven Engineering (\emph{\gls{mde}}) technique which is meta-modeling. \emph{\gls{daf}} covers all data architecture components and their best practices \cite{dataengineers} \cite{warren2015big}.

We explored the main data architecture elements and challenges in our previous work \cite{10092710} \cite{abughazala2022dat}. Based on the study objectives.

The questions and contributions that we address in this paper are:


\begin{itemize}

 \item \textbf{RQ1.} What essential components are needed for data-intensive applications to model a data-intensive architecture? For this, we reviewed the relevant literature to understand the needs and potential scientific value of data-intensive (Section \ref{sec:LTRBackground}).

 \item \textbf{RQ2.} How can a comprehensive model, which includes main concerns in various data-intensive related domains, be designed? We conducted an exploratory study \cite{10092710} that includes interviews with domain experts from academia and industry to answer this question. From this, we designed a well-grounded architecture description framework. The framework includes information about trends, design, and challenges, and future solutions (Sections \ref{sec:meta-model}).
\item \textbf{RQ3.} How can the \emph{\gls{daf}} be used in practice in software design? We used DAF to model real systems (Section \ref{sec:DAF_casestudy}, Section \ref{sec:DAF_evalSec}).

 \end{itemize}

  The rest of this paper is organized as follows. The Background is presented in Section 2, and the methodology in Section 3. The application of \emph{\gls{daf}} to a real case study is described in Section 4. The \emph{\gls{daf}} evaluation is presented in Section 5. Related work is discussed in Section 7, while conclusions are drawn in Section 8.

\section{Methodology}
\label{sec:LTRBackground}

   \begin{table*}[!ht]
\centering
\caption{{Summary of the DAML meta-classes and related references }}
\makebox[\linewidth]{
\begin{tabular}{|p{10cm}|p{5.2cm}|}
\hline
\textbf{ Concept } & \textbf{Reference}   \\
 \hline
 
\textbf{ Data Formats} (structured, semi-structured, unstructured) &  \cite{azad2020role}  \cite{erraissi2018data} \cite{mohammed2016review} \cite{alexandru2016big} \cite{alabdullah2018rise}\\
 \hline
\textbf{ Data Processing Type} (Batch, Real-Time)  & \cite{alexandru2016big} \cite{mohammed2016review} \cite{balusamy2021big} \cite{alabdullah2018rise}\\
 \hline
\textbf{ Data Storage} (File System, NoSQL, NewSQL, ...) &  \cite{cavanillas2016new} \cite{kalna2019meta} \cite{elomari2016data} \cite{verma2018comparative} \cite{borelli2020architectural} \cite{alexandru2016big} \cite{azad2020role} \cite{alabdullah2018rise}\\
 \hline
\textbf{ Data Location} (Cloud, on-Premise) & \cite{balusamy2021big} \cite{mohammed2016review}\\
\hline
\textbf{ Data Generation} (data sources(sensors, camera, API, ...)) & \cite{alexandru2016big} \cite{mohammed2016review} \cite{raj2020modelling}\\
\hline
\textbf{ Data Ingestion} (identify, validate, compress, ...) & \cite{erraissi2018data} \cite{sawant2013big} \cite{borelli2020architectural} \cite{raj2020modelling} \cite{densmore2021data}\\
\hline
\textbf{ Data Processing} (classify, filter, transform, ...) & \cite{balusamy2021big} \cite{borelli2020architectural} \cite{raj2020modelling}\\
\hline
\textbf{ Storing Data} (retrieve, save, archive, ...) & \cite{balusamy2021big} \cite{borelli2020architectural} \cite{raj2020modelling} \\
\hline
\textbf{ Analyzing Data} (describe, diagnose, predict, ...) & \cite{raj2020internet} \cite{balusamy2021big} \cite{ardagna2018model}\\
\hline
\textbf{ Consuming Data} (API, query, visualize) & \cite{alexandru2016big} \cite{ardagna2018model} \\
\hline
\end{tabular}
\label{tab:meta-classes}
}
\end{table*}

The main focus of this paper is to provide an architecture description language \emph{\gls{daml}} that describes the data architecture of data-intensive applications. Section \ref{sec:LR} reports some literature review we used to identify the \emph{\gls{daml}} meta-model. Section \ref{sec:meta-model} reports the meta-model and related meta-classes.

\subsection{Literature Review}
\label{sec:LR}

This section presents an overview of the primary studies found in the literature to discover the data architecture elements. 
Table \ref{tab:meta-classes} shows the main meta-classes composing the \emph{\gls{daml}} and related references.

To make informed decisions, high-quality data is essential \cite{cai2015challenges}. It's crucial to ensure that data is monitored, traced, validated, and transformed from the source to its final destination. One way to manipulate data is through a data pipeline, which automates data flow from one node to another. 
Data pipelines consist of nodes and connections. Nodes carry out tasks such as generation, collection, processing, and storage, while connections transmit data between them \cite{raj2020modelling}. Data can be structured, unstructured, or semi-structured and may require transformation after processing. Data speed and processing method (batch or real-time) are important. Storage type and location (Cloud or on-premise) must be determined for effective retrieval. Data Analytics extracts insights from stored data for informed decisions using descriptive, diagnostic, predictive, and prescriptive techniques like quantitative, qualitative, and statistical methods. To interpret data for business use, it must be presented to users for analysis. Analysts alone cannot fully interpret the data without visualization.
In order to enable the meta-model to depict two tiers of architecture, namely the \emph{\gls{hla}} and \emph{\gls{lla}}, we must incorporate additional information into the primary processes. For instance, data processing may encompass a range of tasks, including classification, filtering, sorting, transformation, cleaning, and so on. This additional detail helps to illustrate the data's behavior.

\subsection{Meta-Model}\label{sec:meta-model}
\begin{center}
  \begin{figure*}[!h]
	\centering
	\makebox[\textwidth]
	{
	    \includegraphics[width=0.95\paperwidth]
	    {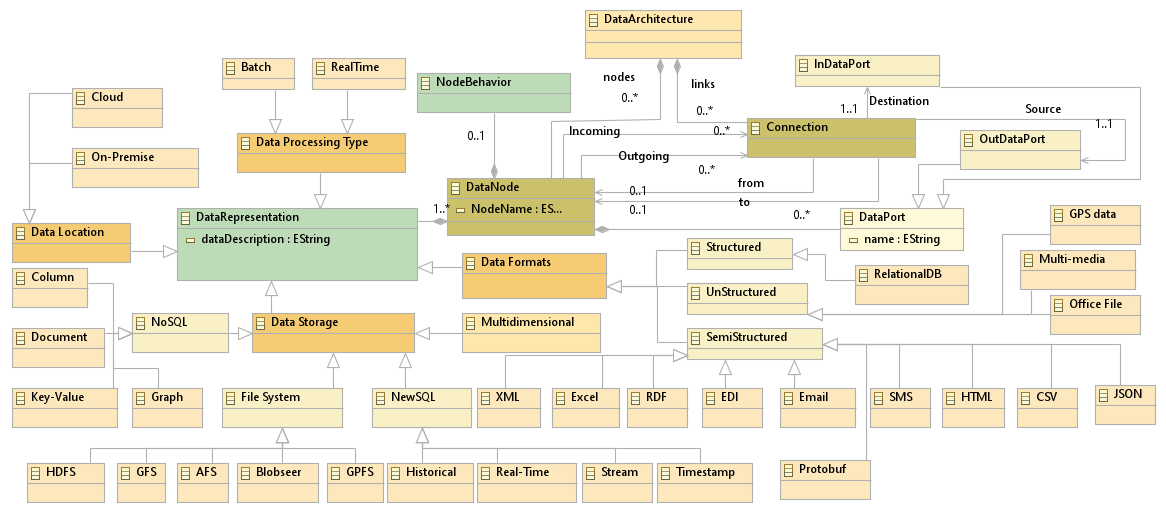}
	}
	\caption{Meta-model: structural concepts}
	\label{fig:dv_mm_s}
    \end{figure*}
\end{center}

\begin{center}
 \begin{figure*}[!ht]
	\centering
	\makebox[\textwidth]
	{   	\includegraphics[width=0.95\paperwidth]{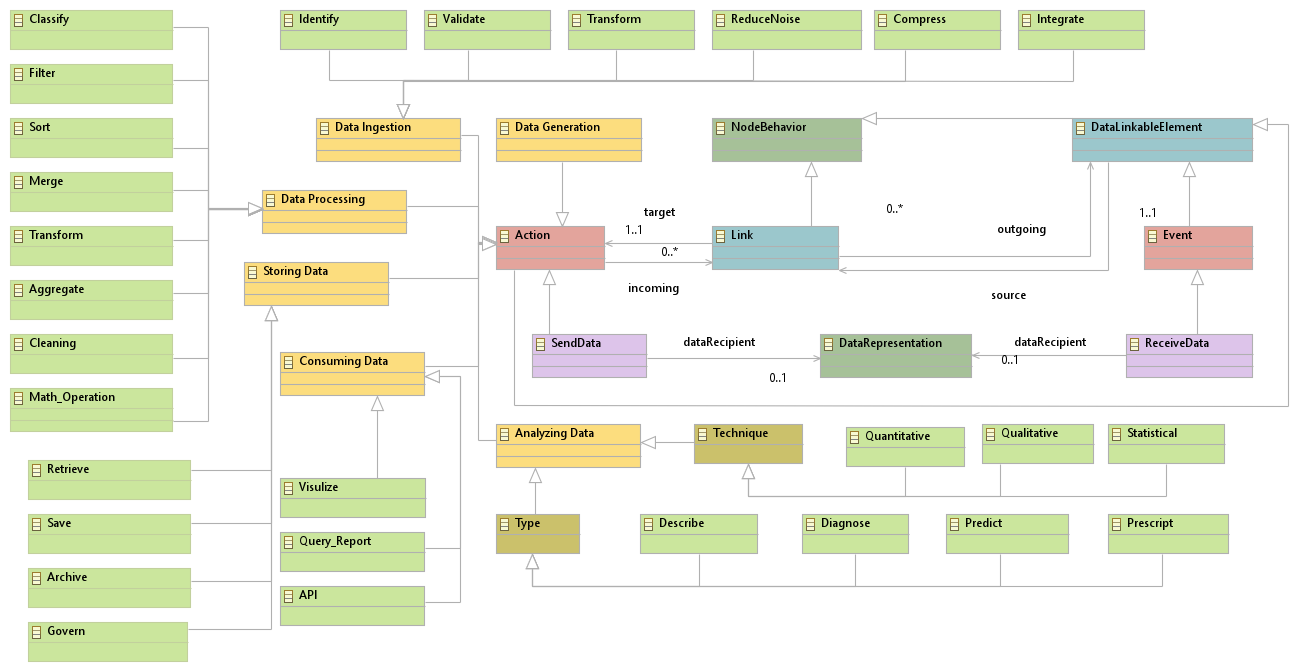}
	}
	\caption{Meta-model: behavioral concepts}
	\label{fig:dv_mm_b}
    \end{figure*}
\end{center}

By defining the underlying meta-model for \emph{\gls{daml}}, then we can
formalize the structure and constructs of the \emph{\gls{daml}} language. Figure \ref{fig:dv_mm_s} and Figure \ref{fig:dv_mm_b} show the parts of the \emph{\gls{daml}} meta-model related
to its structural and behavioral concepts, respectively.

Any {\tt DataArchitecture} of IoT can contain a set of {\tt DataNodes} (components) and {\tt Connections}. A Component is considered a computational unit with an internal state and a known interface \cite{component}. The internal data state of a component is denoted by the current behavior of data representation and its values. A {\tt DataRepresentation} can be described as a list of internal data elements; data representation is represented by actions and events defined in the component behavior, such as {\tt SendData}, {\tt ReceiveData}, etc.
 Data representation includes data formats, data storage technologies, location and processing type.
   
Our meta-model covers all {\tt DataFormats} to describe data diversity of different data sources, {\tt Structured} ({\tt RelationalDB}) , {\tt Semi-Structured}  ({\tt Email, SMS, CSV, JSON, XML} etc.) and {\tt Unstructured} ( {\tt GPS data}, {\tt Multi-media} and {\tt Office Files}).
Data {\tt ProcessingType} to describe how data is going to be processed as a {\tt Batch} or {\tt Real-time}. To describe where data will be stored, we use data storage technologies {\tt DataStorageTech.} that includes {\tt NoSQL} Databases ({\tt Document, Key-value, graph, and column}), {\tt NewSQL} Databases ({\tt Historical}, {\tt Real-Time}, {\tt Stream}, {\tt Timestamp}), and {\tt File System}  ({\tt HDF, GFS, AFS, GPFS and Blobseer}).
Moreover, {\tt Location} describes the location of the data nodes on the cloud or locally.
   
   A {\tt NodeBehavior} indicates the current status of the component that describes the data in a specific data node. For example in a generation data node, we can find elements that describe the source of the data and the format. Every NodeBehavior has a set of behavioral elements denoted by actions, and events that all together depict the data flow within the component.
   Data nodes can interact by passing data through data ports ({\tt DataPort}).  For receiving incoming data input data ports ({\tt InDataPort}) are used while output data ports ({\tt OutDataPort}) are used for sending outgoing data. The Actual communication methods of a message  are shown in Figure \ref{fig:dv_mm_b}. In this context, a connection represents a unidirectional communication channel between two data ports of two different components.
  An action is considered an important behavioral element that represents an atomic task that could be performed inside the data node. This element can be executed when a previous action in the behavioral data flow has been achieved or it could be triggered by an event like {\tt ReceiveData}. Figure \ref{fig:dv_mm_b} shows the main actions of data behavioural elements to describe a data behavior inside the data node. For example, Generation represents the source of the data. Ingestion describes how data can move from source to data lake. Processes include a list of sub-processes that could be used to describe a complete processing node. Store shows the main tasks to save, retrieve, archive, and govern the data. Analyze, is used to describe which type or technique could be used for analyzing the data. Consume to show how data could be consumed  like visualize, reports and API.
  An {\tt Event} is triggered in response to the external
stimulus of the component (e.g., the data reception on an
input data port). To show the data flow and connection between the events and actions , we use {\tt links}, and we could use them to decide the order in which actions can be performed and which one must be executed directly after an event.

\section{Application of \emph{\gls{daf}} Model to the Case Study}
\label{sec:DAF_casestudy}

This section will discuss one of the five evaluated case studies and their corresponding applications.

\begin{center} 
   \begin{figure*}[!h]
	\centering
	\makebox[\textwidth]
	{
	    	\includegraphics[width=0.85\paperwidth]{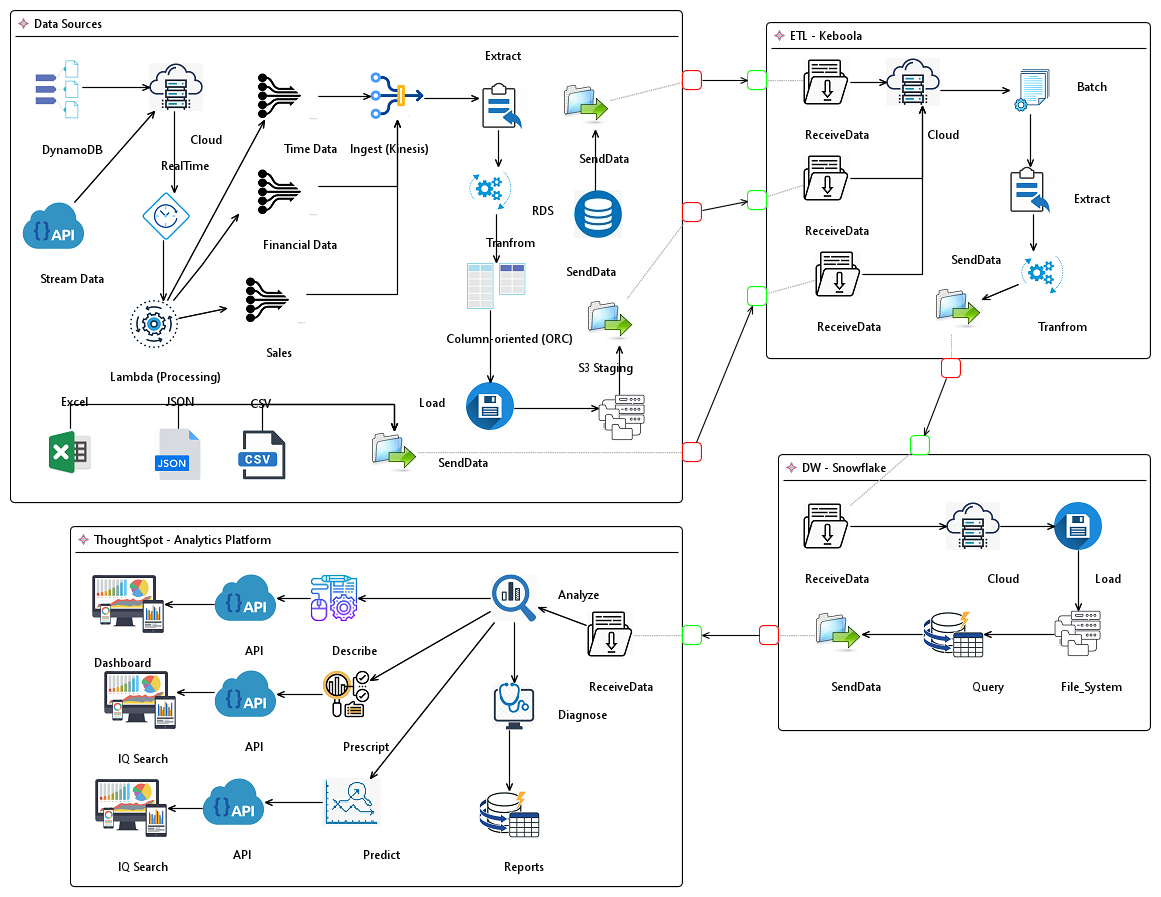}
	}
	\caption{Modeling of the \emph{\gls{adw}} }
	\label{fig:DAF_ADW}
    \end{figure*}
\end{center}

\subsection{Analytical Data Warehouse}
The \emph{\gls{adw}} \ref{section:RM_Cases} collects data from diverse sources, including AWS RDS instances, staged ORC files on S3, CSV files, worksheets, and JSON objects to enhance analytics.

Keboola Connectors is a cloud-based tool with connectors for various data sources. It also provides a suitable environment for data transformations and scheduling of extractions. Keboola prepares data for the Snowflake data warehouse, which is used by the analytics framework ThoughtSpot for creating reports, dashboards, and KPIs.

ThoughtSpot provides customized reports and dashboards as well as AI-powered data insights and searches. These seamlessly integrate with the Company's system, offering a unified working environment and white-labeled embedding capabilities.

\subsection{The model applied to the Case Study}

This section will illustrate how \emph{\gls{daf}} was used to model the ADW case study. Figure \ref{fig:ADW} displays the primary 4 data nodes, namely Data Sources, Processing Node (ETL - Keboola), Data warehouse (Snowflake), and Analytics platform (ThoughtSpot), from a structural standpoint. It is important to note that the figure presented here is a screenshot of our \emph{\gls{dat}} tool in action \cite{abughazala2022dat}. \emph{\gls{dat}} is built on the top meta-model of \emph{\gls{daf}}.

Data is collected from various sources and integrated with Time, Financial, and Sales data. It's transformed and saved on Amazon S3's File System using Keboola Node, which handles Excel, JSON, and CSV files.
This tool connects to different data sources, extracts and prepares data, stores it in a cloud-based warehouse, and generates reports, dashboards, and KPIs for customers.
 
\section{Evaluation} 
\label{sec:DAF_evalSec}



The purpose of Industrial Evaluations is to assess the effectiveness of \emph{\gls{daf}} in accurately describing real-world industrial scenarios and to develop a useful tool for industrial purposes. To achieve this, we involve companies in the evaluation process to gather their feedback.

We have successfully modeled \textit{five distinct cases} for various companies, and the feedback we received has allowed us to improve our tool and make it more efficient for industrial usage. In some instances, we discovered that a higher level of architecture was necessary. For example, instead of utilizing the sub-action of ingestion, it was sufficient to have ingestion as the primary action. This led us to develop two levels of architecture, namely the\textbf{ High Level of Architecture (\emph{\gls{hla}})} and the \textbf{Lower Level of Architecture (\emph{\gls{lla}})}. We also received feedback regarding \textbf{messaging patterns}, and we realized that we needed to incorporate additional messaging patterns into the \emph{\gls{daf}}. These patterns included request/response, publish/subscribe, asynchronous messaging, and synchronous messaging, among others.

These modeled cases were a data pipeline, operational data warehouse, analytical data warehouse, and big data architectures like Lambda and Kappa. 

we have gotten feedback from different stockholders, Big data team leaders, Big data Architects, and Big data engineers, table \ref{tab:DAF_evaluation} shows more details about the evaluation.



\begin{flushleft}
\label{tab:DAF_evaluation}
\begin{table*}[!ht]
	\centering
	\caption{Outline of use cases and roles of the evaluators}
	\label{tab:DAF_evaluation}
	\begin{tabular}{|p{2.5cm}|p{6cm}|p{4cm}|} 
	\hline
	\textbf{Company} & \textbf{Use cases} & \textbf{Experts Roles} \\
	\hline
	\multirow{4}{*}{\textbf{Harri }} & \multirow{2}{*}{Operational Data Warehouse} & Big Data Team Lead \\
	\cline{3-3}
	& & Big Data Architect\\
	\cline{2-3}
	& \multirow{2}{*}{Analytical Data Warehouse} & Big Data Engineer\\
	\cline{3-3}
	&  & Big Data Architect\\
	\hline
	\multirow{3}{*}{\textbf{HP Co.}} & \multirow{3}{*}{Data Pipeline} & Big Data Team Lead \\
	\cline{3-3}
	& & Big Data Architect\\
	\cline{3-3}
	&  & Big Data Engineer\\
	\hline
	
	\textbf{Hydre} & Data Architectures (Lambda, ...) & Researcher \\
	\hline
	
	\textbf{Locally} & NdR Data Architecture & Students \\
	\hline
	\end{tabular}
\end{table*}
\end{flushleft}

Table \ref{tab:daf_cases} presents an overview of the evaluated real cases, including their main components, concerns, and corresponding models.

\begin{table*}[!ht]
\centering
\caption{Real Case Evaluation: Main Components}
\label{tab:daf_cases}
\makebox[\linewidth]{
\begin{tabular}{|l|l|l|l|l|l|l|}
\hline
\multicolumn{1}{|c|}{\textbf{Cases}} &
  \multicolumn{1}{c|}{\textbf{\begin{tabular}[c]{@{}c@{}}Data \\ Formats\end{tabular}}} &
  \multicolumn{1}{c|}{\textbf{\begin{tabular}[c]{@{}c@{}}Processing \\ Type\end{tabular}}} &
  \multicolumn{1}{c|}{\textbf{\begin{tabular}[c]{@{}c@{}}Storage \\ Type\end{tabular}}} &
  \textbf{Location} &
  \multicolumn{1}{c|}{\textbf{\begin{tabular}[c]{@{}c@{}}Data \\ Stages\end{tabular}}} &
  \textbf{Figure} \\ \hline
\textbf{ADW} &
  \begin{tabular}[c]{@{}l@{}}Structured,   \\ Semi-structured\end{tabular} &
  \begin{tabular}[c]{@{}l@{}}RealTime\\  Batch\end{tabular} &
  \begin{tabular}[c]{@{}l@{}}NoSQL \\ NewSQL \\ File\end{tabular} &
  Cloud &
  \begin{tabular}[c]{@{}l@{}}Generation   \\ Ingestion \\ Processing \\ Storing\\ Analyze \\ Visualize \\ Share\end{tabular} &
  \ref{fig:DAF_ADW} \\ \hline
\textbf{ODW} &
  \begin{tabular}[c]{@{}l@{}}Structured,   \\ Semi-structured\end{tabular} &
  \begin{tabular}[c]{@{}l@{}}RealTime\\  Batch\end{tabular} &
  \begin{tabular}[c]{@{}l@{}}NoSQL \\ NewSQL\\  File\end{tabular} &
  Cloud &
  \begin{tabular}[c]{@{}l@{}}Generation   \\ Ingestion \\ Processing \\ Storing \\ Analyze \\ Visualize \\ Share\end{tabular} &
  \ref{fig:DAF_odw} \\ \hline
\textbf{Hydre} &
  \begin{tabular}[c]{@{}l@{}}Structured,   \\ Semi-structured\end{tabular} &
  \begin{tabular}[c]{@{}l@{}}RealTime\\  Batch\end{tabular} &
  \begin{tabular}[c]{@{}l@{}}NoSQL \\ NewSQL\\  File\end{tabular} &
  Cloud &
  \begin{tabular}[c]{@{}l@{}}Generation   \\ Ingestion \\ Processing \\ Storing\\ Analyze \\ Visualize\end{tabular} &
  \ref{fig:DAF_Hydre} \\ \hline
\textbf{Data Pipeline} &
  Semi-structured &
  Batch &
  \begin{tabular}[c]{@{}l@{}}NoSQL \\ NewSQL\\  File\end{tabular} &
  Cloud &
  \begin{tabular}[c]{@{}l@{}}Generation   \\ Ingestion\\ Processing \\ Storing\\ Visualize\end{tabular} &
  \ref{fig:DAF_hp_err} \\ \hline
\textbf{NDR} &
  \begin{tabular}[c]{@{}l@{}}Structured, \\ Un-structured\end{tabular} &
  \begin{tabular}[c]{@{}l@{}}RealTime\\  Batch\end{tabular} &
  \begin{tabular}[c]{@{}l@{}}NoSQL \\ NewSQL\\  File\end{tabular} &
  Clould &
  \begin{tabular}[c]{@{}l@{}}Generation   \\ Ingestion \\ Processing \\ Storing\\ Analyze \\ Visualize \\ Share\end{tabular} &
  \ref{fig:DAF_ndr-hla} \\ \hline
\end{tabular}
}
\end{table*}

The \emph{\gls{daf}} framework is an exceptional solution that accurately models real-life industry cases like data pipelines, data warehouses, and big data architectures such as Lambda and Kappa.
\begin{center} 
   \begin{figure*}[!h]
	\centering
	\makebox[\textwidth]
	{
	    	\includegraphics[width=0.80\paperwidth]{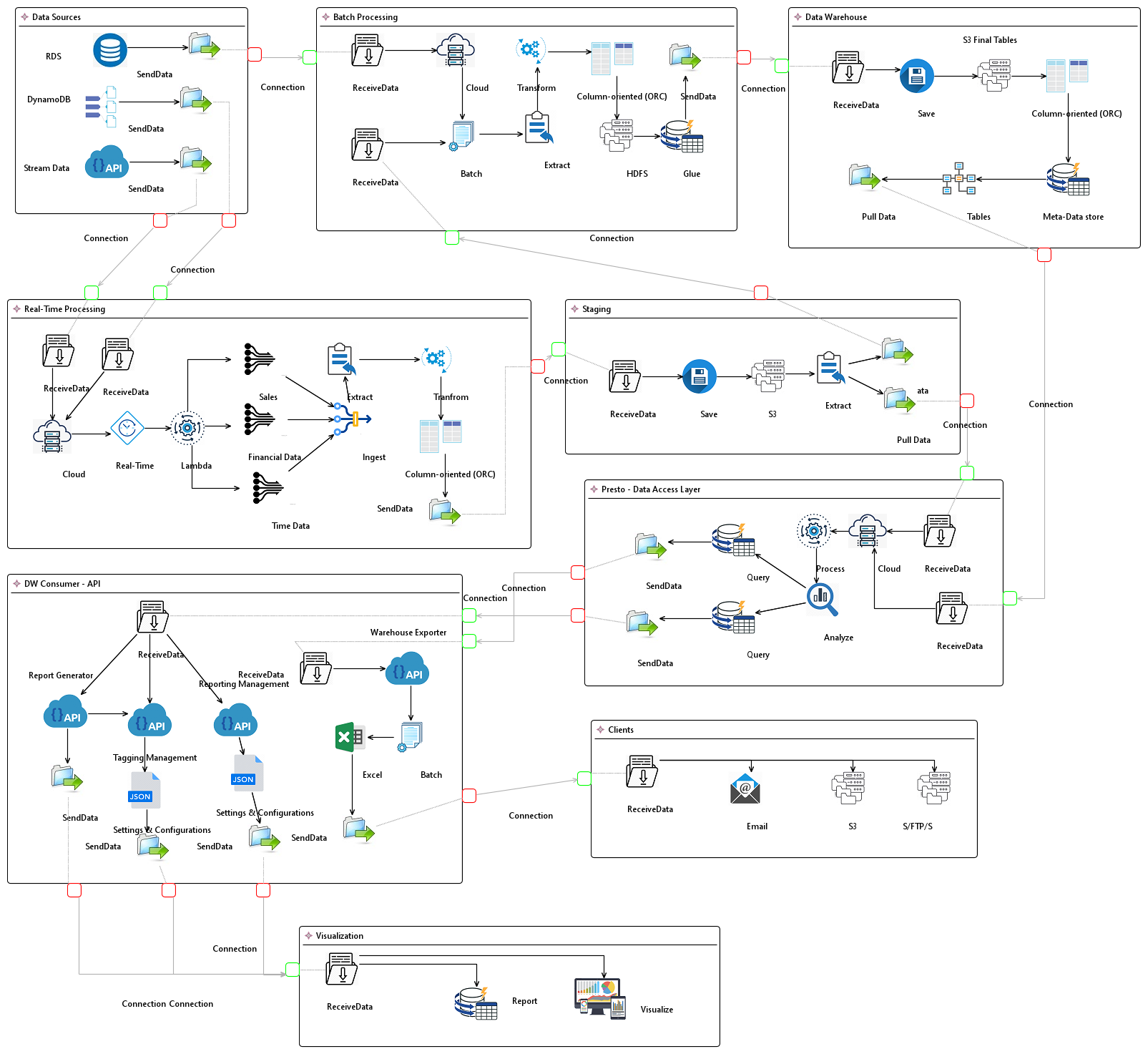}
	}
	\caption{Operational Data Warehouse}
	\label{fig:DAF_odw}
    \end{figure*}
\end{center}

\begin{center} 
   \begin{figure*}[!h]
	\centering
	\makebox[\textwidth]
	{
	    	\includegraphics[width=0.80\paperwidth]{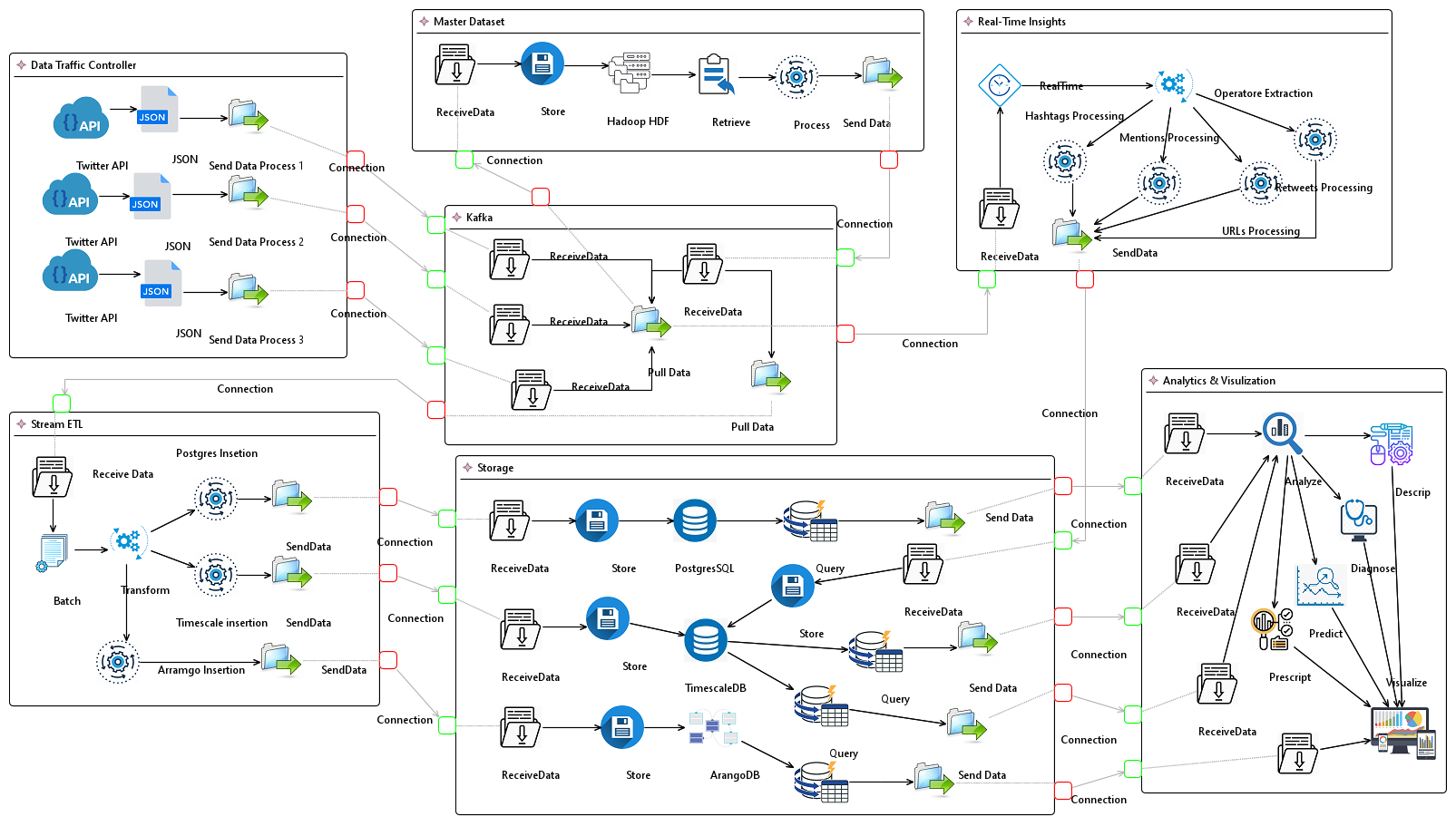}
	}
	\caption{Hydre (Lambda+ Example)}
	\label{fig:DAF_Hydre}
    \end{figure*}
\end{center}

\begin{center} 
   \begin{figure*}[!h]
	\centering
	\makebox[\textwidth]
	{
	    	\includegraphics[width=0.80\paperwidth]{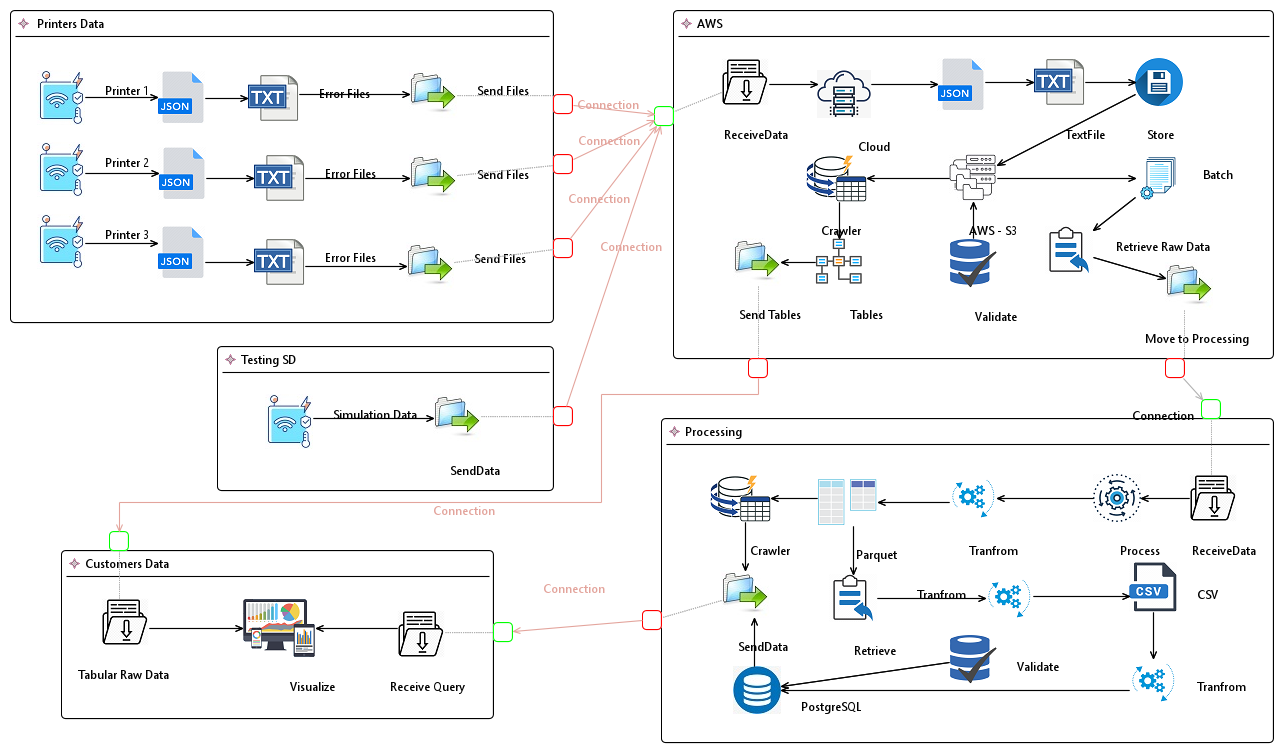}
	}
	\caption{Errors Data Pipeline}
	\label{fig:DAF_hp_err}
    \end{figure*}
\end{center}

\begin{center} 
   \begin{figure*}[htbp]
	\centering
	{
	    	\includegraphics[width=0.9\textwidth, height=7cm]{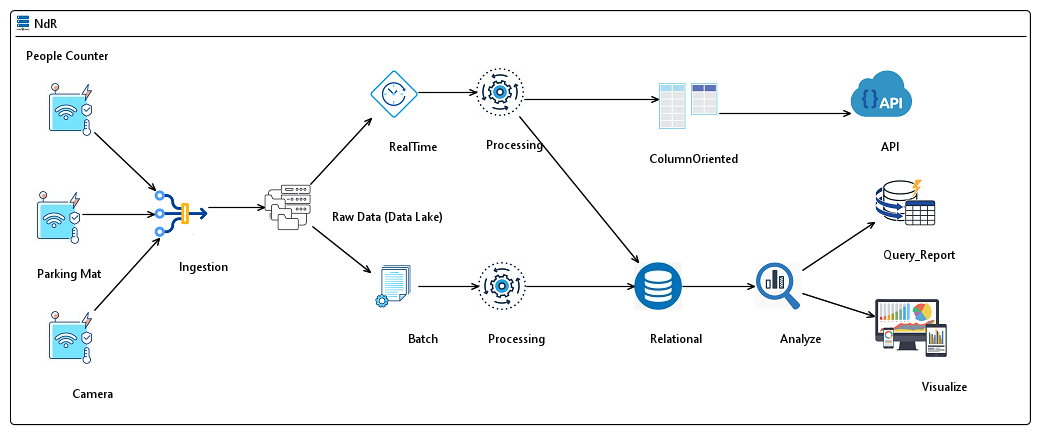}
	}

	\caption{NDR case study - (HLA)}
	\label{fig:DAF_ndr-hla}
    \end{figure*}
\end{center}

\section{Threats To Validity}
The research was conducted with reference to the \emph{\gls{adw}}, which is an analytics architecture that was created by multiple data teams within a single company, situated in different regions across the globe.
The model was validated by an experienced practitioner and other team members not involved in the study to address internal validity. 

Furthermore, the study was validated again by the Hayder (From Lambd+ Author) case study Fig. \ref{fig:DAF_Hydre}, Error Data Pipeline Fig. \ref{fig:DAF_hp_err}, \emph{\gls{odw}}, in the validation process by one more external company from the telecommunications domain (Internet Service Provider). 

\section{Related Work}

This section reviews relevant studies that are related to 
exploiting the most related research to data-intensive IoT. Raj and Bosch \cite{raj2020modelling} proposed a conceptual model for a data pipeline, which contains two main components (nodes and connectors); the node represents the main abstract data node, and the connection represents the way to carry and transmit the data between nodes. \emph{\gls{daf}} includes all data life cycle phases to be able to provide two levels of architecture, \emph{\gls{hla}} (High-Level Architecture) and \emph{\gls{lla}} (Low-Level Architecture); \emph{\gls{lla}} gives the ability to describe the behavior of data nodes( different data formats from different sources, processing types, operations on the data, location, and etc.). 
Borelli  \cite{borelli2020architectural} 
A proposed classification for main software components and their relationships to model a software architecture for particular IoT applications. These components represent the abstract components. \emph{\gls{daml}} has the ability to represent all of the mentioned components and their behavior too. 
Erraissi \cite{erraissi2018data} \cite{erraissi2019big} proposed a meta-model for data sources, ingestion layers, and Big Data visualization layer. \emph{\gls{daml}} has the ability to describe the data in each layer (Source, Ingestion, Processing, Store, Analyze, and Consume). 
 Nesi \cite{nesi2018auditing} provided a solution based on a set of instruments to collect the data in real-time, store, and audit data flow for IoT smart City architecture. Using \emph{\gls{daf}}, we can describe data flow from the source to the final destination at an abstract concept level; It is not technologies-based. 
Bashir \cite{bashir2020big} proposed a meta-model for Data Management and Analytics for IoT smart building. Our meta-model provides a details description of ingestion, data processing, storing data, analyzing, and visualizing the data.

\section{Conclusion and Future Work}

Our paper introduced the \emph{\gls{daf}}, a framework for engineering Data Architecture for Data-Intensive applications. With the \emph{\gls{daf}}, stakeholders can easily describe both High-Level Architecture (\emph{\gls{hla}}) and Low-Level Architecture (\emph{\gls{lla}}). The framework encompasses all essential components of Data Architecture and follows industry best practices. The \emph{\gls{daf}} can model data pipelines, data warehouses, and big data architectures such as Lambda and Kappa.

This is an initial starting point for our future work plan, which can be extended to include. First, finishing the current running evaluations with other industrial cases. Second, extending this work to focus more on data quality. Third, integrating the \emph{\gls{daf}} with other existing technologies and tools.

\chapter{DAT: Data Architecture Modeling Tool for Data-Driven
Applications}
\thispagestyle{plain}

\newacronym{dd}{DD}{Data-Driven}
\newacronym{emf}{EMF}{Eclipse Modeling Framework}
Data is the key to success for any \emph{\gls{dd}} Organization, and managing it is considered the most challenging task. Data Architecture (\emph{\gls{da}}) focuses on describing, collecting, storing, processing, and analyzing the data to meet business needs.
In this tool demo paper, we present the \emph{\gls{dat}}, a model-driven engineering tool enabling data architects, data engineers, and other stakeholders to describe how data flows through the system and provides a blueprint for managing data that saves time and effort dedicated to Data Architectures for IoT applications.
We evaluated this work by modeling five case studies, receiving expressiveness and ease of use feedback from two companies, more than six researchers, and eighteen undergraduate students from the software architecture course.

\section{Introduction}

The International Data Corporation (IDC) \cite{idc} expects that by 2025 there will be more than 175 zettabytes of valuable data for a compounded annual growth rate of 61\%. Ninety zettabytes of data will be from IoT devices, and 30\% of the data generated will be consumed in real-time. 
A {\em data architecture \emph{\gls{da}}} is an integrated set of specification artifacts used to define data requirements, guide integration, control data assets, and align data investments with business strategy. It also includes an integrated collection of master blueprints at different levels of abstraction \cite{10.5555/3165209}.

This tool demo paper presents the {\em Data Architecture Modeling Tool (\emph{\gls{dat}})}, an architecture modeling tool for the model-driven engineering of data architecture for data-driven applications, DAT is a graphical modeling tool for DAF \cite{abughazala2023architecture}.  

\emph{\gls{dat}} is a data architecture modeling tool for IoT applications that shows how data flows through the system and provides a blueprint for it. It allows the stakeholders to describe two levels of data architecture: high-level Architecture (\emph{\gls{hla}}) and Low-Level Architecture (\emph{\gls{lla}}). It focuses on representing the data from source to destination and shows formats, processing types, storage, analysis types, and how to consume it.

The rest of this tool demo paper is organized as follows. The methodology is presented in Section 2. The application of \emph{\gls{dat}} to a real case study is described in Section 3. The \emph{\gls{dat}} evaluation is presented in Section 4. Related work is discussed in Section 5, while conclusions are drawn in Section 6. 

\section{Background}
\label{sec:Background}

The main focus of this paper is to describe the data architecture of IoT applications through the {\em Data Modeling Language (\emph{\gls{daml}})}. 
CAPS Framework in Section \ref{caps}. Section \ref{DATTool} shows \emph{\gls{daml}} and reports on the technologies used to implement the DAT.

\subsection{The CAPS Modeling Framework}
\label{caps}
CAPS \cite{muccini2017caps} is an environment where Situational Aware Cyber-Physical Systems (SiA-CPS) can be described through software, hardware, and physical space models. The CAPS found three main architectural viewpoints of extreme importance when describing a
SiA-CPS: the software architecture structural and behavioral view (SAML), the hardware view (HWML), and the physical space view
(SPML).

This environment is composed of the CAPS modeling framework\footnote{CAPS: http://caps.disim.univaq.it/} and the CAPS code generation framework \cite{sharaf2017architecture} \cite{sharaf2018arduino} that aim to support the architecture description, reasoning, design decision process, and evaluation of the CAPS architecture in terms of data traffic load, battery level and energy consumption of its nodes.

\subsection{The Important of Data Architecture}\label{DATTool}
Data architecture is important because it helps organizations manage and use their data effectively. Some specific reasons why data architecture is important to include: 
\begin{enumerate}
\item  Data quality: it helps to collect, store, and use data consistently and accurately. This is important for maintaining data integrity and reliability and avoiding errors or inconsistencies impacting business operations.

\item Data security: it helps to protect data from unauthorized access or modification and ensure that it is used compliantly and ethically. This is particularly important in industries with strict regulations, such as healthcare or finance.

\item Organizational efficiency: it helps organizations better understand and manage their data, increasing efficiency and productivity. By defining the structures, policies, and standards that govern data within an organization, data architecture can help streamline processes and improve decision-making.

\item Business intelligence and analytics: organizations need to collect, store, and analyze large amounts of data. This can support better decision-making, improve customer relationships, and drive business growth.

\item Scalability and flexibility: A well-designed \emph{\gls{da}} can support the growth and evolution of an organization. It allows organizations to easily add new data sources, incorporate new technologies, and adapt to changing business needs.
\end{enumerate}

\subsection{The DAT Tool}\label{DATTool}
The \emph{\gls{dat}} modeling framework \footnote{DAT Tool Source Code can be found at \url{https://github.com/moamina/DAT}} \footnote{DAT Tool video demo: \url{https://youtu.be/Du0VDg1CLlQ}} gives data architects the ability to define a {\em data view} for \emph{\gls{dd}} IoT Applications through the \emph{\gls{daml}} modeling language \cite{10092710}. 
\begin{center} 
   \begin{figure*}[h!]
	\centering
	\caption{The Data View of CAPS}
	{
	    	\includegraphics[width=1\textwidth]{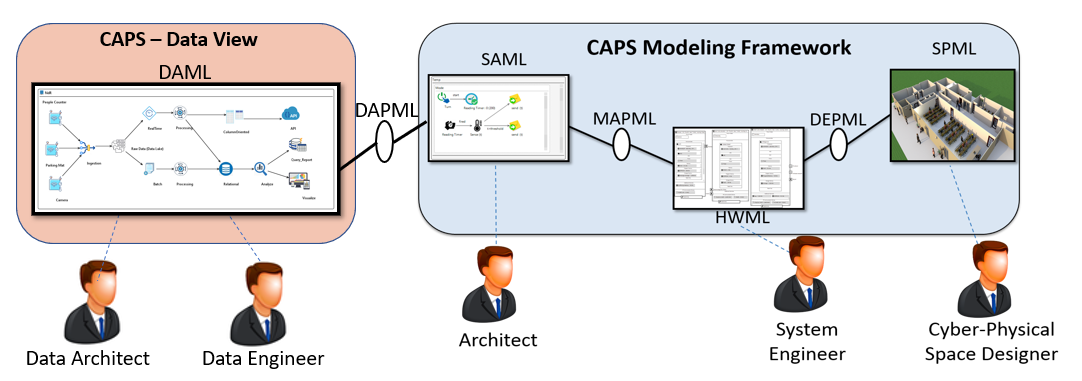}
	}
	\label{fig:dataView_CAPS}
    \end{figure*}
\end{center}
\subsubsection{Technologies.}
Our tool is based on \emph{\gls{mde}}. For that, we use \emph{\gls{emf}} \cite{emf}  for building tools and applications based on the structure data model, which consists of three main parts, \gls{emf} Core, includes a meta-model for describing the models. \gls{emf} Editors contains generic reusable classes for building editors for \gls{emf} models. Eclipse Epsilon \cite{epsilon} is a Java-based scripting language for model-based software engineering tasks (e.g., model-to-model transformation and model validation) which strongly support \gls{emf} and works with UML, XML, Simulink, etc. To create graphical editors and views for the gls{emf} models, we used Eugenia \cite{eugenia}. It is a tool to create a graphical model editor by generating the .gmfgraph, .gmftool, and .gmfmap models that the GMF editor from a single annotated Ecore meta-model needs. 

\gls{dat} implements the \gls{daml} meta-model to be considered a fourth view (data view) for the CAPS, as shown in Figure \ref{fig:dataView_CAPS}. How the \gls{dat} supports the modeling of data views and its application to actual use cases will be presented in Section 3.
\subsubsection{Methodology.} \gls{dat} is built based on a meta-model containing a data architecture as a top root meta-class. Any \textbf{data architecture} of IoT can contain a set of \textbf{DataNodes} (components) and \textbf{connections}. A Component is considered a computational unit with an internal state and a known interface \cite{component}. Data nodes can interact by passing data through \textbf{data ports}. A component's internal state is denoted by the current behavior of data representation and its values. Data representation includes \textit{data formats, storage technologies, location, and processing type}. Every Node Behavior has a set of behavioral elements denoted by actions and events that depict the data flow within the component. This element can be executed when a previous action in the behavioral data flow has been achieved or triggered by an event like \textbf{ReceiveData}. Other main actions are \textbf{Generation}, \textbf{Ingestion}, \textbf{Process},  \textbf{Store}, \textbf{Analyze}, and \textbf{Consume}. An \textbf{event} is triggered in response to the external stimulus of the component. To show the data flow and connection between the events and actions, we use \textbf{links}.

\subsubsection{Steps To Use.}
The architect needs to follow the following steps to use the tool to model any case:
\begin{enumerate}
\item Download the source code from The GitHub \footnote{DAT Tool Source Code can be found at \url{https://github.com/moamina/DAT}} and follow the steps in the tool demo video to lunch the tool \footnote{DAT Tool video demo: \url{https://youtu.be/Du0VDg1CLlQ}}.

\item Define which level of abstraction you need (High-Level or Low-Level). You could use a single Data Node at the High-Level, whereas you need to define the structure and behavior at the low level. 

\item Define the main data nodes and the connection between them to determine the order of each node, such as the data source is the first data node. Ingestion comes the second one, and the connection shows the data flow from the source to the ingestion data node.

\item At the internal behavior, you could use data elements (low-level elements), such as data formats (JSON, XML, Video, ...), and sub-operation, such as (classification, data reduction, cleaning, validation, filter, classification, ...).
\end{enumerate}

\section{Real Use Cases}

\begin{center} 
   \begin{figure*}[!t]
	\centering
	\makebox[\textwidth]
	{
	    	\includegraphics[width=1\textwidth]{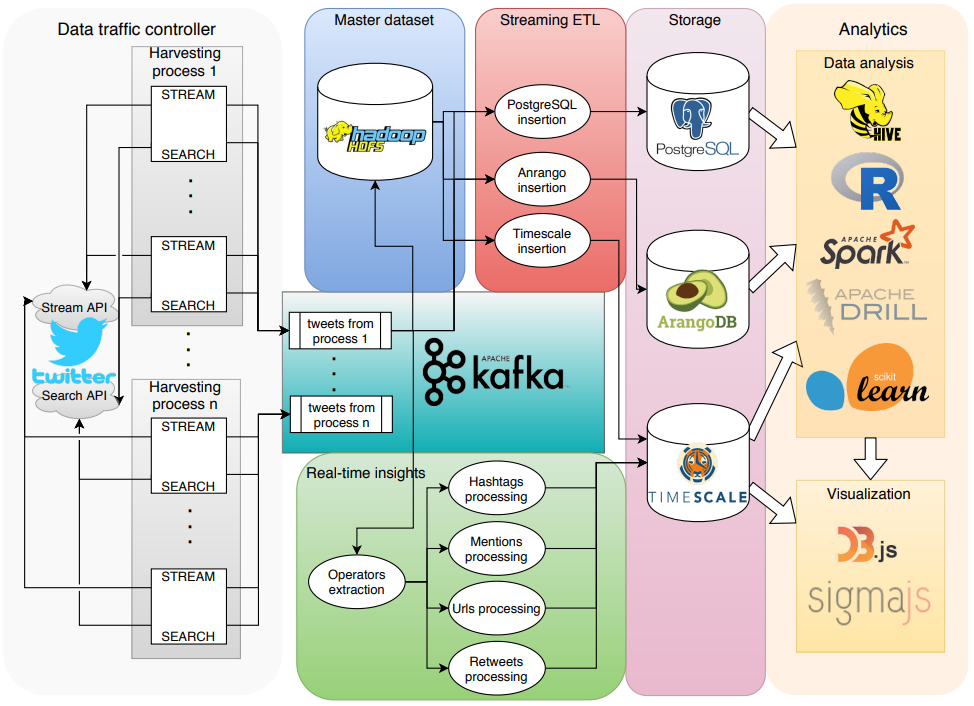}
	}
	\caption{The Hydre architecture}
	\label{fig:DAT_Hydre_realCase}
    \end{figure*}
\end{center}

\begin{center} 
   \begin{figure*}[!t]
	\centering
	\makebox[\textwidth]
	{
	    	\includegraphics[width=0.85\paperwidth]{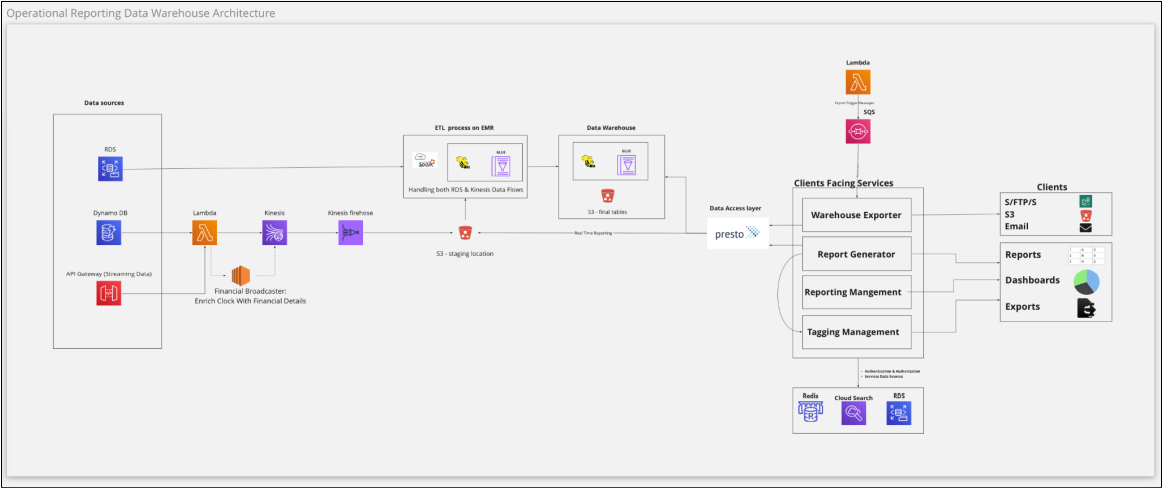}
	}
	\caption{Operational Data-warehouse}
	\label{fig:DAT_ODW}
    \end{figure*}
\end{center}

This section introduces the existing data architecture description used by three companies contributing to the \gls{dat} tool. We have chosen three cases from five cases to present our tool in the tool paper, these cases are from A (\ref{fig:DAT_ODW}),B, and D (\ref{fig:DAT_Hydre_realCase}) from section \ref{section:RM_Cases}.

\section{Evaluation}
The \gls{dat} cases were evaluated through interviews with seven industry professionals from two companies of different domains and different maturity levels and one external researcher; table \ref{tab:EvalTbl} shows more about the roles of the evaluators. The evaluation section will be structured in terms of agreements and suggestions for improvement. 
\label{sec:evaluation}
\begin{table}[h!]
	\centering
	\caption{Outline of use cases and roles of the evaluators}
	\label{tab:EvalTbl}
	\begin{tabular}{|p{2.5cm}|p{6cm}|p{4cm}|} 
	\hline
	\textbf{Company} & \textbf{Use cases} & \textbf{Experts Roles} \\
	\hline
	\multirow{4}{*}{\textbf{Company 1}} & \multirow{2}{*}{Operational Data Warehouse} & Big Data Team Lead \\
	\cline{3-3}
	& & Big Data Architect\\
	\cline{2-3}
	& \multirow{2}{*}{Analytical Data Warehouse} & Big Data Engineer\\
	\cline{3-3}
	&  & Big Data Architect\\
	\hline
	
	\multirow{3}{*}{\textbf{Company 2}} & \multirow{3}{*}{Data Pipeline} & Big Data Team Lead \\
	\cline{3-3}
	& & Big Data Architect\\
	\cline{3-3}
	&  & Big Data Engineer\\
	\hline
	
	\textbf{Company 3} & Lambda Architecture & Researcher \\
	\hline
	
	\textbf{Locally} & NdR Data Architecture & Students \\
	\hline
	\end{tabular}
\end{table}

\subsection{Errors Data Pipeline}
\begin{center} 
   \begin{figure*}[!h]
	\centering
	\makebox[\textwidth]
	{
	    	\includegraphics[width=0.80\paperwidth]{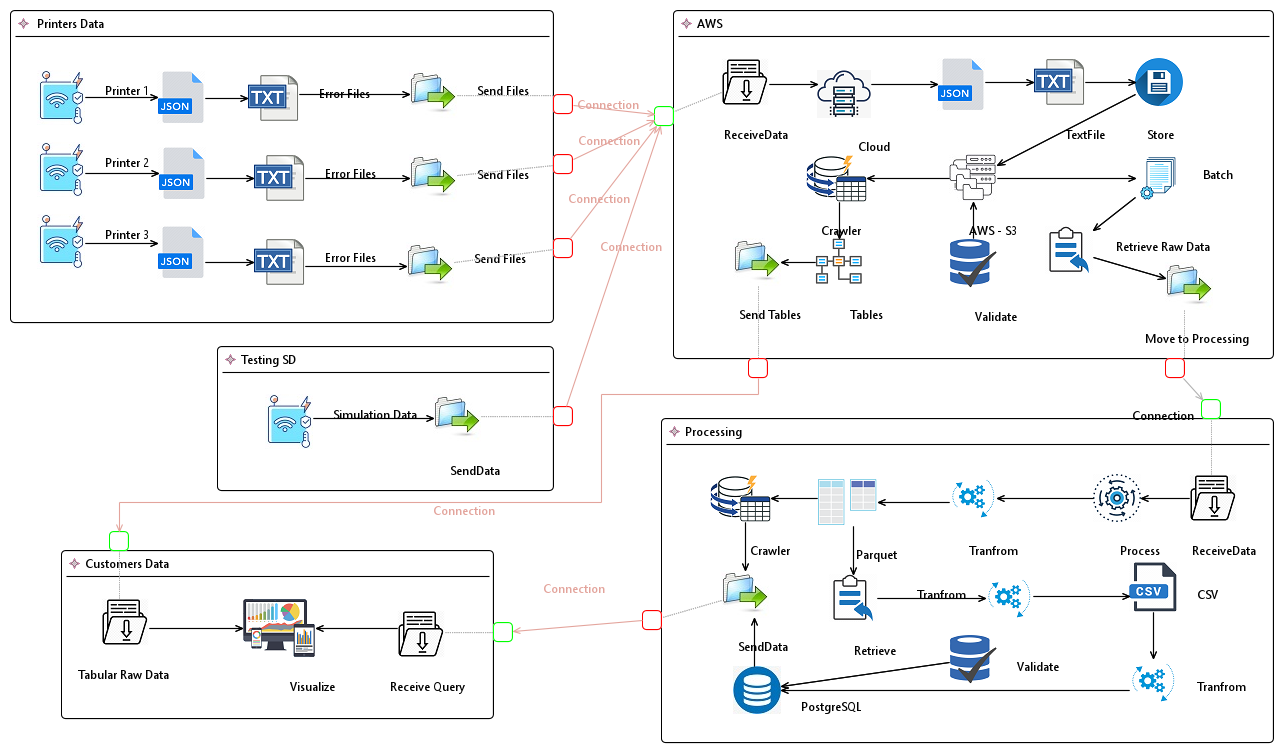}
	}
	\caption{Errors Data Pipeline}
	\label{fig:hp_err}
    \end{figure*}
\end{center}

Figure \ref{fig:hp_err} shows the Errors Data Pipeline using DAT. The first author presented the models and collected the practitioners' feedback. 

\textbf{Agreements:} The model described the data flow from generation to destination. The model was easy to understand for new data engineers. The tool has the flexibility to change and add new nodes.

\textbf{suggestion:} It is good to include the data quality metrics that could apply to the data at each stage. 

\subsection{Hydre (Lambda Architecture)}
\begin{center} 
   \begin{figure*}[!h]
	\centering
	\makebox[\textwidth]
	{
	    	\includegraphics[width=0.80\paperwidth]{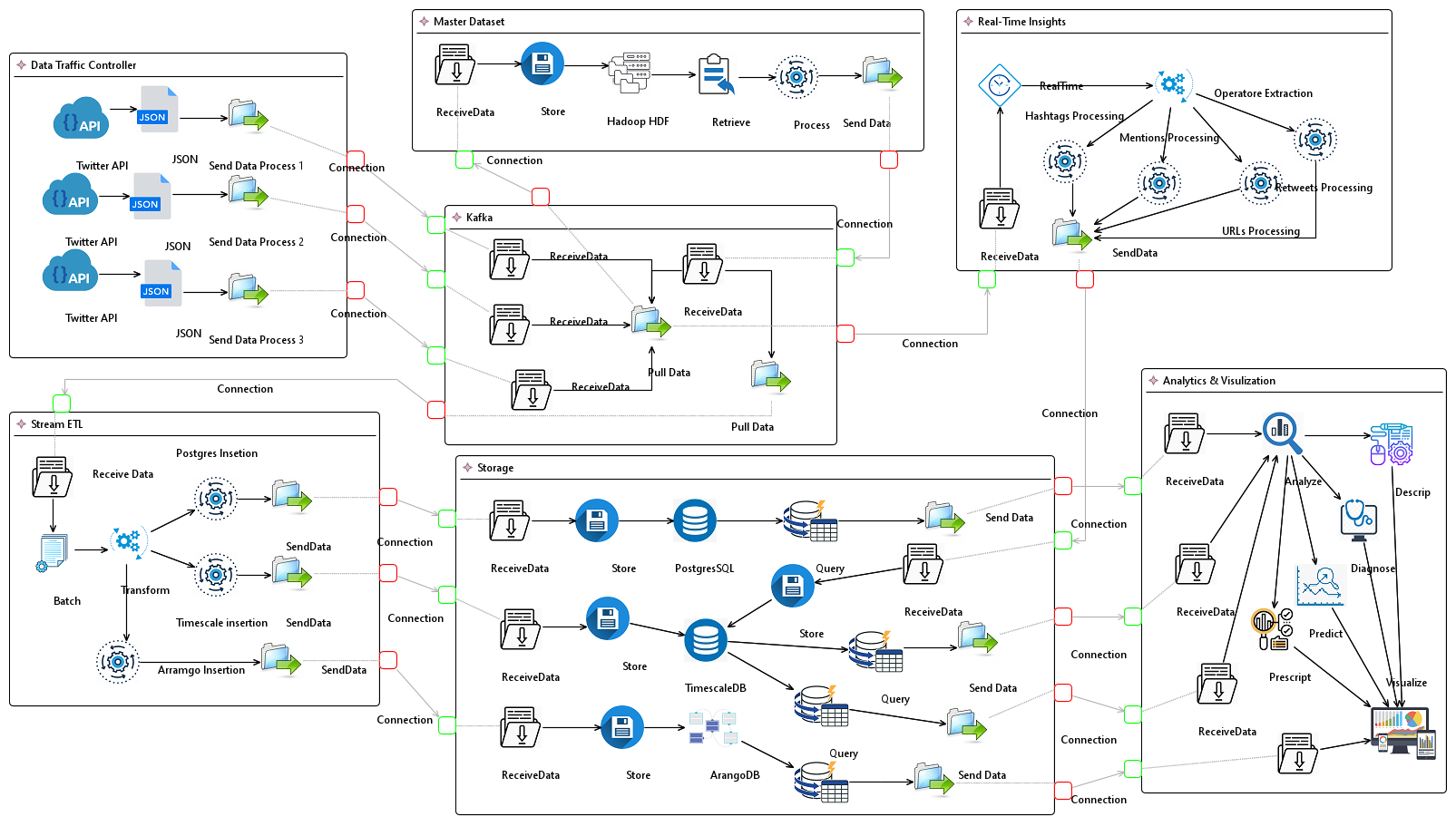}
	}
	\caption{Hydre (Lambda+ Example)}
	\label{fig:Hydre}
    \end{figure*}
\end{center}

Figure \ref{fig:Hydre} shows the Hydre model using \gls{dat}. The first author presented the models and collected feedback from Lambda+'s author. 

\textbf{Agreements:} The model represents very well the Hydre case. The indications of when data are stored on disk are helpful, especially when working on a big data architecture with people who don't know each technology's details. The real-time and batch elements are useful as well.

\textbf{suggestion:} the first suggestion is similar to the first case related to data interaction patterns. The second suggestion was a starting point for us to provide two levels of architecture, High-level architecture (\gls{hla}) and Low-Level architecture (\gls{lla}).
\subsection{Operational Data Warehouse}
\begin{center} 
   \begin{figure*}[!ht]
	\centering
	{
	    	\includegraphics[width=0.80\paperwidth]{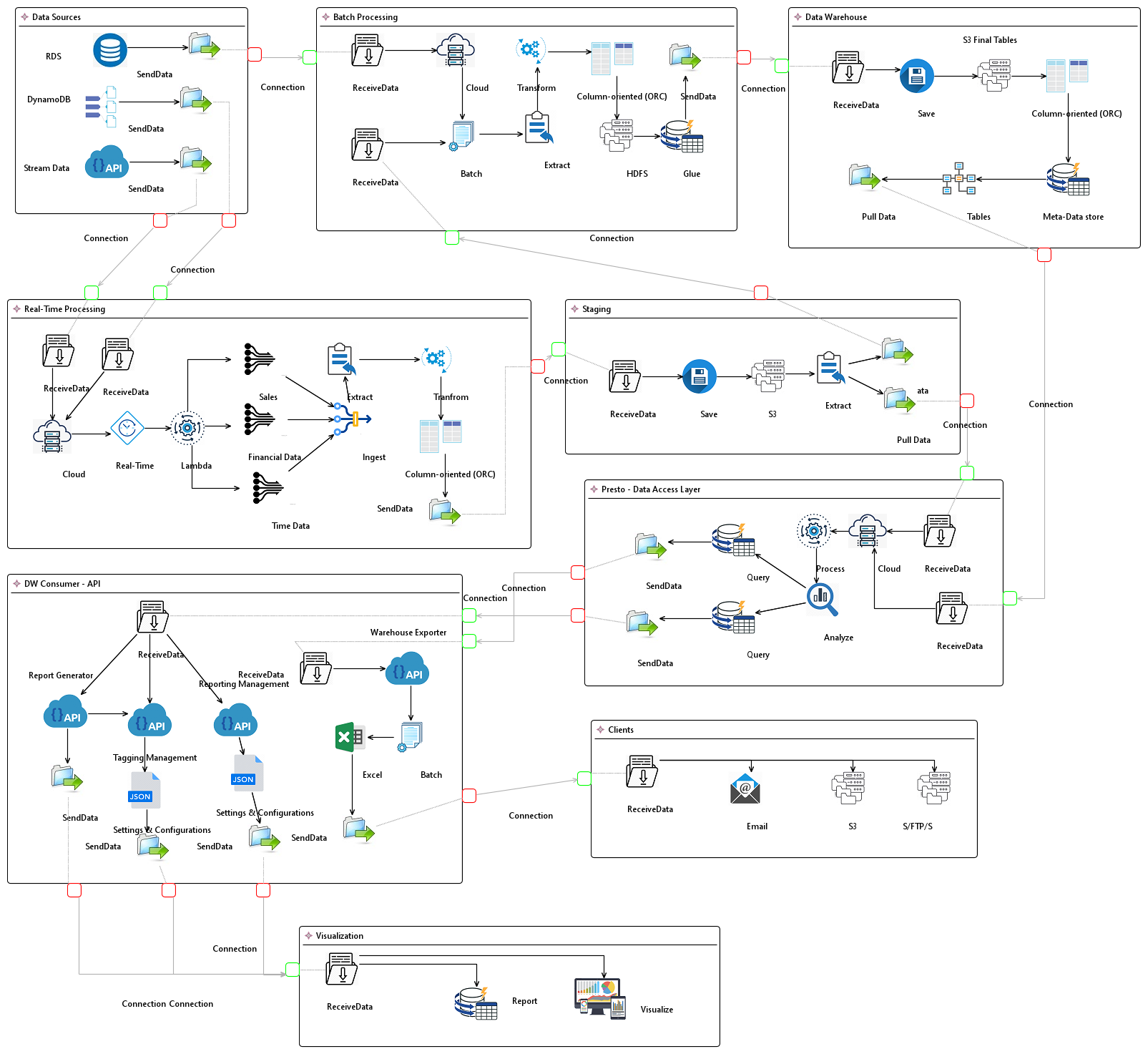}
	}
	\caption{Operational Data Warehouse}
	\label{fig:odw}
    \end{figure*}
\end{center}

Figure \ref{fig:odw} shows the \gls{odw} model using \gls{dat}. The first author presented the models and collected the practitioners' feedback. 

\textbf{Agreements:} The model was able to describe the details of the case and was easy to share and understand by different teams in other parts of the world. The model was a good communication language between team members, which means easy to avoid misinterpretations.

\textbf{suggestion:} In the current version of the \gls{dat}, the only way to show how different components interact with each other is to send/receive data. The suggestion was to include all data interaction patterns (request/response, publish/subscribe, pull/push, and others).


\section{Conclusion and Future Work}

This tool demo paper has presented the \gls{dat}, an architecture description, and the associated modeling platform for the model-driven engineering of Data Architecture for IoT. It is implemented on top of f the Eclipse Modeling Framework. It can allow the stack-holders to describe two levels of data architectures, high-level Architecture (\gls{hla}) and Low-Level Architecture (\gls{lla}).

This is an initial starting point for our future work plan, which can be extended to include finishing the current running evaluations with other companies and trying to model different big data patterns and architectures. Second, try to integrate the DAT with other existing technologies and tools.

\section{Acknowledgment}
The authors would like to thank Prof. Giovanni Stilo, Prof. Annabelle Gillet (Lambda+), Prof. Karthik Vaidhyanathan, Mostafa Shaer, Itay, and Roi from HP Team, Mustafa Tamim and Anas Eid from Harri Team, Mudassir Malik, Apurvanand Sahay, and Arsene Indamutsa as a researcher for their contributions in the evaluation.
 
\chapter{Modeling Data Analytics Architecture for
Data-Driven Applications using DAT}
\thispagestyle{plain}

\newacronym{daa}{DAA}{Data Analytics Architecture}


Data analysis plays a significant role in extracting meaningful information from big data. Data analysis consists of acquisition, storage, management, analytics, and visualization. Providing an abstract view of data analytics applications is crucial to ensure that the data will transfer into meaningful information. Data Architecture (\emph{\gls{da}}) is one of the ways to
provide that.
This article shows industrial experiences in building \emph{\gls{daa}}. We use model-driven engineering to model a data analytics architecture for applications using \gls{dat}.
We evaluated this work by modeling Analytics Data Warehouse (\emph{\gls{adw}})  as a case study from one company, receiving feedback.

\section{Introduction}


The International Data Corporation (IDC) \cite{idc} expects that by 2025 there will be more than 175 zettabytes of valuable data for a compounded annual growth rate of 61\%. Ninety zettabytes of data will be from IoT devices, and 30\% of the data generated will be consumed in real-time.

Data analytics for the Internet of Things (IoT) involves using a range of techniques and technologies to gather, store, and analyze data generated by connected devices and sensors \cite{ahmad2021review} \cite{simmhan2016big}. This can help better to understand the performance and behavior of the IoT systems and to identify patterns and trends in the data that can be used to improve their operations.

There are several fundamental IoT data analytics approaches, including real-time, predictive, and prescriptive analytics \cite{elijah2018overview} \cite{ahmad2021review} \cite{balusamy2021big}. Real-time analytics \cite{gillet2021lambda+} involves using algorithms and software to analyze data generated by IoT devices to identify and act on data trends and patterns in near-real time. Predictive analytics uses historical data and machine learning algorithms to forecast future outcomes and behaviors. In contrast, prescriptive analytics uses optimization algorithms to recommend actions and decisions that can help businesses to achieve their goals.

In the IoT domain, data analytics can be used for various applications, including \textit{ predictive maintenance, energy management, supply chain optimization, and customer behavior analysis} \cite{elijah2018overview}. By leveraging data analytics, we can gain valuable insights into the performance and behavior of their IoT systems. We can use this information to improve their \textit{operations, reduce costs, and drive innovation}.

\textit{Analytics architectures} are the systems and technologies used to collect, store, process, and analyze data to gain insights and make data-driven decisions \cite{zschornig2020iot}. These architectures typically involve using distributed computing systems, such as clusters of servers or cloud-based platforms, to handle the processing and storage of data. In addition to these systems, analytics architectures typically include specialized software tools and algorithms for analyzing and gaining insights from the data. These tools may consist of data mining, machine learning algorithms, and visualization and reporting tools to present the analysis results. An analytics architecture's specific components and design can vary depending on the organization's needs and goals.

Based on our meetings with researchers and practitioners from different companies, we found that there is an urgent need to have a way to provide a blueprint for Analytics Applications.

To support the engineering of data analytics IoT applications, we use an architectural description designed according to the IEEE/ISO/IEC 42010 standard \cite{42010}. The framework supports the architecture description, reasoning, and design decision process applications.

The framework is \textit{\emph{\gls{dat}}}\footnote{DAT: \textbf{D}ata \textbf{A}rchitecture Framework for Io\textbf{T}}. It is based on an architectural approach \cite{42010} to provide a data architecture for IoT applications. It is considered as a data view for CAPS \cite{muccini2017caps} \cite{sharaf2017architecture} \cite{sharaf2018arduino}. 
The DAT \cite{abughazala2022dat} \cite{abughazala2023architecture} is realized by exploiting the advanced Model-Driven Engineering (\emph{\gls{mde}}) technique which is meta-modeling. 

The main contributions of this paper can be summarized as follows:
\begin{enumerate}
\item Using An architecture description of IoT analytics applications that can provide analytics architectures.
\item Industrial experience in building IoT analytics applications.
\item The application of the modeling languages to an Analytics Data Warehouse case study is outlined.
\end{enumerate}

The rest of this paper is organized as follows. The background is presented in Section 2. The research methodology is presented in Section 3. Data analytics architecture challenges are presented in Section 4. The application of DAT to a real case study is described in Section 5. The DAT evaluation is presented in Section 6. Threats to validity in Section 7. Related work is discussed in Section 8, while conclusions are drawn in Section 9.

\section{Background}
    
This section provides an overview of big data analytics and the DAT framework.
\subsection{Big Data Analytics}
Big Data analytics (BDA) is the process of analyzing the massive raw data by data scientists to uncover unidentified  relationships and associations, market trends, and valuable information to make business decisions \cite{Rawat_2021} \cite{balusamy2021big} \cite{zschornig2020iot}. Business intelligence (BI) is more focused and refers to the technologies, applications, and practices used to gather, store, access, and analyze data to help them make more informed business decisions. BI can help businesses identify trends and data patterns and forecast future performance and outcomes \cite{balusamy2021big}.

BDA differs from Data Warehouse (DW); it is considered a storage architecture for the data that various organizations and business enterprises collect \cite{balusamy2021big}.   
BI data consists of data from the storage (previously stored data) and streaming data, supporting the organizations in making strategic decisions. Big data analytics is the science of handling (examining or analyzing) large data sets with a different format, that is, structured, semi-structured, or unstructured data,  which may be streaming or batch data.

\subsection{DAT: Data Architecture Modeling Tool }
\emph{\gls{dat}} is a data architecture modeling tool for IoT applications \cite{abughazala2022dat} that shows how data flows through the system and provides a blueprint for it \cite{10092710}. It allows the stakeholders to describe two levels of data architecture: high-level Architecture (\emph{\gls{hla}}) and Low-Level Architecture (\emph{\gls{lla}}). It focuses on representing the data from source to destination; it shows formats, processing types, storage, analysis types, and how to consume it. It is built based on  a structural and behavioral meta-model. 
The \emph{\gls{dat}} supports the understanding and documentation of \emph{\gls{dd}} applications. For this purpose, it provides a data-view architecture modeling approach designed according to the IEEE/ISO/IEC 42010 standard \cite{42010},
based on modeling language to describe a \emph{\gls{dd}} application: the Data architecture structural and behavioral view (\emph{\gls{daml}}).

\section{RESEARCH METHODOLOGY}
\label{sec:Background}
The objective of this study is to understand the data analytics architecture as well as the challenges experienced at the case company. Based on the study objectives, we formulated the following research question:
\begin{itemize}

\item \textbf{RQ1:} What challenges related to data analytics architecture do practitioners in the case company experience?
\end{itemize}

\begin{center} 
   \begin{figure*}[!ht]
	\centering
	\makebox[\textwidth]
	{
	    	\includegraphics[width=0.70\paperwidth]{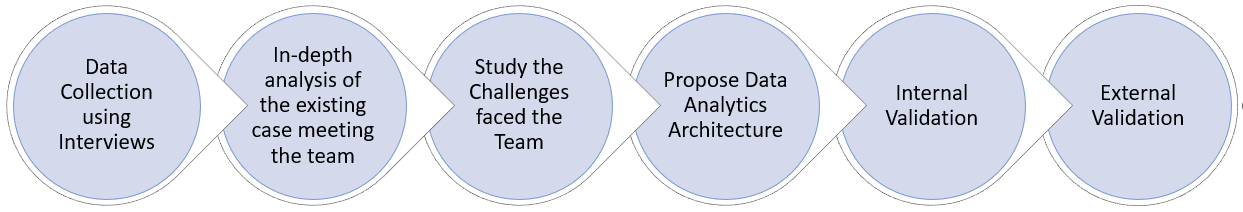}
	}

	\caption{{Research Methodology}}
	\label{fig:dda_study}
    \end{figure*}
\end{center}

The research methodology adopted for the study is illustrated in Figure \ref{fig:dda_study}.

\subsection{About The Company}

The company offers a next-generation frontline employee experience platform built for companies that have served at the heart of their business. The Company suite of talent attraction, workforce management, employee engagement, and compliance technologies enable organizations to intelligently attract, manage, engage, and retain the best talent to run and improve their business. It serves over 20,000 restaurant and hotel locations and four million hospitality employees globally, with emerging growth in retail and healthcare.
\subsection{Exploratory Case Study}
A qualitative approach was chosen; it is a research method that collects and analyzes non-numerical data, such as words, images, and observations. It is often used to explore the experiences, perspectives, and motivations of individuals or small groups in-depth and to gain a deep understanding of real-world cases \cite{verner2009guidelines}. In an exploratory case study, the researcher may use various methods to gather data, such as interviews, observations, and document analysis. The aim is to collect as much rich and detailed information as possible about the case to understand better the context and the factors that influence it.

\subsection{Data Collection} The data was collected through interviews and meetings \cite{singer2008software}. Based on the research objective to study and explore the applications consuming data in the company, an interview guide with 30 questions categorized into three sections was formulated. The first author prepared and reviewed the interview guide with the second author. Each interview lasted 50 to 60 minutes.
The first section focused on the background and experience of the interviewees, table \ref{tab:dda_roles} shows the role of each interviewee. The second section focused on the primary data components of data collection, storage, processing, analysis, and visualization. The last section inquired about best practices and the obstacles faced during each step of building the application.
The interviews were recorded with the interviewees' permission and transcribed later for analysis. Data collected through the meetings and discussions are also incorporated.

\begin{table}[h!]

	\centering
 
	\caption{The Roles of the Interviewees}
	\label{tab:dda_roles}
	\begin{tabular}{|p{2cm}|p{7cm}|p{3.5cm}|} 
	\hline
	\textbf{Company} & \textbf{Use cases} & \textbf{Experts Roles} \\
	\hline
	\multirow{3}{*}{\textbf{Case A}} & \multirow{3}{*}{Analytical Data Warehouse } & Big Data Team Lead \\
	\cline{3-3}
	& & Big Data Architect\\
	\cline{3-3}
	&  & Big Data Engineer\\
	\hline

	\end{tabular}
 
\end{table}

\subsection{Data Analysis}
The first author prepared a summary of all recorded audio. The transcripts were investigated for relations, similarities, and dissimilarities. The first author prepared notes during the meetings with the team and analyzed them further. The main contact point helped with interpreting the parts of the Data Architecture. These notes and the codes from transcripts were further analyzed to obtain an end-to-end view of different use cases. It also helped to understand the parts common to all use cases. After careful analysis of collected data and based on the inputs from the author, the first author developed the first model for Data Analytics Architecture, which got refined through iterations.

The practitioners shared the case with a detailed description. The first author prepared notes during the meetings with the practitioners and analyzed them further, which required more in-depth interviews and additional questions to understand the architecture better.
After deep analysis of collected data and based on the inputs from the second author and the practitioners, the first author enhanced DAT to provide the Data Analytics Architecture model.

\subsection{Validation Study}
The validation study is performed internally and externally through qualitative interviews and feedback sessions. 
First, The first author presented the Data Analytics Architecture model of the case to the teams inside the case company. The reflections about the model, overall opinion, agreements, and suggestions for improvement were collected from every practitioner. These reflections are considered internal validation.
One telecommunication company was selected for external validation of our findings and is in the validation process. another case is a Lambda+ \cite{gillet2021lambda+} as another external validation Fig. \ref{fig:lambda} shows Data Analytics Architecture for the provided case.

The architecture model was then modified to address issues raised during the talks. Also, the inputs from the practitioners are incorporated into the architecture model. The remaining issues will be addressed in future works.

\section{DATA Analytics Architecture Challenges}
\label{dda_IndusExp}

In this section, insights into data analytics architecture challenges are presented based on the findings from our case analysis. We have identified eight major challenges with data analytics architecture used in the company :
\begin{enumerate}
\item \textit{Big Data} analytics architectures should be able to handle (process, analyze, and combine) a huge amount of data from different sources at high velocity.
 \begin{enumerate}
 \item \textit{Volume} the sheer size and scale of the data sets that are used in big data analytics. These data sets can be massive, often consisting of millions or billions of records, and may be generated from various sources, from the case we have weblogs, social media, and transactional systems. The volume of big data can present challenges for data processing and analysis techniques, which may not be able to handle such large volumes of data efficiently. As a result, the company uses distributed computing systems, such as clusters of servers or cloud-based platforms, to process and store big data. 
 \item \textit{Variety} the wide range of data types included in big data sets. These data sets may consist of structured data, such as records in a database, and unstructured data, such as text, audio, and video. They may also include semi-structured data, such as XML or JSON, which has some inherent structure but is not organized similarly to structured data. From the case we have Column-oriented, Excel, JSON, CSV, and relation database formats. The variety of data types in the case can present challenges for data processing and analysis techniques, which may not be designed to handle such a wide range of data types. As a result, the company uses specialized software tools and algorithms, such as natural language processing and machine learning, to analyze and gain insights from big data. In addition, data integration and management strategies may be used to combine and organize the different data types to support analysis and decision-making, {such as Extract, Transform, Load (ETL) strategy: This involves extracting data from various sources, transforming it to fit a common data model, and loading it into a central repository for analysis and reporting}.
 
 \item \textit{Velocity} the speed at which data is generated and processed in big data analytics. This can be a significant challenge, as big data sets may be generated at very high rates, often in real-time, and may need to be processed and analyzed quickly to extract valuable insights. From the case we data in different speed. The velocity can present challenges for data processing and analysis techniques. As a result, the company uses distributed computing systems, such as clusters of servers or cloud-based platforms, to process and store big data. In addition, specialized software tools and algorithms, such as stream processing and in-memory databases, are often used to analyze and gain insights from big data in real-time.

 \end{enumerate}

\item \textit{Data Collection and Storage} refers to gathering and storing data from various sources, such as sensors, databases, or user input. Once the data has been collected, it needs to be stored in a suitable location for further processing and analysis. There are several options for storing data, Flat files, Relational databases, NoSQL databases, Local storage, and Cloud storage. 

\item \textit{Data Processing}  is related to processing large datasets. Converting the data from different forms into one form for analysis is challenging.
\begin{enumerate}
\item \textit{Data Processing Type} Stream processing; This type of processing involves analyzing data as various sources generate it, in real-time or near real-time. This contrasts batch processing, where data is collected and processed in batches at regular intervals. This approach is suitable for processing data that is not time-sensitive and does not need to be processed in real-time.

\item \textit{Scalable data processing} refers to the ability of a data processing system to handle increasing amounts of data and maintain or improve its performance as the volume of data grows. Scalability is an essential consideration in the design of data processing systems, mainly when dealing with large and complex data sets, such as those encountered in big data analytics. A scalable data processing system can handle data volume and complexity increases without significant performance degradation. This may be achieved through distributed computing systems, such as clusters of servers or cloud-based platforms, which can provide additional resources and processing power as needed. In addition, scalable data processing systems may use specialized software tools and algorithms, such as map-reduce and parallel processing, to distribute and manage data processing across multiple computing resources.

\end{enumerate}

\item \textit{Personalization of analytics} refers to the use of data and algorithms to adapt the analysis and presentation of data to individual users or groups of users. This may involve considering factors such as the user's role, preferences, and past behavior, as well as the specific goals and objectives of the user or organization. Personalization in analytics aims to provide users with a more adapted and relevant experience, which can help improve the usefulness and impact of the data and insights generated from the analysis. The Personalization of analytics can be achieved through machine learning algorithms, which can learn from user data and interactions to make personalized recommendations or predictions. It may also involve data visualization and reporting tools that support the customization of data views and presentations for different users.

\item \textit{Share Data} is the practice of making data available to multiple users or systems for analysis and decision-making. This may involve sharing data within the company, or with external parties such as researchers or other stakeholders. Sharing data for analytics can help to improve the accuracy and relevance of the data being analyzed, as well as the insights and decisions that are made based on the analysis. It can also help to promote collaboration and innovation by making data available to a broader community of users and systems. To share data for analytics, companies may use various techniques, such as data warehousing, data lakes, and data sharing platforms, to store and manage the data in a way that allows for easy access and analysis. Data sharing may also be subject to legal and regulatory requirements, such as privacy laws and data governance policies, which must be considered when sharing data for analytics.

\item \textit{Security and Privacy} as data sets may contain sensitive or confidential information that must be protected from unauthorized access or disclosure. To ensure the security and privacy of data used for analytics, organizations may implement various measures, such as encryption, access controls, and data anonymization. These measures can help prevent unauthorized access to data and protect the privacy of individuals whose data is being analyzed. In addition, organizations may also develop data governance policies and procedures to ensure that data is collected, processed, and shared in compliance with applicable laws and regulations. Ensuring the security and privacy of data used for analytics can help protect the rights of individuals and maintain stakeholders' trust and confidence in the data and insights generated from the analysis.

\item \textit{Analyzing data} is examining data to draw conclusions and make informed decisions. There are several different types and techniques of data analytics.
The four types of analytics are: 
\begin{enumerate}
\item \textit{Descriptive analytics} This type of analytics involves summarizing and describing data, often using statistical techniques such as mean, median, and standard deviation.
\item \textit{Diagnostic analytics} This type of analytics involves drilling down into data to understand the root cause of problems or issues.
\item \textit{Predictive analytics} This type of analytics involves using data to predict future outcomes.
\item \textit{Prescriptive analytics} This type of analytics involves using data to identify the best action to take in a given situation.
\end{enumerate}
 The Three techniques of analytics are:
 \begin{enumerate}
  \item \textit{Quantitative analysis}; the data based on numbers. it can be performed as (Nominal, Ordinal, Interval, or Ratio data). For Example, financial ratios evaluate the performance of a company. Ratios such as the price-to-earnings ratio or the debt-to-equity ratio provide a way to compare different companies and make decisions about investing in them.
 \item \textit{Qualitative analysis}; basically answers to “how,” 
“why” and “what" questions can be performed by one of two approaches (deductive and inductive). For Example, Analyzing Customer reviews on a website.

 \item \textit{Statistical analysis}; that uses statistical methods for analyzing data such as (A/B testing, Correlation, and Regression). An example is a survey of customer satisfaction with a product. The survey data is collected, and then statistical methods are used to analyze the data and make inferences about the population of all customers. 
\end{enumerate} 

\item \textit{Data visualization} uses visual representations, such as charts, graphs, and maps, to make complex data sets more straightforward to understand and interpret. Visualization can be a powerful tool for exploring and analyzing big data. It can help identify patterns, trends, and relationships in the data that may not be immediately apparent from raw data. In addition, visualization can make it easier for stakeholders, such as business leaders and data scientists, to communicate and share insights from extensive data analysis. Many visualization tools and techniques can be used for big data, including 2D and 3D charts, geographic maps, and interactive dashboards. These tools may be integrated into specialized big-data analytics platforms or used with general-purpose data visualization software.

\end{enumerate}

\section{Application of DAT Models to the Analytics Data Warehouse Case Study}
\label{sec:casestudy}

This section introduces The Analytics Data Warehouse (\gls{adw}) from company (A)\ref{section:RM_Cases} as acase study and its DAT.

\begin{center} 
   \begin{figure*}[!h]
	\centering
	\makebox[\textwidth]
	{
	    	\includegraphics[width=0.86\paperwidth]{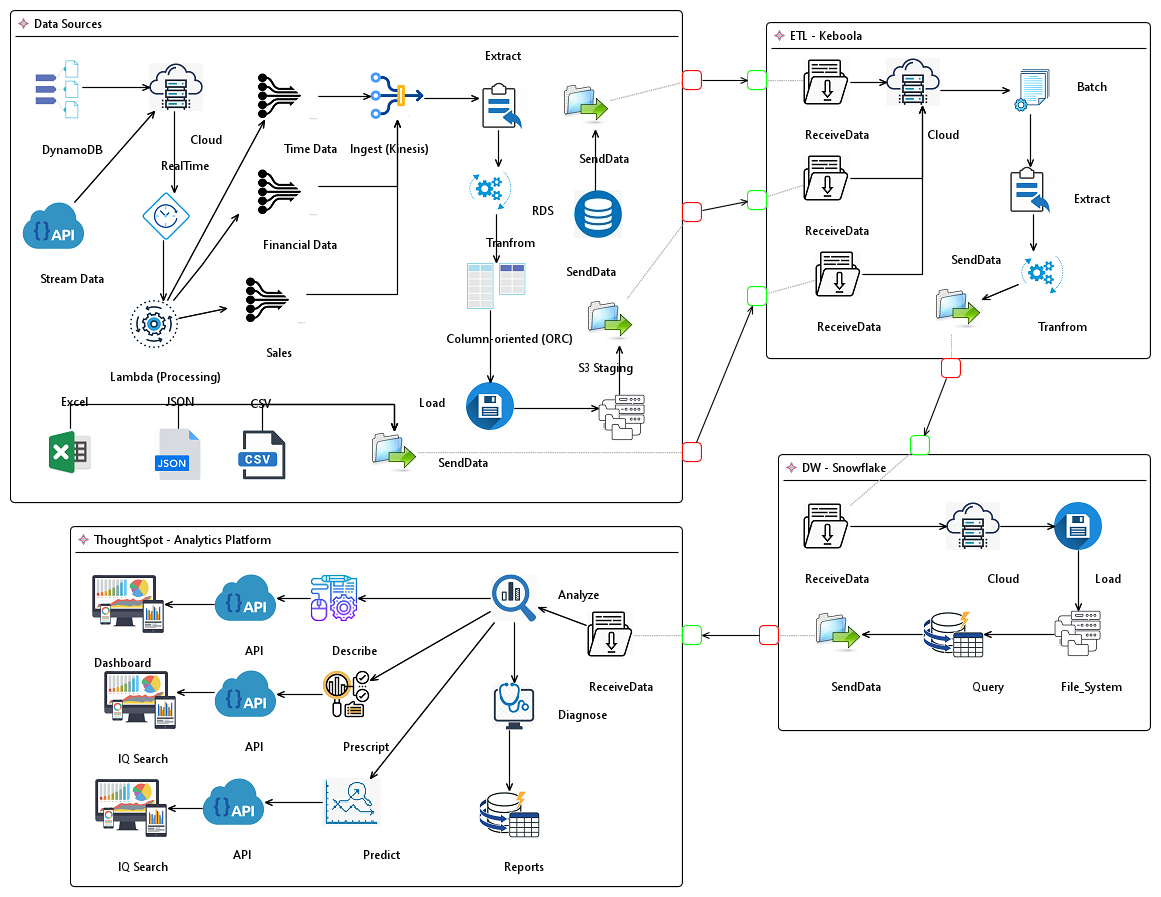}
	}
	\caption{Modeling of the Analytics Data Warehouse }
	\label{fig:ADW}
    \end{figure*}
\end{center}
\subsection{The Analytics Data Warehouse}

The data sources of this Analytics Data Warehouse are the AWS RDS (Relational Database Service) instances, the duplicate staging ORC files saved on S3 that are generated from the data streams. In addition, this data warehouse consumes data from CSV files, worksheets, and JSON objects that augment the data in the warehouse to make the data-rich for analytics purposes. 
The ETL process in this data warehouse is developed and run using a cloud-based third-party tool called Keboola Connectors. This tool provides all the needed connectors for data sources like AWS MySQL RDSs, ORC(Optimized Row Columnar) file extractors, Google Drive sheets, CSV readers, etc. Keboola also offers the required environment and tools to perform data transformations on the extracted data. Moreover, it provides the needed capabilities to schedule executing all extractions and transformation and orchestration between them. In the case implementation of this data warehouse, the data prepared using Keboola is stored on the cloud-based Snowflake data warehouse. The data in the Snowflake data warehouse is consumed by another cloud-based advanced BI and analytics framework called ThoughtSpot. This framework allows the customers to build analytical reports, dashboards, and KPIs.
Through ThoughtSpot, The Company provides customers with an out-of-the-box set of reports, dashboards, and KPIs. Also, customers can build reports and dashboards based on their needs and business inquiries and perform advanced AI-driven data insights and searches. Thoughtspot reports, and dashboards are integrated with company’s system (the Frontend portal and mobile) applications through the SSO mechanism with white-labeled embedding capabilities to offer the customers and end users one integrated and unified working environment and experience.

\subsection{The DAT model applied to the ADW Case Study}
This section shows the modeling of the ADW case study using DAT. From a structural point of view, Figure \ref{fig:ADW}  shows the primary 4 data nodes; Data Sources, Processing Node (ETL - Keboola), Data warehouse (Snowflake), and Analytics platform (ThoughtSpot).

The Data source shows how to collect the data from different data sources in different formats, integrate with (Time, Finacial, and Sales data), and ingest data to be transferred into a Column-oriented format to be saved on File System(Amazon S3). Other data source is Excel, JSON, and CSV files will be sent directly to the Keboola node to be processed in the cloud.  This Node provides all the needed connectors 
with other data sources. It provides the needed capabilities to schedule (Batch) extractions and transformation. The extracted prepared data is stored in the cloud-based Snowflake data warehouse. Then it will be consumed by another cloud-based advanced BI and analytics framework. The analytics node shows the ability of the framework to provide analytical reports, dashboards, and KPIs to the customers. 

\section{Validation Study} 

The Analytical Data Architecture model was validated through interviews with three industry professionals from one company. The first author conducted the interview study for validation. 
The research aimed to provide data analytics architectures that address the challenges in the section \ref{dda_IndusExp}.

This validation of the case will be structured in terms of agreements and suggestions for improvement. Agreements refer to situations where practitioners agree and confirm, while suggestions for improvement refer to situations
in which the interviewees had different opinions. 

\textit{Analytical Data Warehouse}
Figure \ref{fig:ADW} shows the ADW model using DAT. The first author presented the models and collected the practitioners' feedback. 

\textbf{Agreements:} The provided model shows the data flow from different data sources with different formats (Variety) to the final destination,  the real-time data (Velocity),  the type of data processing (batch and real-time), storing data in more than one storage technologies in the cloud (Scalability), ways of analyzing data, and Finally, explains the consuming data by sharing and visualizing it. Moreover, The model was easy to share and understand by different teams in other parts of the world. The model was a good communication language between team members, which means easy to avoid misinterpretations.  

\textbf{Suggestion:} In the provided model, the only way to show how different components interact is to send/receive data. The suggestion was to include all data interaction patterns (request/response, publish/subscribe, pull/push, and others). Current Visualization is enough, but it is better to show how to Visualize data; the suggestion was to show visualization ways such as charts, pies, and others. It is good to provide a way to show sensitive data and a way to protect the data. 

Finally, DAT showed the ability to provide analytics architecture for IoT applications.  

\begin{center}
 \begin{figure*}[!t]
	\centering
	\makebox[\textwidth]
	{   	\includegraphics[width=1\paperwidth]{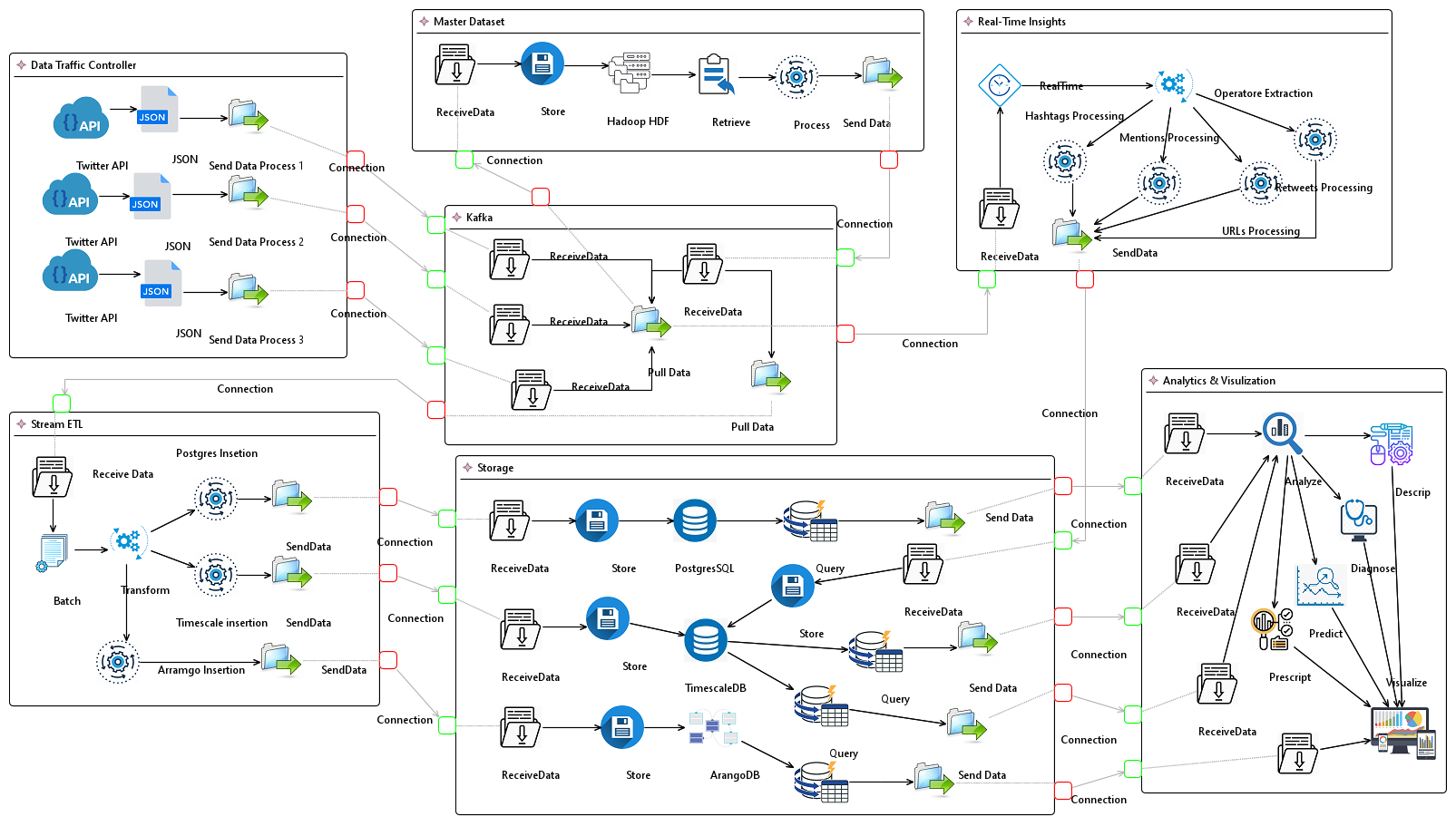}
	}
	\caption{ Lambda+ Model}
	\label{fig:lambda}
    \end{figure*}
\end{center}

\section{Threats to Validity}
This study was based on an existing case (Analytics Data Warehouse) related to analytics architectures, developed by more than one data team from the same company located in different parts of the world. 
To address the internal validity threat, one of the company's practitioners with good experience with data analytics architecting was asked to validate the model. Also, the model was validated with other teams in the company who were not involved in the study. 

{Furthermore, the study was validated again by the Hayder (From Lambd+ Author) case study Fig. \ref{fig:lambda} and in the validation process by an external company from the telecommunications domain (Internet Service Provider). Hyder is from a (Cocktail) research project which aims to study the discourses in two domains in, health and food, and identify weak signals in real-time using social network data. The data come from Twitter, compute real-time insights, and store data for exploratory analysis.}

\section{Related Work}
This section reviews relevant studies that are related to 
exploiting the most related research to data analytics for IoT. 
Mishra and Misra \cite{mishra2017structured} show that structured and unstructured data can be mined to provide meaningful insight into the business. DAT can model all data formats (structured, Semi-structured, and unstructured).

Raj and Bosch \cite{raj2020modelling} proposed a conceptual model for a data pipeline, which contains two principal components (nodes and connectors), the node represents the primary abstract data node, and the connection represents the way to carry and transmit the data between nodes. DAML includes all data life cycle phases and gives the ability to describe the behavior of data nodes( different data formats from other sources, processing types, operations on the data, location, etc.). 

Erraissi \cite{erraissi2018data} \cite{erraissi2019big} proposed a meta-model for  data sources, ingestion layers, and Big Data visualization layer. DAML can describe the data in each layer (Source, Ingestion, Processing, Store, Analyze, and Consume). 

Bashir \cite{bashir2020big} proposed a meta-model for Data Management and Analytics for IoT smart building; our meta-model provides detailed descriptions of ingestion, data processing, storing data, analyzing, and visualizing the data.

\section{Conclusion and Future Work}
{
In the immediate future, it will be inexorable that the daily analysis will not be able to keep up with the daily influx of data. Along with the increased popularity of data and its products, challenges associated with it also increased. Data scientists and other practitioners working with data spend considerable time combating those challenges.

This paper highlighted those challenges and showed the applicability of DAT in providing an analytical architecture by evaluating a real case  (Analytical Data Warehouse).  The modeled case was validated using an exploratory case study where practitioners from one company participated. 

Our future work plan can include finishing the current running evaluations with other companies. 
}
\chapter{Ensuring Data Quality Throughout the Data Journey}
\thispagestyle{plain}

\newacronym{dafqa}{DAFQA}{Data Architecture Quality Framework for Data-Intensive Applications}
\newacronym{ge}{GE}{Great Expectations}
\newacronym{dqd}{DQDs}{Data Quality Dimensions}

In data-intensive applications, ensuring data quality has become increasingly crucial for effective decision-making and operational processes. A comprehensive architecture framework for modeling and monitoring data-quality (\gls{dq}) has been proposed to address this challenge. This framework covers the entire data lifecycle, from data ingestion to consumption. It emphasizes the importance of continuous monitoring and feedback loops to proactively identify and rectify potential data quality issues. By adopting this approach, data-intensive applications can optimize their performance and ensure reliable and high-quality data for decision-making purposes. The paper concludes with real-world case studies and practical recommendations to support the adoption and implementation of the proposed framework.

\section{Introduction} 
More than 175 zettabytes of valuable data are anticipated to exist by 2025, representing a compound annual growth rate of 61\%, according to the International Data Corporation (IDC) \cite{idc}. It's quite concerning that organizations lose an average of \$12.9 million every year due to poor data quality \cite{Gartner}. 
In any enterprise system, data plays a pivotal role as it aids in decision-making, training machine learning (ML)/deep learning (DL) models, generating insights, creating reports, and building smart, sustainable cities \cite{bibri2019anatomy}. The "data-intensive" refers to the strategic process of utilizing insights from data. Adopting the best practices for collecting, storing, analyzing, and safeguarding valuable data is crucial to ensure a systematic data-driven approach.

A Data Architecture (\emph{\gls{da}}) is essential for any data-intensive application. It refers to the collection of models, policies, rules, and standards that regulate data collection, storage, formatting, integration, and utilization in data systems and organizations\cite{10.5555/3165209}.
\emph{\gls{dq}} is of utmost importance for making informed decisions based on data analysis. Data accuracy and reliability directly affect the credibility of the insights and decisions derived from it \cite{karkouch2016data} \cite{ji2020quality}. Incorrect data quality can result in inaccurate conclusions, flawed decision-making, and adverse effects on the organization \cite{cai2015challenges}.

Developing and implementing quality data architecture solutions can pose a significant challenge for organizations that lack effective and adequate data architecture tools. In such cases, manual processes are often relied upon to manage data architecture, which can be time-consuming and prone to errors. The scalability and flexibility of these manual processes may also be limited, making it challenging to address the growing complexity of modern data environments \cite{castellanos2019survey} \cite{chen2016agile}.

Our \textit{DAFQA} \footnote{\emph{\gls{dafqa}}} framework, which adheres to the internationally recognized IEEE/ISO/IEC 42010 standard \cite{42010}, empowers data architects to effortlessly construct data-intensive applications. Through clearly defined data quality metrics and key dimensions, including completeness, accuracy, timeliness, consistency, and conformity \cite{wang2023overview}. Our framework effectively covers all data lifecycle stages for comprehensive monitoring and continual improvement.

The main contributions of this paper can be summarized as follows:
\begin{enumerate}

\item Using An architecture description of data-intensive applications that can provide high-quality architectures.

\item A set of model-to-text transformations that transforms Modeling Language to Python Code that uses \emph{\gls{ge}} Library \cite{greatexpectations} to monitor the data quality.

\item The application of the modeling languages to the Errors Printers Data Pipeline and Telecommunication case studies is outlined.

\end{enumerate}

Automatically generating code for data quality checks is a highly efficient and \textit{time-saving} solution. It ensures \textit{consistency and standardization} of data across various datasets, improving accuracy and reliability. This promotes sound data governance practices, facilitating agility in adapting to changing data sources. Additionally, this method offers documentation for traceability and compliance purposes, simplifying the assessment process and enabling informed decisions based on reliable data.

The rest of the paper is organized as follows. The background is presented in Section 2. The research methodology is presented in Section 3. Data Quality challenges are presented in Section 4. Modeling data-intensive applications in Section 5. Data quality code generation in section 6.
The application of \emph{\gls{dafqa}} to a real case study is described in Section 7. The results are presented in Section 8. Related work is discussed in Section 9, while conclusions are drawn in Section 10.

\section{Background}




\subsection{Great Expectations}

\emph{\gls{ge}} is a data validation and profiling tool that helps organizations maintain data quality and improve team communication \cite{greatexpectations}. It enables data scientists, engineers, and business stakeholders to work together to ensure clean, accurate, and consistent data. Expectations can be defined, data quality can be tracked, and reports and visualizations can be generated. With \gls{ge}, organizations can build trust in their data and make informed decisions based on reliable information.

\section{Research Methodology}
\label{sec:Background}

\subsection{Definition of research questions}
This study aims to understand the industry's data quality metrics for \emph{\gls{da}}. Based on the study objectives, we formulated the following research questions:


\begin{itemize}
\item \textbf{RQ1:}
What critical challenges do practitioners face in maintaining high-quality data for data-intensive applications?

\item \textbf{RQ2:} In order to ensure data quality, what methods can be employed for monitoring? Additionally, what dimensions of data quality may be evaluated in the process?

 \end{itemize}

\begin{center} 
   \begin{figure*}[!ht]
	\centering
	\makebox[\textwidth]
	{
	    	\includegraphics[width=0.70\paperwidth]{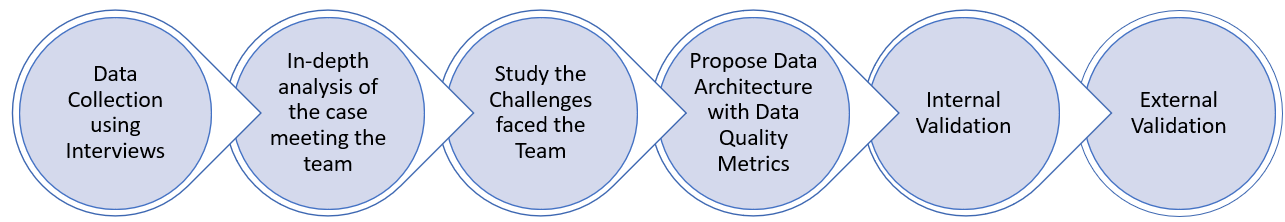}
	}

	\caption{{Research Methodology}}
	\label{fig:study}
    \end{figure*}
\end{center}

 Figure \ref{fig:study} illustrates the research methodology utilized for the study.

\subsection{ Overview of use cases}

\subsubsection{\textbf{ Case A : Errors Data Pipeline}} 
This company is a multinational in the information technology industry and has a significant presence in various technology sectors. One of its areas of specialization is Printers and Imaging Solutions, where it's a prominent player in the printing industry. They provide a variety of inkjet, laser, and multifunction printers designed for home and office use. The case at hand deals with managing error data from these printers.

\subsubsection{\textbf{Case B: ISP - Internet Service Provider} }
\label{sec:CaseE_ISP}
The company is an Internet Service Provider (ISP) telecommunications company that provides individuals and businesses with Internet connectivity. ISPs offer different technologies, such as broadband, wireless, and satellite connections, to facilitate data transmission between users and the global network of servers.

\subsection{Exploratory Case Study}
For this research, a qualitative approach was selected. This method involves gathering and analyzing non-numerical information, such as images, observations, and words. It is particularly useful for investigating individual experiences, perspectives, and motivations and gaining a comprehensive understanding of real-world situations \cite{verner2009guidelines}. When conducting an exploratory case study, the researcher might employ several data collection methods, including interviews, observations, and document analysis. The objective is to gather extensive and detailed information about the case to comprehend better the context and factors that impact it.
\subsubsection{Data Collection} Interviews and meetings were conducted to study and explore data applications in companies \cite{singer2008software}. A guide with 35 questions was used, categorized into three sections. The first section focused on background and experience, the second on data components and data quality metrics, and the last on best practices and obstacles. Interviews lasted 50-60 minutes and were recorded with permission. Data from meetings and discussions were also collected.

\begin{table}[h!]
	\centering
	{\caption{The Roles of the Experts}}
	
	\begin{tabular}{|p{2cm}|p{4cm}|p{5cm}|} 
	\hline
	\textbf{Company} & \textbf{Use cases} & \textbf{Experts Roles} \\
	\hline
	\multirow{3}{*}{\textbf{Case A}} & \multirow{3}{*}{Data Pipeline} & Data Quality Engineer \\
	\cline{3-3}
	& & Big Data Architect\\
	\cline{3-3}
	&  & Big Data Engineer\\
	\hline
	
	\multirow{3}{*}{\textbf{Case B}} & \multirow{3}{*}{ISP} 
        & Big Data Team Lead \\
	\cline{3-3}
	& & Big Data Architect\\
	\cline{3-3}
	&  & Big Data Engineer\\
	\hline

	\end{tabular}
 \label{tab:roles}
\end{table}

\subsubsection{Data Analysis}
The main contact point provided an interpretation of the Data Components. Analyzing the notes and codes from the transcripts gave a comprehensive view of different use cases, including the common parts across all cases. The first author used the collected data and inputs to develop the first model for Data Architecture, which underwent several iterations for refinement and included data quality metrics.

By collaborating with productive practitioners, the author enhanced the data architecture model, encompassing quality matrices. The improved architecture model was meticulously designed by incorporating feedback from the author and the practitioners, analyzing the gathered data, and verifying its precision.

\subsubsection{Validation Study}
The model underwent internal and external validation, including interviews and feedback sessions with the case company teams. The practitioners' opinions and suggestions were considered for internal validation.

A telecommunication company was selected to undergo the external validation process to validate our findings further. Any outstanding issues will be addressed in future work.

\section{Data Quality Challenges}
\label{IndusExp}

It is critical to have high-quality data for systems that rely extensively on data, such as Machine Learning and Deep Learning. If these systems provide poor data, the result will also be poor. As a result, ensure the data is of the highest possible quality to obtain the greatest outcomes. In this section, insights into data  Quality challenges based on the findings from our cases analysis mainly affect the quality of the data :


\begin{enumerate}
\item \textit{Data inaccuracy} Several factors can contribute to data inaccuracies. One such factor is human error during the data entry process. Additionally, problems can arise during the processing and integration of different data sets, leading to inconsistencies and inaccuracies.  
\item \textit{Missing Data}
In data transmission, some data can be lost either partially or entirely. Fortunately, there are means of detecting where the loss occurred. However, retrieving the lost data can be a complex process. Unfortunately, the identification of missing data often happens towards the end of the data lifecycle, leading to suboptimal data products. This challenge is common across all use cases and can occur at different points in the data lifecycle.

\item \textit{Duplicated Data} occur when a dataset contains multiple copies of the same data or data records, leading to inefficiency, confusion, and inaccuracies.

\item \textit{Outdated Data} In data analysis, the importance of accurate and up-to-date information cannot be overstated. Data becoming outdated can result in misleading conclusions and flawed decision-making processes. This is because outdated data no longer reflects current affairs or trends and may contain inaccuracies or biases that have since been corrected. Therefore, it is crucial to ensure that data is regularly updated and refreshed to maintain relevance and usefulness. 

\item \textit{Validity Data Errors} Data that does not conform to predetermined rules, constraints, or standards for data integrity can lead to validity data errors. This type of error is also known as a validity or data validity error. It indicates that the data is "invalid" because it fails to meet the established criteria or requirements. These errors can manifest in various ways and arise at different stages of the data life cycle, Examples of Validity Data Errors.

\item \textit{Consistency Data Errors} Minimizing any disparities, deviations, or discrepancies within a dataset or across multiple datasets is paramount. Such variations, commonly known as data consistency errors, can significantly negatively impact the accuracy and effectiveness of data analysis. Therefore, detecting these inconsistencies is critical in optimizing data quality and maximizing the value of its insights.

\end{enumerate}

\section{Modeling High-Quality Data-Intensive Applications}
Formalizing the structure and constructs of the Data Modeling Language (\emph{\gls{daml}}) language is necessary by defining the underlying meta-model for \emph{\gls{daml}}.

\subsection{Structural Meta-model}
The concept of Data Architecture involves a collection of Data Nodes and Connections. A Component is a building block of this architecture, possessing an internal state and a defined interface \cite{component}. The component's current behavior and values indicate its internal data state. Its behavior includes actions like Send Data, Receive Data, and events that define the data representation. Data representation encompasses different aspects such as formats, storage technologies, location, and processing type.

\begin{figure*}[h!]
	\centering
	
	\makebox[\textwidth]
	{
	\includegraphics[scale=0.52]{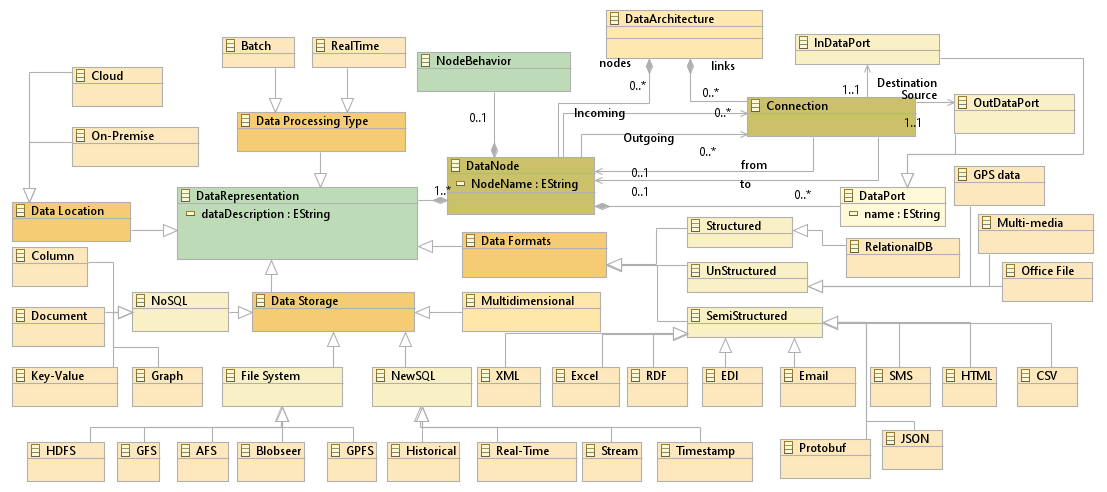}
	}
	\caption{Structural Meta-model}
	\label{fig:DAMLstructure}
\end{figure*}

Our comprehensive meta-model includes various 
{\tt DataFormats} to cater to the diverse data sources, such as {\tt Structured} (RelationalDB), {\tt Semi-Structured} (Email, SMS, CSV, JSON, XML, etc.), and {\tt Unstructured} (GPS data), Multi-media, and Office Files).
We also consider the {\tt ProcessingType} to determine whether the data will be processed as a {\tt Batch} or {\tt Real-time}. To specify the data storage location, we incorporate various {\tt DataStorageTech} such as {\tt NoSQL} Databases (Document, Key-value, graph, and column), {\tt NewSQL} Databases (Historical, Real-Time,  Stream, Timestamp), and {\tt File System} ( HDF, GFS, AFS, GPFS and Blobseer).
Furthermore, we use {\tt Location} to indicate the data node's location, whether it is in the cloud or locally.

\subsection{Behavioral Meta-model}

{\tt NodeBehavior} refers to the current status of a component that describes the data in a specific node. For instance, in a data node that generates data, we can find elements that define the source and format of the data. Every NodeBehavior has a set of behavioral elements that include actions and events, depicting the flow of data within the component. Data nodes can interact by passing data through data ports (DataPort). InDataPort is used to receive incoming data, while OutDataPort sends outgoing data. Figure 3 shows the actual communication methods of a message. In this context, a connection represents a unidirectional communication channel between two data ports of two components.

\begin{center} 
   \begin{figure*}[!h]
	\centering
	\makebox[\textwidth]
	{
	    	\includegraphics[width=0.9\paperwidth]{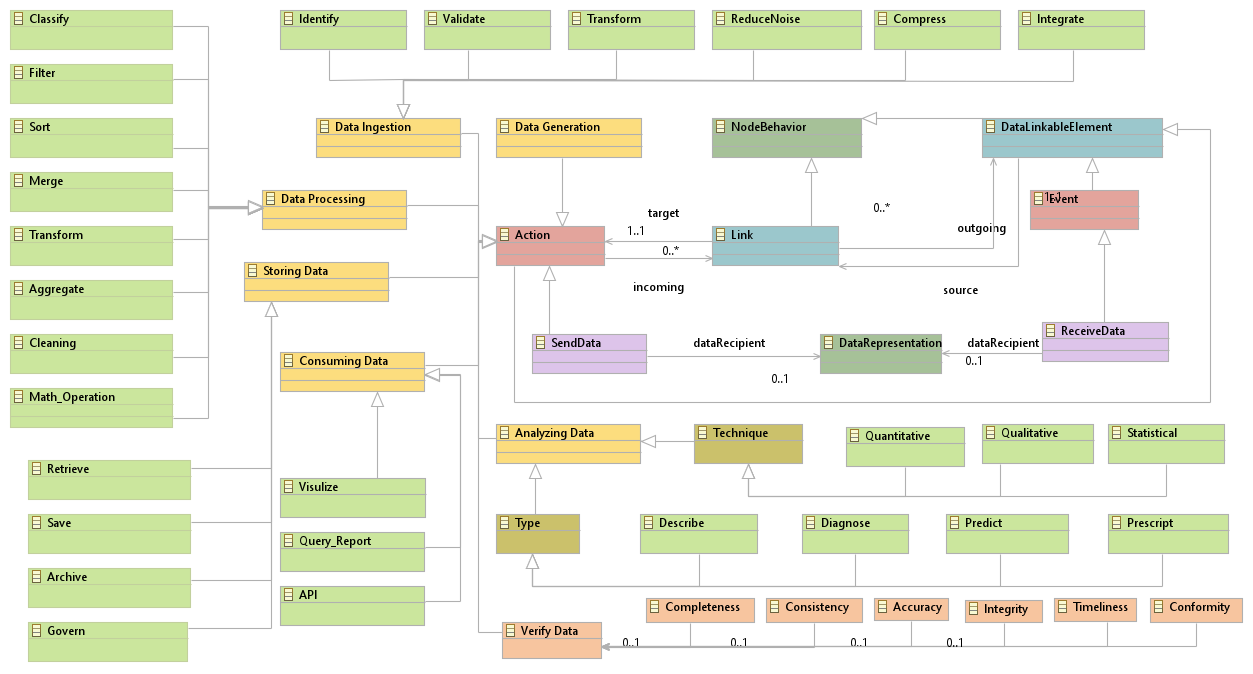}
	}
	\caption{Quality Driven Meta-Model For Data Life Cycle}
	\label{fig:MM_DQM}
    \end{figure*}
\end{center}

Actions in data nodes serve as fundamental behavioral components representing individual tasks. These tasks can be executed once a preceding step in the data flow has been achieved or when an event, such as receiving data, has been triggered. Figure \ref{fig:MM_DQM} illustrates the direct actions of data behavior elements that describe data behavior within the node. For instance, Generation denotes the data source, while Ingestion outlines how data can be transferred from the source to the data lake. Processes comprise a set of sub-processes that can describe a complete processing node, while the store outlines the primary tasks to save, retrieve, archive, and govern data. Analyze describes the type or technique used to analyze data, while Consume shows how data can be consumed, such as through visualization, reports, and APIs.\textit{Verifying data} is a critical step in the data life-cycle that involves checking for accuracy and completeness to ensure that it is valid and reliable. Our selection comprised the primary six data quality dimensions \cite{ramasamy2020big}.

  When a component receives an external stimulus, such as data on an input data port, it triggers an {\tt Event}. {\tt Links} are used to visualize the flow of data and connections between events and actions. These links can also determine the order in which steps should be taken and which must be executed immediately after an event.

\section{Data Quality Code Generation}
\label{sec:Code_Generation}

Our framework streamlines the process of conducting data quality checks and produces Python code that identifies potential issues with data quality, such as missing values, outliers, and incorrect data types. \emph{\gls{ge}} offers comprehensive features to define and execute \emph{\gls{dq}} checks, and we utilize them to generate the code for these checks. The code generated can be easily integrated into existing data pipelines, allowing for a seamless assessment of data correctness and usefulness. By utilizing \emph{\gls{dafqa}} and \emph{\gls{ge}}, you can significantly improve the accuracy and reliability of your data analysis, as depicted in Figure~\ref{fig:CG}.

\begin{figure*}[!t]
	\centering
	{
	\includegraphics[scale=0.50]{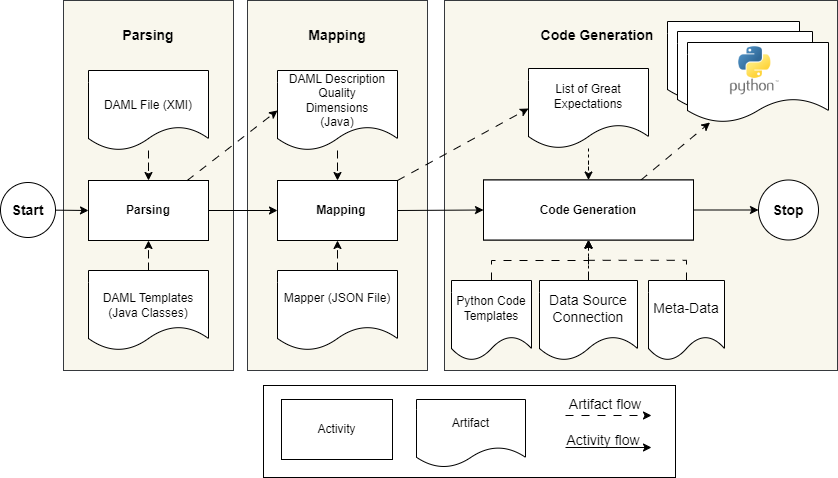}
 }
	\caption{Code Generation Framework}
	\label{fig:CG}
\end{figure*}

\subsection{Modeling}
A " Verify Data" data action includes six data quality metrics. This action can be attached to any data source. Figure \ref{fig:MM_DQM} shows the part of the behavioral meta-model.

\subsection{Parsing}
To extract the appropriate subset of \gls{daml} model values that adhere to the \gls{daml} meta-model, it is typical for the \gls{daml} model to be saved in an XMI (XML Metadata Interchange) file. Nevertheless, this file frequently includes various optional data that must be removed through filtering. When converting a \gls{daml} model into code, we must define templates for the different \gls{daml} sub-models like data nodes, connectors, ports, and data. These templates outline the structure and logic of the code, following the \gls{daml} meta-model. The parser scans the \gls{daml} XMI file and extracts relevant information based on defined templates. This involves generating objects for each data piece and a list of data quality characteristics that can be linked to any data source. The resulting data architecture object contains a \gls{daml} model description in Java, which can be utilized to generate code. Implementing code generators frequently follows the pattern of template-based code generating \cite{voelter2003catalog}. It entails creating templates that define the logic and code structure based on the meta-model and utilizing a parser to extract the information required to instantiate these templates from the input model. Accordingly, the data architecture object includes The \gls{dqd} related to each data source. 

\subsection{Mapping}
Through the Mapper activity, we can effortlessly map the quality dimensions to expectations for each data source. This is made possible by the Great Expectations feature that provides a detailed list of expectations. The result is a comprehensive list of data sources, including their respective quality dimensions with mapped expectations, which can be used for code generation. The major six \gls{dqd} from the model are displayed in table  \ref{tab:mapper} along with the mapped expectation from the Great Expectations tool.


\begin{table}[]
\centering
\caption{Part Of Mapper}
\label{tab:mapper}
\begin{tabular}{|l|l|}
\hline
\textbf{DQDs}          & \textbf{Great Expectations}                   \\ \hline
\textbf{Uniqueness}                    & expect\_column\_values\_to\_be\_unique          \\ \hline
\multirow{2}{*}{\textbf{Completeness}} & expect\_column\_values\_to\_not\_be\_null       \\ \cline{2-2} 
                                       & expect\_column\_values\_to\_be\_null            \\ \hline
\multirow{3}{*}{\textbf{Validity}}     & expect\_column\_values\_to\_be\_of\_type        \\ \cline{2-2} 
                                       & expect\_column\_values\_to\_be\_in\_type\_list  \\ \cline{2-2} 
                                       & expect\_column\_values\_to\_be\_increasing      \\ \hline
\multirow{4}{*}{\textbf{Consistency}}  & expect\_column\_value\_lengths\_to\_be\_between \\ \cline{2-2} 
                                       & expect\_column\_value\_lengths\_to\_equal       \\ \cline{2-2} 
                                       & expect\_column\_values\_to\_match\_regex        \\ \cline{2-2} 
                                       & expect\_column\_values\_to\_match\_regex\_list  \\ \hline
\multirow{2}{*}{\textbf{Timeliness}}   & expect\_column\_min\_to\_be\_between            \\ \cline{2-2} 
                                       & expect\_column\_max\_to\_be\_between            \\ \hline
\multirow{2}{*}{\textbf{Accuracy}}     & expect\_column\_values\_to\_be\_in\_set         \\ \cline{2-2} 
                                       & expect\_column\_values\_to\_not\_be\_in\_set    \\ \hline
\end{tabular}
\end{table}

\subsection{Generate Data Quality Check Code}
 This task is reliant on four specific items:
 \begin{enumerate}
     \item Output from earlier activities, a collection of data sources with associated expectations.
     \item Code templates in Python.
     \item The connection details for the data source.
     \item The meta-data of the data source.
 \end{enumerate}
 
When generating code, we utilize a list of Python code templates. Connection details and metadata determine which data needs to undergo quality checks. This results in a list of expectations. Figure \ref{fig:GCEx} displays an example of generated code that connects to a MySQL database using the table "userinfo" to obtain the batch of data needed to run the test; lines 22 to 26 demonstrate two expectations that will be applied to the "username" column ("expect\_column\_values\_to\_be\_unique").
\begin{figure*}[!ht]
	\centering
 \makebox[\textwidth]
	{
	   \includegraphics[scale=0.65]{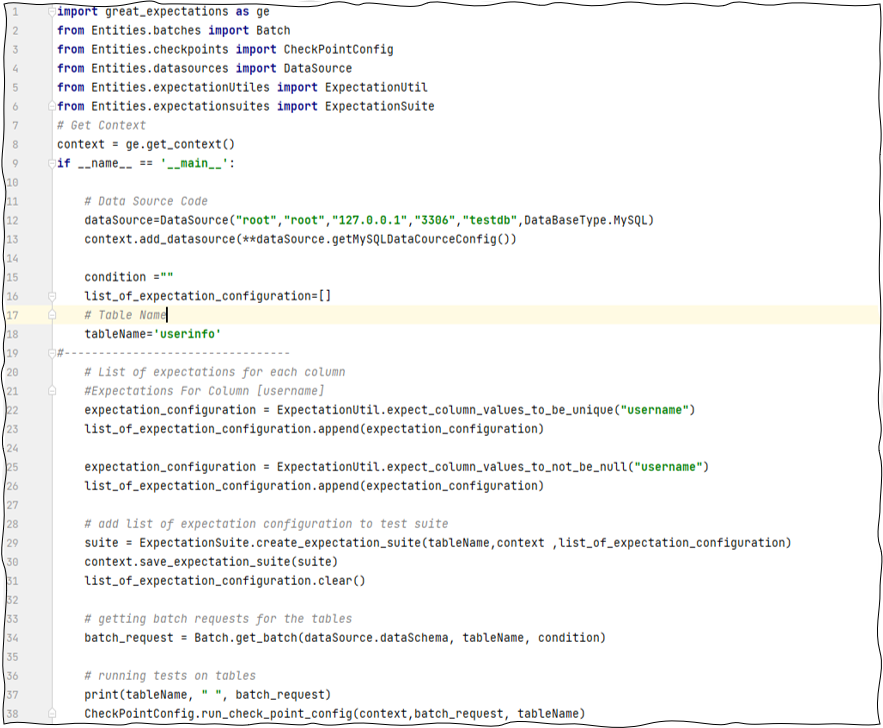}
    }
	\caption{Example Of Generated Code}
	\label{fig:GCEx}
\end{figure*} Runnable Python files that use the Great Expectation library to validate, document, and profile the data are the end product of the code creation activity. This helps the team communicate more effectively and maintains the quality of the data.


\section{Results}
Industry professionals have expertly validated proposed data architecture models to create a framework supporting exceptional data architecture. This blueprint is invaluable for organizations looking to efficiently organize and manage their data assets while ensuring top-notch quality. Through
feedback from professionals, the framework has been tailored to meet real-world requirements and
address unique industry challenges. Validation from experts further enhances the framework’s credibility, making it an essential resource for maintaining high-quality data architecture.

To maximize opportunities in data management, one can either establish a \textbf{proactive design data architecture} or use the generated \textbf{Python code (Great Expectation) to monitor data quality} in any data pipeline efficiently. For the best outcome, combining both approaches is also a viable option.

\begin{figure*}[!t]
	\centering
	{
	\includegraphics[scale=0.45]{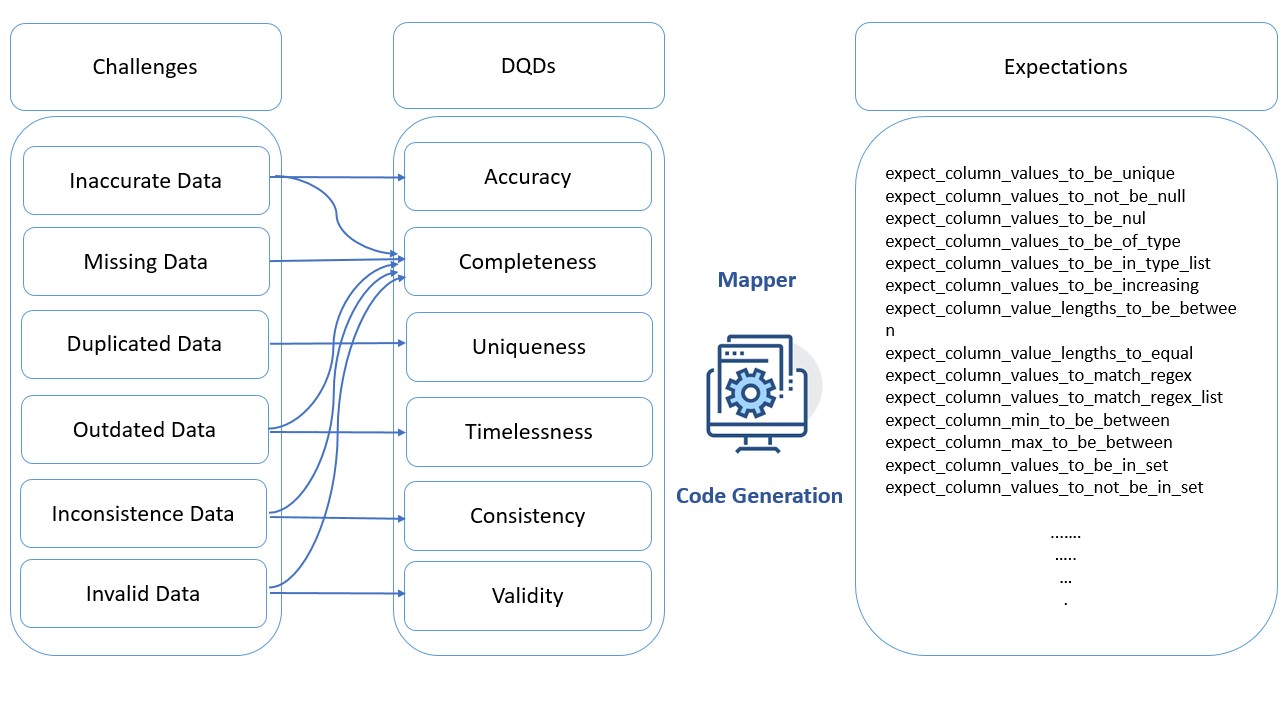}
 }
	\caption{Challenges Mapped to DQDs and Great Expectations}
	\label{fig:Soluation}
\end{figure*}

Figure \ref{fig:Soluation} outlines the challenges of developing data-intensive applications and ensuring the required data quality dimensions are met during the code generation process.



For both cases, we engaged in the process of manually
constructing Python code and integrating Great Expectations.
Based on the evaluations conducted, it has been observed that: 

\begin{itemize}
    \item Embedding data quality dimensions right from the outset ensures consistent data quality maintenance throughout the entire data lifecycle, from data collection, storage, and processing to analysis and reporting. This proactive approach \textit{prevents any possibility of error propagation}, thereby making sure that the data remains accurate and reliable at every stage.
    \item By implementing data quality measures at the architectural level, \textit{proactive error detection} and resolution can be achieved. This allows for identifying and resolving data quality issues as they arise, which eliminates the need for complex and costly issue resolution after the fact.
    \item By addressing data quality at the source, organizations can greatly improve their efficiency and cut costs associated with correcting data errors. This approach ensures smooth operations and can have a substantial impact on reducing overall expenses.
    \item Organizations can rely on high-quality data at each stage of the lifecycle to ensure accurate analytics and informed decision-making. By instilling confidence in the insights derived from their data, businesses can make better decisions that positively impact their bottom line.
    \item Designing systems with an architectural approach to data quality is essential for supporting scalability and future growth. As data-intensive systems become more complex and larger in scale, ensuring data quality can pose significant challenges. However, an architectural approach to data quality from the outset guarantees that systems are equipped to handle these challenges, making them highly adaptable and scalable to confidently meet any future needs.
    
\end{itemize}




\section{Related Work}

This section will examine studies pertinent to utilizing the most related research for the data quality framework for data-intensive applications.

\gls{dafqa} automates generating data quality checks for any data source, mapping them to predefined expectations from \gls{ge}, generating runnable Python code, and monitoring data quality. It can also generate Data Docs reports and take customizable actions based on evaluation outcomes to save time and effort while ensuring high-quality data.

BIGQA \cite{fadlallah2023bigqa} simplifies data quality assessment for experts and management specialists across domains and contexts. It generates tailored reports and runs smoothly on parallel or distributed computing frameworks. It uses simple operators for large datasets and facilitates incremental assessment to save time. BIGQA creates plans for assessing data quality, while DAFQA generates Python code for data quality that can be easily integrated into CI/CD processes.
In \cite{soni2021improving} proposed A new method to measure dataset reliability introduces a "believability factor" by sampling a portion of the dataset and calculating execution time and Mean Absolute Error.
DAFQA can handle all data formats and creates Python code to verify the six primary data quality dimensions.
  In \cite{elouataoui2022advanced} proposed a Big Data Quality Assessment Framework with 12 metrics, including four new ones: Integrity, Accessibility, Ease of Manipulation, and Security, and measured data quality with weights at three levels and five quality aspects for a macro view.
 In \cite{taleb2021big} proposed a Big Data Quality Management framework that works on the entire Big Data lifecycle. The framework relies on a Data Quality Profile enriched with valuable information as it progresses through various stages, such as Big Data project parameters, quality requirements, quality profiling, and quality rule proposals. DAFQA uses a meta-model that can model the entire data life cycle. This allows DAFQA to work on every stage and storage, regardless of format.

 \section{Conclusion and Future Work}

In This paper, we proposed the \gls{dafqa}, an architecture description and associated modeling framework for High-quality data-intensive applications. \gls{dafqa} can give a solution to many data management challenges like Data Silos. It provides a data quality code generation to monitor the quality that could integrated easily with any data pipeline. The \gls{dafqa} has been evaluated by two companies, with six practitioners in the evaluation phase.

This is an initial starting point for our future work plan, which can be extended to include. Data Contracts to ensure the provider can provide the quality of data for consumers based on the provided quality metrics.

\chapter{PyDaQu: Python Data Quality Code Generation Based on Data Architecture}
\thispagestyle{plain}

\newacronym{pydaqu}{PyDaQu}{Python Data Quality Code Generation Based on Data Architecture}

Accurate and dependable data is critical when making crucial business decisions. However, verifying the accuracy of complex and extensive datasets can be both error-prone and time-consuming when done manually. We developed a PyDaQu, our automated framework that creates data quality checks code based on a standardized template. PyDaQu offers a variety of quality assurance measures, including validation, completeness, and consistency checks. These measures ensure exceptional data quality while simultaneously streamlining your data management processes. With PyDaQu, creating data quality checks requires significantly less time and effort. We have thoroughly evaluated PyDaQu using data from two different industry domains.

\section{Introduction} 
The significance of ensuring high data quality cannot be overstated for data-driven applications that seek to thrive in today's data-driven business landscape \cite{karkouch2016data} \cite{ji2020quality}. Reliable data is a critical asset that enables us to make well-informed decisions, boost operational efficiency, enhance customer satisfaction, comply with regulations, manage risks, save costs, facilitate data integration, and maintain trust and credibility. Accurate data is essential in streamlining processes, improving customer experiences, mitigating compliance risks, avoiding unnecessary expenses, and building stakeholder trust \cite{cai2015challenges}. 



Great Expectations \gls{ge}\cite{ge}, Deequ \cite{deequ}, and Dbt \cite{dbt} are powerful data quality enhancement tools. However, configuring and setting up the environment can be time-consuming and prone to human error. Furthermore, utilizing their predefined methods (expectations, constraints,  ...) may be challenging, requiring additional code writing.

This paper introduces  \textit{PyDaQu}\footnote{\emph{\gls{pydaqu}}} \cite{abughazala2023pydaqu}, a cutting-edge code generation framework that leverages the powerful \emph{\gls{dat}} modeling framework, including data quality metrics. With \emph{\gls{pydaqu}}, developers can easily convert their \emph{\gls{dat}} models into Python code while incorporating the popular Great Expectation library. This results in a more streamlined workflow and the ability to efficiently produce high-quality code that meets their expectations.

Generating code for data quality checks automatically is a \textit{ time-saving } and highly efficient solution. It guarantees \textit{ consistency and standardization } across many datasets, thus enhancing accuracy and reliability. Promoting agility in adapting to changing data sources improves data governance. Moreover, it offers \textit{ documentation for traceability and compliance purposes}, streamlining the assessment process and enabling informed decisions based on reliable data.

The rest of this tool demo paper is organized as follows. The background is presented in Section 2. The related work is in Section 3.  The Use Case of \emph{\gls{pydaqu}} is described in Section 4. The \emph{\gls{pydaqu}} tool, approach, and architecture are presented in Section 5. Evaluation is discussed in Section 6, while conclusions are drawn in Section 7.

\section{Background}
\label{sec:Background}

This section provides an overview of the \gls{dat} modeling framework and \gls{ge}.

\subsection{DAT Modeling Framework}
\gls{dat} is a modeling tool for Data-Driven applications \cite{abughazala2022dat} \cite{abughazala2023architecture} that helps visualize the flow of data through the system and create a blueprint for it. Stakeholders can use it to describe high-level (\gls{hla}) and low-level data architecture (\gls{lla}) \cite{10092710}, including formats, processing types, storage, analysis types, and consumption methods. The tool is based on a structural and behavioral meta-model and supports the understanding and documentation of data-driven applications; it includes metrics for data quality. It uses a modeling language called Data Architecture Structural and Behavioral View (\gls{daml}), designed according to the IEEE/ISO/IEC standard \cite{42010}.




\section{Related Work}
This section will examine studies pertinent to utilizing the most related research for the \emph{\gls{dq}} framework.

\emph{\gls{pydaqu}} automates generating data quality checks for any data source, mapping them to predefined expectations from Great Expectations, generating runnable Python code, and monitoring data quality. It can also generate Data Docs reports and take customizable actions based on evaluation outcomes to save time and effort while ensuring high-quality data.

BIGQA \cite{fadlallah2023bigqa} simplifies data quality assessment for experts and management specialists across domains and contexts. It generates tailored reports and runs smoothly on parallel or distributed computing frameworks. It uses simple operators for large datasets and facilitates incremental assessment to save time. BIGQA creates plans for assessing data quality, while PyDaQu generates Python code for data quality that can be easily integrated into CI/CD processes.
In \cite{soni2021improving} proposed A new method to measure dataset reliability introduces a "believability factor" by sampling a portion of the dataset and calculating execution time and Mean Absolute Error. PyDaQu can handle all data formats and creates Python code to verify the six primary data quality dimensions.

 In \cite{taleb2021big} proposed a Big Data Quality Management framework that works on the entire Big Data lifecycle. The framework relies on a Data Quality Profile enriched with valuable information as it progresses through various stages, such as Big Data project parameters, quality requirements, quality profiling, and quality rule proposals. PyDaQu uses DAT and can model the entire data life cycle. This allows PyDaQu to work on every stage and storage, regardless of format.


\section{Use Cases}
This section provides a brief case description utilized by PyDaQu to test the generated code on actual industry data, these cases from companies B and C in section \ref{section:RM_Cases}.

\subsection{Case A: ISP - Internet Service Provider}

As part of our research project, we carried out an investigation utilizing the raw data provided by an ISP Telecommunication company. The dataset featured a vast array of details about the internet usage patterns of their customers, all of which were stored in a complex relational database. Our experiment was conducted to check the quality of the data, which could be used to inform better decision-making strategies.   

\subsection{Case B: Errors Data Pipeline}
This case outlines the data pipeline for error data from different printers. Files come in JSON format, including printer version, location, ink type, software version, and time. The data is saved on Amazon S3 and can be accessed using a query engine. After processing, the data is converted to parquet format, CSV, and relational database format for customer queries. We were able to run the quality check on raw data (same format of generation) and relational databases.

\section{PyDaQu Framework}

\subsection{Approach}
\emph{\gls{pydaqu}} 
 \footnote{PyDaQu Tool Source Code : \url{https://github.com/khitam90/PyDaQu}}
 \footnote{PyDaQu Tool Demo Video : \url{https://youtu.be/697F6h0q7ss}} 
is a framework that automates the process of \emph{\gls{dq}} checks and generates Python code to check for \emph{\gls{dq}} issues such as missing values, outliers, and incorrect data types. 
Great Expectations is a tool that provides features to define and perform data quality checks, and \emph{\gls{dat}} uses Great Expectations to generate the code for these checks. The \emph{\gls{dat}} models are transformed into Great Expectations code, which can be integrated into existing data pipelines. By running this code, \emph{\gls{dat}} can assess the correctness and usefulness of the data, providing insights into the quality of the data being analyzed. Overall, \emph{\gls{dat}} and Great Expectations offer a powerful solution for automating data quality checks and improving the accuracy and reliability of data analysis.
\subsection{Architecture}
To fully comprehend the architecture of \emph{\gls{pydaqu}}, it is crucial to examine each software component that comprises the system. Figure \ref{fig:CG} offers an overview of the tool as an excellent reference point. In-depth explanations and demonstrations for each component will be provided to ensure a comprehensive understanding of \emph{\gls{pydaqu}}.

\begin{figure*}[!ht]
	\centering
	 \makebox[\linewidth]
	{
	\includegraphics[scale=0.52]{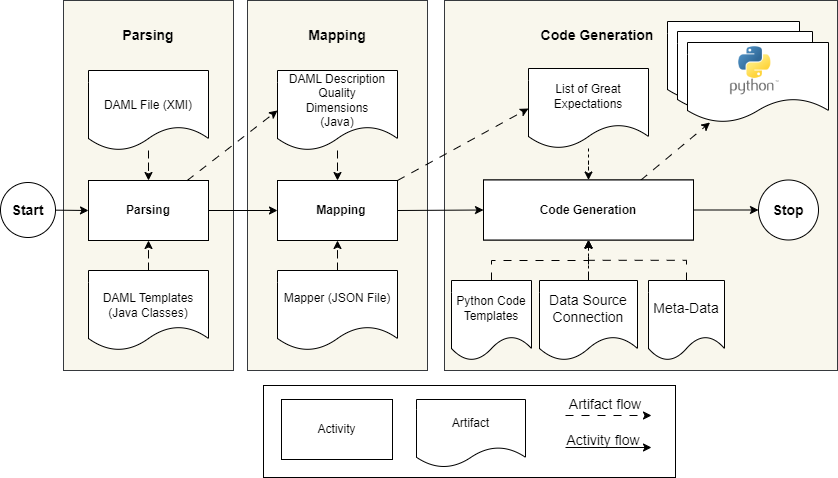}
 }
	\caption{PyDaQu automatic code generation framework}
	\label{fig:CG}
\end{figure*}

\begin{figure*}[!t]
	\centering
   \makebox[\textwidth]
	{
	   \includegraphics[scale=0.85]{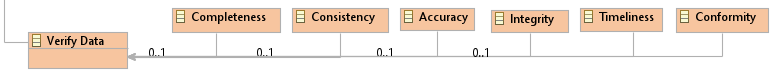}
    }
	\caption{"Verify Data" Action}
	\label{fig:verifydata}
\end{figure*}

\textbf{DAT Model}
A new data action " Verify Data" has been added to ensure the model includes data quality metrics. This action can be attached to any data source. Figure \ref{fig:verifydata} shows the part of the behavioral meta-model.

\textbf{Parsing}, The \gls{daml} model is typically stored in an XMI (XML Metadata Interchange) file, which contains various optional information that needs to be filtered out to extract a relevant subset of \gls{daml} model values conforming to the \gls{daml} meta-model.

To generate code from the DAML model, templates are defined for different \gls{daml} sub-models, such as data nodes and connectors, ports, and data. These templates define the code structure and logic based on the \gls{daml} meta-model.

The parser reads the \gls{daml} XMI file and extracts the necessary information based on the defined templates. This process involves creating objects representing each data element and a list of data quality dimensions linked to any data source. The resulting object of data architecture carries the \gls{daml} model description in Java, which can be used to generate code.

Template-based code generation is a common pattern in implementing code generators \cite{voelter2003catalog}. It involves defining templates that describe the code structure and logic based on the meta-model and using a parser to extract the necessary information from the input model to instantiate these templates.
Developing a \textit{ coding template} involves creating reusable code snippets that can be saved as plain text files. The process includes identifying common patterns, crafting concise templates with placeholders, categorizing and organizing them, thoroughly testing, and keeping them up-to-date as necessary \cite{SYRIANI201843}.

Accordingly, the resulting data architecture object contains the data quality dimensions linked to each data source. 

\textbf{Mapping}, This activity inputs the data architecture object and the \gls{daml} model description in Java, resulting from the previous parsing activity. This Mapper represents the core section of the code generation process and is responsible for mapping quality dimensions to their corresponding expectations for each data source. Table ~\ref{tab:mapper} shows the list of quality dimensions and corresponding expectations provided by \gls{ge}.

\begin{table}[]
\centering
\caption{Part Of Mapper}
\label{tab:mapper}
\begin{tabular}{|l|l|}
\hline
\textbf{Quality   Dimensions}          & \textbf{Great   Expectations}                   \\ \hline
\textbf{Uniqueness}                    & expect\_column\_values\_to\_be\_unique          \\ \hline
\multirow{2}{*}{\textbf{Completeness}} & expect\_column\_values\_to\_not\_be\_null       \\ \cline{2-2} 
                                       & expect\_column\_values\_to\_be\_null            \\ \hline
\multirow{3}{*}{\textbf{Validity}}     & expect\_column\_values\_to\_be\_of\_type        \\ \cline{2-2} 
                                       & expect\_column\_values\_to\_be\_in\_type\_list  \\ \cline{2-2} 
                                       & expect\_column\_values\_to\_be\_increasing      \\ \hline
\multirow{4}{*}{\textbf{Consistency}}  & expect\_column\_value\_lengths\_to\_be\_between \\ \cline{2-2} 
                                       & expect\_column\_value\_lengths\_to\_equal       \\ \cline{2-2} 
                                       & expect\_column\_values\_to\_match\_regex        \\ \cline{2-2} 
                                       & expect\_column\_values\_to\_match\_regex\_list  \\ \hline
\multirow{2}{*}{\textbf{Timeliness}}   & expect\_column\_min\_to\_be\_between            \\ \cline{2-2} 
                                       & expect\_column\_max\_to\_be\_between            \\ \hline
\multirow{2}{*}{\textbf{Accuracy}}     & expect\_column\_values\_to\_be\_in\_set         \\ \cline{2-2} 
                                       & expect\_column\_values\_to\_not\_be\_in\_set    \\ \hline
\end{tabular}
\end{table}

Finally, the result of Mapper is a list of data sources defined with their list of quality dimensions and mapped predefined expectations (from Great Expectations). These objects will be an entry to the following code generation activity.

\textbf{Generate Python Data Quality Checkpoints Code}, 
 This activity depends on four artifacts:
 \begin{enumerate}
     \item Previous activity's output, a list of data sources with related expectations.
     \item Python code templates.
     \item The connection details for the data source.
     \item The meta-data of the data source.
 \end{enumerate}
 
\begin{figure*}[!ht]
	\centering
	 \makebox[\textwidth]
	{
	\includegraphics[scale=0.60]{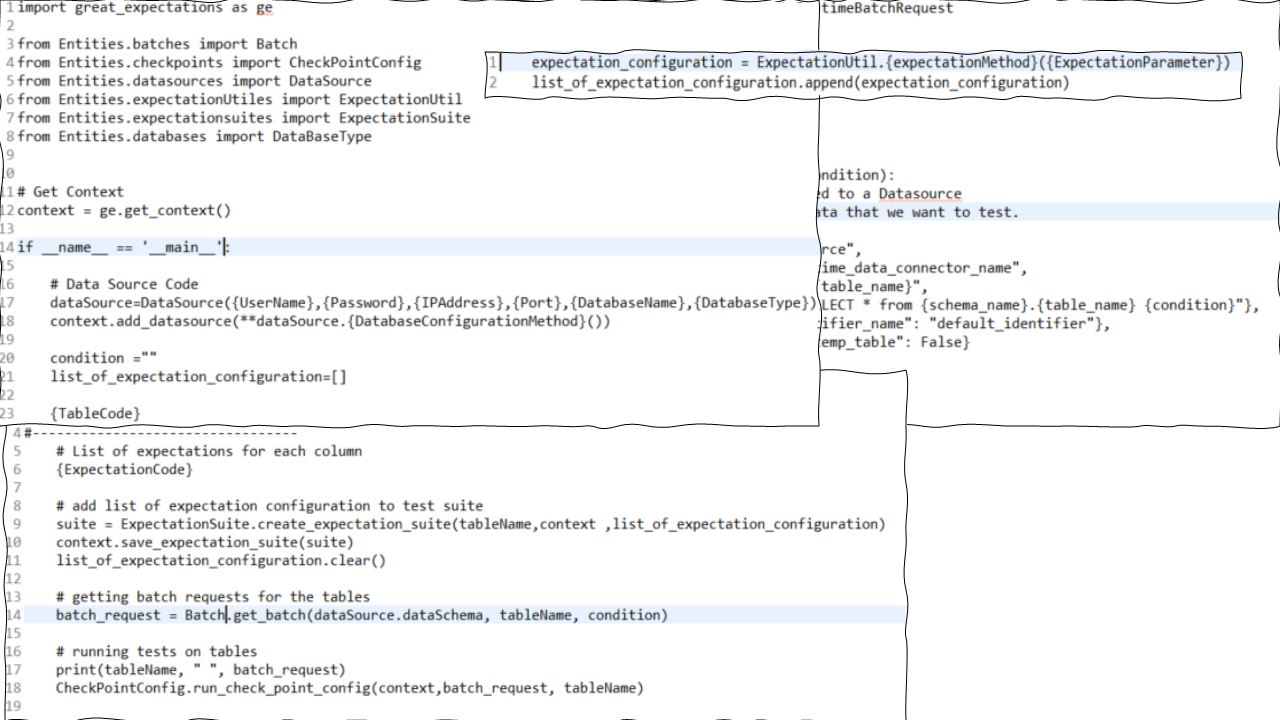}}
	\caption{Templates Example}
	\label{fig:TempEx}
\end{figure*}

During the code generation, we use the list of Python code templates. 
Every part of the code within braces "\{\}" will be replaced with values or generated code see Figure \ref{fig:TempEx}. We use the connection details and the meta-data to select which part of the data we need to run the quality checks (the list of expectations).

Figure \ref{fig:GCEx}, shows an example of generated code that connects to MySQL database, using table "userinfo" to get the batch of data to run the test; lines 22 show two of expectations 
("expect\_column\_values\_to\_be\_unique") will be applied to "username" column.

\begin{figure*}[!ht]
	\centering
  \makebox[\textwidth]
	{
	   \includegraphics[scale=0.68]{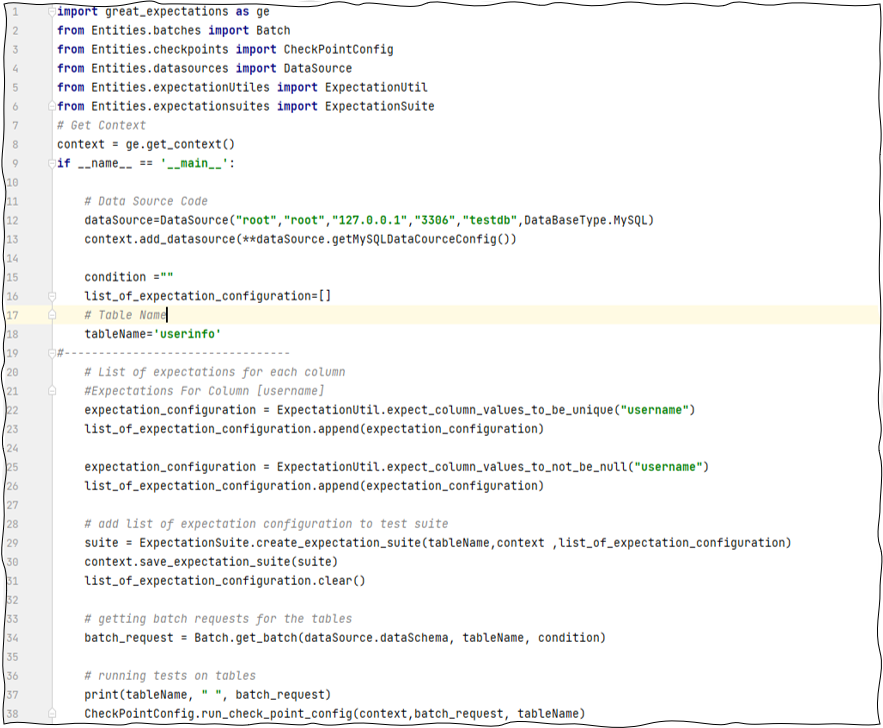}
    }
	\caption{Example Of Generated Code}
	\label{fig:GCEx}
\end{figure*}

\begin{figure*}[!ht]
	\centering
  \makebox[\textwidth]
	{
	   \includegraphics[scale=0.62]{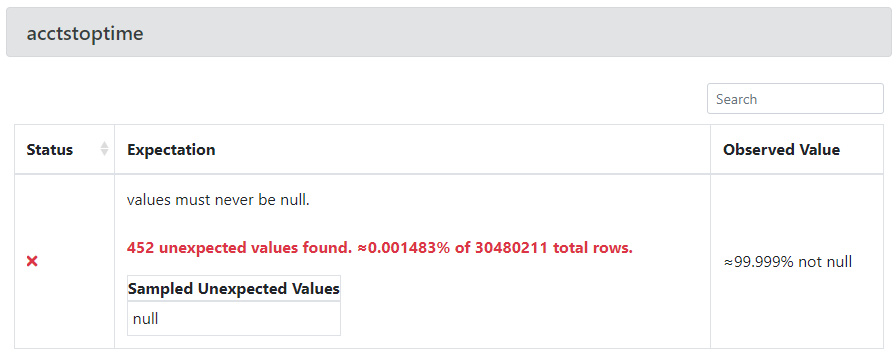}
    }
	\caption{Results of running the generated code}
	\label{fig:GE_Result}
\end{figure*}

The results of code generation activity are runnable Python files that use the \gls{ge} library to validate, document, and profile the data to maintain quality and improve team communication. Figure \ref{fig:GE_Result} shows the results of running the generated Python files. 

\section{Evaluation}



Our recent publication \cite{abughazala2022dat} encompasses a thorough assessment of \gls{dat} in diverse domains. We have included one case study from our previous research and another from the telecommunications industry to enhance our argument. The cases were selected based on the data available to execute the code produced by \gls{pydaqu} to check its effectiveness.

For both cases, we engaged in the process of manually constructing Python code and integrating Great Expectations. After conducting a thorough evaluation, we found a significant difference in the time required to complete a data quality check when comparing manual Python coding versus utilizing \gls{pydaqu} to create and execute the code.

The table  \ref{tab:CodeGen_timeR} shows the estimated time required to manually generate the code based on the engineer's seniority level. Seniors are expected to complete the task within three days, while juniors may require up to five days. On the other hand, it takes only ten minutes to generate and run the code using \gls{pydaqu}.

\begin{table}[]
\centering
\caption{Comparing the Time Required for Manual and Automated Code}
\label{tab:CodeGen_timeR}
\begin{tabular}{|c|c|c|c|}
\hline
\textbf{ Cases }                   & \textbf{ Engineer Level } & \textbf{ Manual } & \multicolumn{1}{l|}{\textbf{PyDaQu}} \\ \hline
\multirow{2}{*}{\textbf{ Case A }} & Fresh                   & 5 days          & \multirow{3}{*}{\textbf{10 minutes}}                              \\ \cline{2-3}
                & Senior & 3 days &  \\ \cline{1-3}
\textbf{ Case B } & Senior & 3 days &  \\ \hline
\end{tabular}
\end{table}

Utilizing code generation templates is a practical approach to building code consistently aligned with the established coding standards. This practice enables developers to streamline their coding processes and produce high-quality code in a structured and systematic manner. By following these templates, developers can ensure that their code is error-free, optimized for performance, and easily maintainable. This ultimately leads to faster development cycles, improved code quality, and enhanced software reliability. 


\section{Conclusion}

In this tool demonstration paper, we have presented \gls{pydaqu}, an excellent Python code generator that employs the \gls{ge} Library to assess data quality. \gls{pydaqu} proves to be an indispensable tool for engineers, as it helps them save time while developing data-checking code, regardless of their level of expertise. The generated code seamlessly integrates with any data pipeline, which makes it a valuable asset for teams working with data. Extensive testing on real-world data across various domains has demonstrated \gls{pydaqu}'s efficacy in delivering reliable and accurate evaluations. \gls{pydaqu} sets the benchmark for quality data assessment and is a must-have tool for anyone working with data.
Engineers can confidently create high-quality, error-free code efficiently and systematically by utilizing code generation templates. This approach results in faster development, better code, and more reliable software.

\chapter{Results And Evaluation}
\thispagestyle{plain}

In this section, we aim to demonstrate the effectiveness of our proposed approach from two distinct angles. Firstly, we will compare our approach to existing studies to highlight our novel approach's unique advantages and necessity. Secondly, we will present the industry feedback to showcase how our approach aligns perfectly with the needs and requirements of the industry.

\section{Comparative to the existing approaches}
In section \ref{sec:Dat_state_Of_the_Art}, we thoroughly examined various existing studies for engineering data-intensive systems, including modeling, development, and analysis. However, none of these approaches provided a comprehensive modeling solution to the challenges of designing, analyzing, and developing high-quality data-intensive systems. That’s where DAT comes in. Our proposed approach offers a novel solution to these challenges and is specifically designed to address the shortcomings of existing techniques. Our evaluation conclusively establishes that DAT is a novel approach superior to existing approaches.

\begin{center}
 \begin{table*}[h!]
\centering
\caption{Comparative table on supporting different engineering capabilities}
\label{tab:comp1}
 \makebox[\linewidth]{
\centering
\begin{tabular}{llllll}
\hline
\multirow{2}{*}{\textbf{Related Works}} &
  \multirow{2}{*}{\textbf{\begin{tabular}[c]{@{}l@{}}Abstract \\ Level\end{tabular}}} &
  \multirow{2}{*}{\textbf{\begin{tabular}[c]{@{}l@{}}Graphical  \\ Tool\end{tabular}}} &
  \multirow{2}{*}{\textbf{\begin{tabular}[c]{@{}l@{}}
  Code \\ Generation\end{tabular}}} &
  \multirow{2}{*}{\textbf{Analysis}} &
  \textbf{\begin{tabular}[c]{@{}l@{}} Empirical \\ Assessment \end{tabular}} \\  \cline{6-6} 
 &
   &
   &
   &
   &
  Approach \\ \hline
Accordant \cite{castellanos2021accordant} &
  High &
  No &
  Yes &
  \begin{tabular}[c]{@{}l@{}}Quality of Service\\ (Update time,\\ latency,Deadline)\end{tabular} &
  \begin{tabular}[c]{@{}l@{}}Proof of concept \\and a case study\end{tabular} \\ \hline
DICE \cite{guerriero2016towards} \cite{artac2018infrastructure} \cite{perez2019uml} &
  High &
  No &
  Yes &
  \begin{tabular}[c]{@{}l@{}}Anomalies detection \\ in real-time \\  data streams\end{tabular} &
  \begin{tabular}[c]{@{}l@{}}Proof of concept \\and a case study\end{tabular} \\ \hline
Gribaudo \cite{gribaudo2018performance} &
  High &
  No &
  No &
  \begin{tabular}[c]{@{}l@{}}Performance of \\ running 
 applications\end{tabular} &
  \begin{tabular}[c]{@{}l@{}}Proof of concept\\ and Experiment\end{tabular} \\ \hline
Raj and Bosh \cite{raj2020modelling} &
  High &
  No &
  No &
  No &
  \begin{tabular}[c]{@{}l@{}}Proof of concept and\\ multiple cases study\end{tabular} \\ \hline
Erraissi \cite{erraissi2019big} \cite{erraissi2018data} &
  Low &
  No &
  No &
  No &
  Proof of concept \\ \hline
Bashir \cite{bashir2020big} &
  High &
  No &
  No &
  No &
  \begin{tabular}[c]{@{}l@{}}Proof of concept \\and multiple \\cases study\end{tabular} \\ \hline
\textbf{DAT} &
  \textbf{High \& Low} &
  \textbf{Yes} &
  \textbf{Yes} &
  \textbf{Six DQDs} &
  \begin{tabular}[c]{@{}l@{}}Proof of concept \\and multiple \\cases study\end{tabular} \\ \hline
\end{tabular}%
}
 \end{table*}
\end{center}




Tables \ref{tab:comp1} and \ref{tab:comp2} show the comparative findings from the assessment related to DAT Modeling features and engineering data-intensive systems and present the following features. 

\subsection{Supporting the engineering of data-intensive systems}
In this section, we evaluate the selected data-intensive approaches based on their ability to support various engineering tasks in developing, analyzing, and deploying data-intensive systems.
The assessment results are summarized in Table \ref{tab:comp1}. The following section presents several exciting findings related to the engineering support provided by DAT.
\begin{enumerate}
    \item All selected studies provide High-Level Architecture, while DAT provides both high and low levels of architecture. Erraissi \cite{erraissi2019big} \cite{erraissi2018data} is the only study that provides low-level architecture.
    \item It is evident that none of the selected studies offer a graphical tool, whereas DAT is the only framework that provides a graphical modeling tool.
    \item DIC \cite{guerriero2016towards} \cite{artac2018infrastructure} \cite{perez2019uml} provides a code for real-time data monitoring. Accordant \cite{castellanos2021accordant} generates functional Java code, which serves as a Spark driver program. DAT generates Python code that utilizes the Great Expectations Library to check data quality dimensions in any data source.
    \item DAT analyzes six data quality dimensions: completeness, consistency, accuracy, integrity, timeliness, and conformity. Accordant focuses on quality of service, including update time, latency, and deadline. DIC provides performance, reliability, and monitoring data analysis.
\end{enumerate}

\subsection{Supporting the modeling of data-intensive systems}
This section compares different modeling approaches for data-intensive applications. The findings from the assessment presented in Table \ref{tab:comp2} are compared to the modeling features supported by DAT, which will be detailed in the next section:

\begin{enumerate}
     \item Most approaches don't support storage type modeling except DIC, which has limited database support. It's crucial to model all storage types and keep varying data formats and speeds in mind. DAT includes all storage types in its modeling framework.
    \item Monitoring data formats is crucial as it moves from source to destination. Data is generated in various formats and needs to be converted into a unified form. DIC supports limited data formats, while DAT classifies them into three categories: structured, semistructured, and unstructured.
    \item Previous studies on data life cycle modeling have focused on specific phases rather than modeling the entire life cycle. Meta-models have been proposed for certain phases, such as Generation, Ingestion, Visualization, Sink, Estimation, and Transformation. DAT has developed an environment that allows for modeling all phases and their sub-details.
\end{enumerate}

It is clear that the current literature on modeling data architecture is limited and incomplete, with some phases of the data life cycle, data formats, storage types, and data processing speeds being overlooked. However, DAT provides a comprehensive and efficient solution to bridge this gap. DAT ensures a smooth and successful data journey with its advanced approach and fast-generated data quality checks.

\begin{table*}[h!]
\centering
\caption{Comparative table on supporting different data-intensive modeling features}
\label{tab:comp2}
\makebox[\linewidth]{
\begin{tabular}{llllll}
\hline
\multirow{2}{*}{\textbf{Related Works}} &
 \multirow{2}{*}{ \textbf{\begin{tabular}[c]{@{}l@{}}Storage\\Type\end{tabular}}} &
 \multirow{2}{*}{ \textbf{\begin{tabular}[c]{@{}l@{}}Data \\Format \end{tabular}}} &
  \multicolumn{2}{c}{\textbf{Data Life Cycle}} &
  \multirow{2}{*}{\textbf{Main Focus}} \\ \cline{4-5}
 & 
   &
  &
  \multicolumn{1}{c}{\begin{tabular}[c]{@{}c@{}}Main Phases\end{tabular}} &
  Details &
   \\ \hline
Accordant \cite{castellanos2021accordant} &
  No &
  No &
  \begin{tabular}[c]{@{}l@{}}Partially \end{tabular} &
  No &
  \begin{tabular}[c]{@{}l@{}}Designing and deploying \\ big data analytics (BDA)\end{tabular} \\ \hline
DICE \cite{guerriero2016towards} \cite{artac2018infrastructure} \cite{perez2019uml} &
  \begin{tabular}[c]{@{}l@{}} Basic \end{tabular} &
  \begin{tabular}[c]{@{}l@{}} Basic \end{tabular} &
  Partially &
  No &
  \begin{tabular}[c]{@{}l@{}}Modeling Big Data Tech. \\ (Apache Hadoop, MapReduce \\ and Apache Storm)\end{tabular} \\ \hline
M. Gribaudo \cite{gribaudo2018performance} &
  No &
  No &
  No &
  No &
  \begin{tabular}[c]{@{}l@{}}modeling framework for\\ performance evaluation of\\systems based on\\lambda architecture\end{tabular} \\ \hline
A. Raj, J. Bosh \cite{raj2020modelling}&
  No &
  No &
  \begin{tabular}[c]{@{}l@{}} Yes \end{tabular} &
  No &
  \begin{tabular}[c]{@{}l@{}}A conceptual model of\\the data pipeline\end{tabular} \\ \hline
A. Erraissi \cite{erraissi2019big} \cite{erraissi2018data} &
  No &
  No &
  \begin{tabular}[c]{@{}l@{}} Partially \end{tabular} &
  Yes &
  \begin{tabular}[c]{@{}l@{}}Meta-model for specific stages\\ of the data life cycle\end{tabular} \\ \hline
M. Bashir \cite{bashir2020big} &
  No &
  No &
  Partially &
  No &
  \begin{tabular}[c]{@{}l@{}}Proposed a metamodel for \\Big Data Management \\and Analytics\end{tabular} \\ \hline
  \textbf{DAT} &
  \textbf{Yes} &
  \textbf{Yes} &
  \textbf{Yes} &
  \textbf{Yes} &
  \begin{tabular}[c]{@{}l@{}}A framework offers \\system and software design,\\cutting-edge code generation,\\and data quality analysis \\ for data-intensive systems.\end{tabular} \\ \hline
\end{tabular}%
}
\end{table*}

\section{Supporting High-quality data-intensive applications}

This section demonstrates how to deliver high-quality applications from the data architecture to the final implementation. First: modeling high-quality architecture. Second: Using model-to-text transformation using PyDaQu to generate data quality checks code.

\begin{itemize}
    \item With DAT, we can offer a comprehensive and well-structured data architecture solution to tackle common data architecture challenges like data silos in new or legacy systems. Our DAT is designed to address these challenges effectively, ensuring seamless and secure data integration across all your systems.
    \item Managing data efficiently and competently is vital to any organization. Every data lifecycle stage must be handled with utmost care, from creation to consumption. This requires establishing clear policies and procedures for data lifecycle management. At DAT, we provide a holistic framework that helps maintain the quality dimensions for every phase of the data lifecycle.
    \item It's possible to create data quality checks for any data source at any point in the data lifecycle. This innovative method is significantly faster than the traditional approach, taking only a third of the time and requiring fewer resources. So, with PyDaQu, you can ensure the quality of your data more efficiently and effectively.
\end{itemize}

\section{Industry Evaluation}

Industrial Evaluations are a crucial step towards developing an effective tool for industries. With our focus on assessing the accuracy of DAT in describing real-world industrial scenarios, we involve companies in the evaluation process to gather their feedback. Our efforts have led to the successful modeling of five distinct cases for various companies. Thanks to the valuable feedback from these companies, we have improved our tool and made it more efficient for industrial usage. 

\subsection{Evaluating The Modeled Cases}

\begin{table*}[!ht]
\centering
\caption{Real Case Evaluation: Main Components}
\label{tab:daf_cases}
\makebox[\linewidth]{
\begin{tabular}{|l|l|l|l|l|l|l|}
\hline
  \multicolumn{1}{|c|}{\textbf{Cases}} &
  \multicolumn{1}{c|}{\textbf{\begin{tabular}[c]{@{}c@{}}Data \\ Formats\end{tabular}}} &
  \multicolumn{1}{c|}{\textbf{\begin{tabular}[c]{@{}c@{}}Processing \\ Type\end{tabular}}} &
  \multicolumn{1}{c|}{\textbf{\begin{tabular}[c]{@{}c@{}}Storage \\ Type\end{tabular}}} &
 
  \multicolumn{1}{c|}{\textbf{\begin{tabular}[c]{@{}c@{}} Data Life-cycle\\  Phases\end{tabular}}} &
  \multicolumn{1}{c|}{\textbf{\begin{tabular}[c]{@{}c@{}} Case \\ Complexity\end{tabular}}}\\ 
  \hline
\textbf{A} &
  \begin{tabular}[c]{@{}l@{}}Structured,   \\ Semi-structured\end{tabular} &
  \begin{tabular}[c]{@{}l@{}}RealTime\\  Batch\end{tabular} &
  \begin{tabular}[c]{@{}l@{}}NoSQL \\ NewSQL \\ File System\end{tabular} &
  \begin{tabular}[c]{@{}l@{}}Generation, Ingestion \\ Processing, Storing\\ Analyze, Visualize \\ Share\end{tabular} &
  \begin{tabular}[c]{@{}l@{}} 4 ( 48 ) \end{tabular}\\ \hline
\textbf{B} &
  \begin{tabular}[c]{@{}l@{}}Structured,   \\ Semi-structured\end{tabular} &
  \begin{tabular}[c]{@{}l@{}}RealTime\\  Batch\end{tabular} &
  \begin{tabular}[c]{@{}l@{}}NoSQL \\ NewSQL\\  File System\end{tabular} &
  \begin{tabular}[c]{@{}l@{}}Generation, Ingestion \\ Processing, Storing\\ Analyze, Visualize \\ Share\end{tabular} &
  \begin{tabular}[c]{@{}l@{}} 9 ( 72 ) \end{tabular}\\
 \hline
\textbf{C} &
  \begin{tabular}[c]{@{}l@{}}Structured,   \\ Semi-structured \\Unstructured \end{tabular} &
  \begin{tabular}[c]{@{}l@{}}RealTime\\  Batch\end{tabular} &
  \begin{tabular}[c]{@{}l@{}}NoSQL \\ NewSQL\\  File System\end{tabular} &
  \begin{tabular}[c]{@{}l@{}}Generation, Ingestion \\ Processing, Storing\\ Analyze, Visualize\end{tabular} &
  \begin{tabular}[c]{@{}l@{}} 7 ( 67 ) \end{tabular}\\ \hline
\textbf{D} &
  Semi-structured &
  Batch &
  \begin{tabular}[c]{@{}l@{}}NoSQL \\ NewSQL\\  File System\end{tabular} &
  \begin{tabular}[c]{@{}l@{}}Generation, Ingestion\\ Processing, Storing\\ Visualize\end{tabular} &
  \begin{tabular}[c]{@{}l@{}} 5 ( 34 ) \end{tabular}\\ \hline
\textbf{E} &
  \begin{tabular}[c]{@{}l@{}}Structured, \\ Un-structured \\Unstructured\end{tabular} &
  \begin{tabular}[c]{@{}l@{}}RealTime\\  Batch\end{tabular} &
  \begin{tabular}[c]{@{}l@{}}NoSQL \\ NewSQL\\  File System\end{tabular} &
  \begin{tabular}[c]{@{}l@{}}Generation, Ingestion \\ Processing, Storing\\ Analyze, Visualize \\ Share\end{tabular} &
  \begin{tabular}[c]{@{}l@{}} 8 ( 75 ) \end{tabular}\\ \hline
\end{tabular}
}
\end{table*}

The \textit{versatility and adaptability} of the framework in question have been proven through various real-case scenarios. Each case represents a unique set of data management challenges, from data formats and processing types to storage strategies and lifecycle phases.
Table \ref{tab:daf_cases} illustrates that the framework can provide architectural solutions that include structured, semi-structured, and unstructured data. Additionally, the framework's versatility is demonstrated by its ability to handle real-time and batch-processing methods, making it suitable for different data types and operational requirements. Furthermore, the framework provides comprehensive support for various storage strategies, including NoSQL, NewSQL, and traditional file systems, which enhances data retrieval and management.


DAT can ensure a holistic approach towards data management throughout its entire lifecycle - from creation to sharing. It is designed to tackle challenges of varying complexities, as demonstrated by the complexity scores in table \ref{tab:daf_cases}. This indicates our framework's scalability and robustness.

\subsection{Evaluating Addressed Challenges}

With DAT, you can rest easy knowing there is a compelling architectural solution to tackle DA challenges. By leveraging the powerful features of DAT, you can address the key challenges associated with data architecture,
as outlined in the previous section (related to the real case study). After conducting an in-depth analysis of various industrial cases, It became evident that utilizing the DAT can provide significant advantages to organizations.
DAT can provide:
\begin{itemize}
    \item Complex and scalable architecture solutions that could help model multiple data sources with different formats, speeds, and stages of the data life cycle, including different storage types and locations ( see table \ref{tab:daf_cases}). 
    \item The Hyder case is a perfect example of how batch and real-time processing can be integrated using Lambda architecture. By modeling this case, we can understand the significance of utilizing both approaches and how they can enhance the overall data processing capabilities.
    \item A common language, organizations can establish a standardized framework to facilitate seamless communication and comprehension of data assets by implementing DAT. This enables teams across regions to share and interpret data easily, promoting collaboration and improving outcomes.
    \item Data quality, The key to ensuring high-quality data lies in establishing rigorous \textit{standards and procedures} for data collection, management, and usage. Additionally, implementing quality checks can help ensure that data is accurate, complete, and consistent. By running \textit{data quality checkpoints}, you can automate processes and reduce the need for manual data processing. This can improve operational efficiency and productivity, saving valuable time and effort in development. 
\end{itemize}

In conclusion, the framework's proven \textit{ robustness and flexibility} in designing complex data environments make it a powerful solution for modern data-intensive industries. Its comprehensive approach to data format diversity, processing requirements, storage systems, and lifecycle management phases ensures that data is stored, processed, and presented efficiently, meeting the multifaceted demands of today's industries.

\chapter{Concluding Remarks and Future Work}
\thispagestyle{plain}

This research examines data architecture practices, data-intensive models, and quality monitoring. Its primary objective is to identify the challenges faced in developing data-intensive systems and propose an innovative approach to modeling and monitoring high-quality data-intensive applications.

This thesis outlines the research on data architecture and quality for data-intensive systems. The primary focus of the research was on the challenges related to data architecture and the development of a robust data architecture framework, which is a crucial step in creating data-intensive applications. Through an exploratory case study, the key challenges at each stage of the data architecture were identified.

Our research has brought to light the importance of enhancing data architecture practices. This realization has led us to develop a comprehensive framework for modeling data architecture. The framework offers a standardized protocol and specialized terminology for data architecture to facilitate effective communication among teams within and among organizations.

We created a state-of-the-art code generation framework for ensuring data quality in data-intensive applications. This framework effectively tackles the challenges that arise during the implementation and maintenance of such applications.

Finally, We validated our work by testing it on an industrial case from a different domain company. Our methods improved data quality during modeling and code generation.

\section{Key contributions}
The key contributions of this thesis are listed below.

\textbf{Objective 1: To model data-intensive applications}

\begin{itemize}
    \item Identification of the key challenges related to data architecture during the development of data-intensive models.    
    \item Identify the key elements for Data Architecture.
    \item Defines a meta-model for Data Architecture data-intensive applications.
    \item Develop graphical model language on top of the meta-model.
    \item Provide real industry cases as an evaluation for the meta-mode.
\end{itemize}

\textbf{Objective 2: To facilitate the monitoring of data quality}
    \begin{itemize}
        \item Identification of the key challenges related to \gls{dq}.
        \item Enhance the meta-model to contain  \gls{dqd}.
        
        \item Develop code generation framework to monitor the \gls{dq}.
    \end{itemize}

\section{Future Work}

In this dissertation, we explore the topic of data quality from two different perspectives. Firstly, we develop a model that incorporates various quality dimensions. Secondly, we integrate these dimensions into the code generation process to enable the monitoring of data quality. Our research emphasizes the importance of maintaining high-quality data in both business and academic settings and proposes a solution-oriented approach to address the challenges that arise.

Our future work aims to streamline the process of communicating data details and quality requirements from data consumers to data providers. Our solution will ensure that the provided data meets the necessary quality standards, thereby benefiting all parties involved.

\cleardoublepage
\addcontentsline{toc}{chapter}{bibliography}
\bibliography{include/3-back/bibliography}

\cleardoublepage
\appendix 
 \renewcommand\chaptername{Appendix}
\chapter{Appendix}

\begin{longtable}[c]{@{}|l|l|@{}}
\caption{Mapper Table from QD to GE}
\label{tab:mapper}\\
\toprule
\textbf{Quality   Dimensions} & \textbf{Great Expectations}                                 \\* \midrule
\endfirsthead
\multicolumn{2}{c}%
{{\bfseries Table \thetable\ continued from previous page}} \\
\toprule
\textbf{Quality   Dimensions} & \textbf{Great Expectations}                                 \\* \midrule
\endhead
\textbf{Uniqueness}           & expect\_column\_values\_to\_be\_unique                      \\* \midrule
\multirow{2}{*}{\textbf{Completeness}} & expect\_column\_values\_to\_not\_be\_null       \\* \cmidrule(l){2-2} 
                              & expect\_column\_values\_to\_be\_null                        \\* \midrule
\multirow{17}{*}{\textbf{Validity}}    & expect\_column\_values\_to\_be\_of\_type        \\* \cmidrule(l){2-2} 
                              & expect\_column\_values\_to\_be\_in\_type\_list              \\* \cmidrule(l){2-2} 
                              & expect\_column\_values\_to\_be\_in\_set                     \\* \cmidrule(l){2-2} 
                              & expect\_column\_values\_to\_not\_be\_in\_set                \\* \cmidrule(l){2-2} 
                              & expect\_column\_values\_to\_be\_between                     \\* \cmidrule(l){2-2} 
                              & expect\_column\_values\_to\_be\_increasing                  \\* \cmidrule(l){2-2} 
                              & expect\_column\_values\_to\_be\_decreasing                  \\* \cmidrule(l){2-2} 
                              & expect\_column\_distinct\_values\_to\_equal\_set            \\* \cmidrule(l){2-2} 
                              & expect\_column\_distinct\_values\_to\_contain\_set          \\* \cmidrule(l){2-2} 
                              & expect\_column\_mean\_to\_be\_between                       \\* \cmidrule(l){2-2} 
                              & expect\_column\_median\_to\_be\_between                     \\* \cmidrule(l){2-2} 
                              & expect\_column\_stdev\_to\_be\_between                      \\* \cmidrule(l){2-2} 
                              & expect\_column\_unique\_value\_count\_to\_be\_between       \\* \cmidrule(l){2-2} 
                              & expect\_column\_most\_common\_value\_to\_be\_in\_set        \\* \cmidrule(l){2-2} 
                              & expect\_column\_sum\_to\_be\_between                        \\* \cmidrule(l){2-2} 
                              & expect\_column\_min\_to\_be\_between                        \\* \cmidrule(l){2-2} 
                              & expect\_column\_max\_to\_be\_between                        \\* \midrule
\multirow{12}{*}{\textbf{Consistency}} & expect\_column\_value\_lengths\_to\_be\_between \\* \cmidrule(l){2-2} 
                              & expect\_column\_value\_lengths\_to\_equal                   \\* \cmidrule(l){2-2} 
                              & expect\_column\_values\_to\_match\_regex                    \\* \cmidrule(l){2-2} 
                              & expect\_column\_values\_to\_match\_regex\_list              \\* \cmidrule(l){2-2} 
                              & expect\_column\_values\_to\_match\_like\_pattern            \\* \cmidrule(l){2-2} 
                              & expect\_column\_values\_to\_not\_match\_like\_pattern       \\* \cmidrule(l){2-2} 
                              & expect\_column\_values\_to\_match\_like\_pattern\_list      \\* \cmidrule(l){2-2} 
                              & expect\_column\_values\_to\_not\_match\_like\_pattern\_list \\* \cmidrule(l){2-2} 
                              & expect\_column\_values\_to\_match\_strftime\_format         \\* \cmidrule(l){2-2} 
                              & expect\_column\_values\_to\_be\_dateutil\_parseable         \\* \cmidrule(l){2-2} 
                              & expect\_column\_values\_to\_be\_json\_parseable             \\* \cmidrule(l){2-2} 
                              & expect\_column\_values\_to\_match\_json\_schema             \\* \midrule
\multirow{6}{*}{\textbf{Timeliness}}   & expect\_column\_values\_to\_not\_be\_null       \\* \cmidrule(l){2-2} 
                              & expect\_column\_min\_to\_be\_between                        \\* \cmidrule(l){2-2} 
                              & expect\_column\_max\_to\_be\_between                        \\* \cmidrule(l){2-2} 
                              & expect\_column\_values\_to\_be\_in\_set                     \\* \cmidrule(l){2-2} 
                              & expect\_column\_values\_to\_not\_be\_in\_set                \\* \cmidrule(l){2-2} 
                              & expect\_column\_values\_to\_be\_between                     \\* \midrule
\multirow{34}{*}{\textbf{Accuracy}}    & expect\_column\_values\_to\_not\_be\_null       \\* \cmidrule(l){2-2} 
                              & expect\_column\_values\_to\_be\_null                        \\* \cmidrule(l){2-2} 
                              & expect\_column\_values\_to\_be\_of\_type                    \\* \cmidrule(l){2-2} 
                              & expect\_column\_values\_to\_be\_in\_type\_list              \\* \cmidrule(l){2-2} 
                              & expect\_column\_values\_to\_be\_in\_set                     \\* \cmidrule(l){2-2} 
                              & expect\_column\_values\_to\_not\_be\_in\_set                \\* \cmidrule(l){2-2} 
                              & expect\_column\_values\_to\_be\_between                     \\* \cmidrule(l){2-2} 
                              & expect\_column\_values\_to\_be\_increasing                  \\* \cmidrule(l){2-2} 
                              & expect\_column\_values\_to\_be\_decreasing                  \\* \cmidrule(l){2-2} 
                              & expect\_column\_value\_lengths\_to\_be\_between             \\* \cmidrule(l){2-2} 
                              & expect\_column\_value\_lengths\_to\_equal                   \\* \cmidrule(l){2-2} 
                              & expect\_column\_values\_to\_match\_regex                    \\* \cmidrule(l){2-2} 
                              & expect\_column\_values\_to\_not\_match\_regex               \\* \cmidrule(l){2-2} 
                              & expect\_column\_values\_to\_match\_regex\_list              \\* \cmidrule(l){2-2} 
                              & expect\_column\_values\_to\_not\_match\_regex\_list         \\* \cmidrule(l){2-2} 
                              & expect\_column\_values\_to\_match\_like\_pattern            \\* \cmidrule(l){2-2} 
                              & expect\_column\_values\_to\_not\_match\_like\_pattern       \\* \cmidrule(l){2-2} 
                              & expect\_column\_values\_to\_match\_like\_pattern\_list      \\* \cmidrule(l){2-2} 
                              & expect\_column\_values\_to\_not\_match\_like\_pattern\_list \\* \cmidrule(l){2-2} 
                              & expect\_column\_values\_to\_match\_strftime\_format         \\* \cmidrule(l){2-2} 
                              & expect\_column\_values\_to\_be\_dateutil\_parseable         \\* \cmidrule(l){2-2} 
                              & expect\_column\_values\_to\_be\_json\_parseable             \\* \cmidrule(l){2-2} 
                              & expect\_column\_values\_to\_match\_json\_schema             \\* \cmidrule(l){2-2} 
                              & expect\_column\_distinct\_values\_to\_equal\_set            \\* \cmidrule(l){2-2} 
                              & expect\_column\_distinct\_values\_to\_contain\_set          \\* \cmidrule(l){2-2} 
                              & expect\_column\_mean\_to\_be\_between                       \\* \cmidrule(l){2-2} 
                              & expect\_column\_median\_to\_be\_between                     \\* \cmidrule(l){2-2} 
                              & expect\_column\_stdev\_to\_be\_between                      \\* \cmidrule(l){2-2} 
                              & expect\_column\_unique\_value\_count\_to\_be\_between       \\* \cmidrule(l){2-2} 
                              & expect\_column\_most\_common\_value\_to\_be\_in\_set        \\* \cmidrule(l){2-2} 
                              & expect\_column\_sum\_to\_be\_between                        \\* \cmidrule(l){2-2} 
                              & expect\_column\_min\_to\_be\_between                        \\* \cmidrule(l){2-2} 
                              & expect\_column\_max\_to\_be\_between                        \\* \cmidrule(l){2-2} 
                              & expect\_column\_kl\_divergence\_to\_be\_less\_than          \\* \bottomrule
\end{longtable}

\cleardoublepage
\pagenumbering{gobble}

\end{document}